%% file: 2ndorder.tex
\documentclass[11pt,a4paper]{article}
\usepackage{jcappub}
\usepackage{times}
\usepackage{bm}
\usepackage{import}
\usepackage{xifthen}
\usepackage{pdfpages}
\usepackage{transparent}
\usepackage{graphicx}
\usepackage{float}
\usepackage{multicol}
\usepackage{amsmath}
\usepackage{mathtools}
\usepackage{multirow}
\usepackage{dcolumn}
\usepackage{longtable}
\usepackage{comment}
\usepackage{caption}

\newcommand{\up}[1]{{\rm #1}}

\newcommand{\etc}{\noindent $\bullet$ }

\newcommand{\Ietc}[1]{\etc \textit{\textbf{#1}}.---}

\newcommand{\Dquad}{\qquad\qquad}

\newcommand{\nnn}{\nonumber \\}
\newcommand{\nn}{\nonumber }
\newcommand{\beeq}{\begin{equation}}
	\newcommand{\eneq}{\end{equation}}
\newcommand{\bear}{\begin{eqnarray}}
	\newcommand{\enar}{\end{eqnarray}}

\newcommand{\LL}{\mathcal{L}}    % spatial gauge transformation
\renewcommand{\AA}{\mathcal{A}}  % delta g_00 component
\newcommand{\BB}{\mathcal{B}}    % delta g_oa component
\newcommand{\CC}{\mathcal{C}}    % delta g_ab component
\newcommand{\rbar}{\bar r}       % comoving line-of-sight distance

\newcommand{\BBP}{\BB_\parallel}

\newcommand{\AAP}{\AA_{,\parallel}}
\newcommand{\CCPP}{\CC_{\parallel\parallel}}

\newcommand{\UU}{u}              % four velocity
\newcommand{\VV}{\mathcal{U}}    % spatial component of four velocity
\newcommand{\dUU}{\delta\UU}     % perturbation in 0-component of four velocity
\newcommand{\VL}{\mathcal{V}}

\newcommand{\zz}{z}                 % observed redshift
\newcommand{\ttt}{\theta}           % observed angle theta
\newcommand{\pp}{\phi}              % observed angle phi
\newcommand{\CG}{\hat g}         % conformally transformed metric
\newcommand{\CK}{\hat k}         % conformally transformed wavevector
\newcommand{\cc}{\lambda}        % conformally transformed affine parameter
\newcommand{\dea}{{\delta n}}    % delta e        
\newcommand{\eP}{\dea_\parallel} % delta e parallel
\newcommand{\dnu}{\delta\nu}     % delta_nu
\newcommand{\NN}{n}              % observed photon direction in FRW
\newcommand{\NC}{\mathbb{C}}     % normalization amplitude
\newcommand{\eTT}{\dea_\ttt}     % tangential direction
\newcommand{\ePP}{\dea_\pp}      % tangential direction
\newcommand{\dhnu}{\widehat{\Delta\nu}}

\newcommand{\deaP}{\dea_\parallel}

\newcommand{\drr}{\delta r}     % radial distortion
\newcommand{\dtt}{\delta\ttt}   % angular distortion
\newcommand{\dpp}{\delta\pp}    % angular distortion

\newcommand{\dT}{\delta\eta}    % time lapse 
\newcommand{\dz}{\delta z}      % lapse in observed redshift
\newcommand{\HH}{\mathcal{H}}   % conformal Hubble
\newcommand{\Dcc}{\Delta\cc}    % difference in affine parameters
\newcommand{\DT}{\Delta\eta}    % perturbation to tau w.r.t. obs z

\newcommand{\DX}{\Delta x}      % perturbation to x w.r.t. obs z

\newcommand{\dV}{\delta V}           % volume distortion
\newcommand{\dD}{\mathcal{D}}        % fluctuation
\newcommand{\ddL}{\delta\mathcal{D}_L} % its perturbation
\newcommand{\dA}{\mathcal{D}_A}      % physical luminosity distance
\newcommand{\DD}{\mathbb{D}}         % deformation matrix
\newcommand{\dDD}{\delta\DD}         % 2nd order part
\newcommand{\dg}{\delta g}           % perturbation of metric determinant
\newcommand{\GG}{\delta\hat\Ga}% conformally transformed Gamma^a_{bc}k^bk^c

\newcommand{\OO}{\mathcal{O}}

\newcommand{\mM}{\mathcal{M}}

\newcommand{\al}{\alpha}
\newcommand{\be}{\beta}
\newcommand{\ga}{\gamma}
\newcommand{\de}{\delta}

\newcommand{\si}{\sigma}

\newcommand{\Om}{\Omega}
\newcommand{\Ga}{\Gamma}
\newcommand{\Si}{\Sigma}

\newcommand{\para}{\parallel}
\newcommand{\pa}{\partial}

\newcommand{\bobs}{\bar {\rm o}}

\begin{document}

	\begin{titlepage}
		
		\setcounter{page}{1} \baselineskip=15.5pt \thispagestyle{empty}
		\pagenumbering{roman}
		
		\bigskip
		
		\vspace{1cm}
		\begin{center}
			{\fontsize{20}{28}\selectfont \bfseries Second-order gauge-invariant formalism for the cosmological observables:\\ Complete verification of their gauge-invariance}
		\end{center}
		
		\vspace{0.2cm}
		
		\begin{center}
			{\fontsize{13}{30}\selectfont Matteo Magi$^a$ and Jaiyul Yoo $^{a,b}$ }
		\end{center}
		
		\begin{center}
			\vskip 8pt
			\textsl{$^a$ Center for Theoretical Astrophysics and Cosmology,
				Institute for Computational Science}\\
			\textsl{University of Z\"urich, Winterthurerstrasse 190,
				CH-8057, Z\"urich, Switzerland}
			
			\vskip 7pt
			
			\textsl{$^b$Physics Institute, University of Z\"urich,
				Winterthurerstrasse 190, CH-8057, Z\"urich, Switzerland}
			
			\vskip 7pt
			
			\today
			
		\end{center}
		
		\makeatletter{\renewcommand*{\@makefnmark}{}
			\footnotetext{matteo.magi@uzh.ch,~~ jyoo@physik.uzh.ch}\makeatother}

		\vspace{1.2cm}
		\hrule \vspace{0.3cm}
		\noindent {\sffamily \bfseries Abstract} \\[0.1cm]
		Accounting for all the relativistic effects, we have developed the fully nonlinear gauge-invariant formalism for describing the cosmological observables and presented the second-order perturbative expressions associated with light propagation and observations without choosing a gauge condition. For the first time, we have performed a complete verification of the validity of our second-order expressions by comparing their gauge-transformation properties from two independent methods: one directly obtained from their expressions in terms of metric perturbations and the other expected from their nonlinear relations. The expressions for the cosmological observables such as galaxy clustering and the luminosity distance are invariant under diffeomorphism and gauge-invariant at the observed position. We compare our results to the previous work and discuss the differences in the perturbative expressions. Our second-order gauge-invariant formalism constitutes a major step forward
		in the era of precision cosmology and its applications in the future will play a crucial role for going beyond the power spectrum and probing the early universe.

		\vskip 10pt
		\hrule
		
		\vspace{0.6cm}
	\end{titlepage}
	
	\noindent\hrulefill \tableofcontents \noindent\hrulefill

	\pagenumbering{arabic}
	
	\newpage
	\section{Introduction}
	Thanks to tremendous technological and observational advances in recent years, 
	we have witnessed an era of precision cosmology, and an unprecedented amount of
	data in large-scale surveys allows us to measure important statistics and extract accurate information 
	about our universe. Furthermore, this trend will accelerate in the coming years, 
	making the near future of cosmology particularly promising. The upcoming large-scale
	surveys on the ground include the Dark Energy Spectroscopic Instrument (DESI) \cite{desi}, the Vera
	C. Rubin Observatory (formerly LSST) \cite{lsst}, the Square Kilometre Array (SKA) \cite{ska}, and the next generation CMB experiment
	CMB-S4 \cite{cmb-4}, and several space-based missions will be operational such as
	Euclid \cite{euclid}, the Nancy Grace Roman Space Telescope (formerly WFIRST) \cite{wide} and the Laser Interferometer Space Antenna (LISA) \cite{LISA}. These surveys will dramatically improve 
	our understanding of cosmology and fundamental physics, thanks to an unprecedented amount of high precision data. Cosmological perturbations on large scales originate 
	from random fields sourced by quantum fluctuations in the early universe. These 
	fields assume random values at each spatial point according to their corresponding 
	probability measure and cosmological information is encoded in the $N$-point 
	correlation functions. In order to take full advantage of the new opportunities 
	in the forthcoming era of precision cosmology, the theoretical predictions for 
	the $N$-point correlation functions should be made at least as accurate as 
	the level of precision set by the upcoming surveys.
	
	The standard inflation model predicts a nearly Gaussian distribution in the early 
	universe \cite{juan}. While the non-Gaussianity we observe in large-scale structure largely
	reflects the nonlinear evolution of gravity, a small amount of the primordial 
	non-Gaussianity might be hidden in our measurements (see, e.g., \cite{Planck:2019kim}). 
	To describe the deviation from Gaussianity, it is necessary to go beyond the 
	linear order in perturbation theory to capture statistics higher than the power
	spectrum. The simplest measure of non-Gaussianity is the three-point correlation 
	function, which corresponds to the bispectrum in Fourier space. The most simple 
	type of non-Gaussianity is the local one (see, e.g., \cite{2004PhR...402..103B} for a review), 
	which peaks in the squeezed limit in Fourier space. Such a squeezed bispectrum is 
	particularly important because it breaks the degeneracy in galaxy bias \cite{2008PhRvD..77l3514D,2008ApJ...677L..77M} and 
	satisfies a particular consistency relation to the power spectrum \cite{juan,creminelli,fizpatrick}. 
	Since the standard inflationary model predicts a vanishingly small non-Gaussianity,
	a detection of such primordial non-Gaussianity could potentially rule out a vast 
	class of inflationary models \cite{gianmassimo,wen}. However, this exciting possibility can 
	only be realized, if we control the systematic errors in our theoretical 
	predictions below the level of signals we are interested\,in.
	
	Cosmological perturbation theory has been extensively studied 
	in literature (see, e.g., \cite{tomita,tomita2,tomita3,nakamura,bruni,hwang,noh}), and it solves the Einstein equation up 
	to the desired order in perturbations, after a gauge choice is often made 
	(see \cite{hwang,noh,2017JCAP...10..027G} for the calculations in a general metric representation). 
	In cosmological observations, there exists another step that relates the physical
	solutions to the observations. Cosmological information is mostly carried by
	photons (or gravitational waves), and they propagate along the past light cone to reach us in our rest frame.
	For example, galaxy clustering is dominated by the matter density fluctuation,
	which is a solution by the Einstein equation, but is not directly what we measure.
	The light propagation and observations in the observer rest frame give rise 
	to extra contributions to galaxy clustering such as the redshift-space distortion, 
	gravitational lensing effects, the gravitational redshift, and other relativistic
	effects (\cite{sachswolfe,kaiser1,kaiser2,2014CQGra..31w4001Y}, see \cite{yoo09} for a complete
	treatment). A full relativistic description of galaxy clustering at the linear
	order was developed \cite{YFZ,yoo10,ruth11,chandler,jeong}. For the same reason, the
	relativistic effects are present in other cosmological probes such as the
	luminosity distance \cite{sasaki,sugiura,
		pyne,kibble,hui, 
		2006PhRvD..73b3523B,2012PhRvD..86b3510D,2015MNRAS.450..883K,biern1,biern2,fulvioyoo}.
	Given the relativistic nature of weak gravitational lensing and cosmic microwave
	background (CMB) anisotropies, their theoretical descriptions were largely 
	complete from the inception for weak lensing \cite{gunn1,gunn2, 1991ApJ...380....1M, 
		kaiser2, 1998MNRAS.301.1064K} and for CMB \cite{sachswolfe,
		bond} (see, e.g., \cite{1998PhRvD..57.3290H,dodelson, 2003AIPC..666...45H, 2006PhR...429....1L} for a review).
	However, extra relativistic contributions were found even at the linear order
	in perturbations for weak gravitational lensing \cite{2003PhRvL..91b1301D, 
		2008PhRvD..78l3530B,2012PhRvD..86h3513S,2012PhRvD..86h3527S, 2012MNRAS.426.1121C, 
		2013JCAP...08..051Y,2016JCAP...01..024A, 2018JCAP...07..067G, thepaper, 2021PhRvD.104h3548G} and for CMB
	\cite{2008PhRvD..78l3529Z,2019arXiv190509288Y, sandra}.
	
	Given these recent developments, it is therefore natural to develop the
	second-order relativistic description of the cosmological observables. With the high 
	precision measurements of CMB anisotropies, the theoretical descriptions of CMB beyond
	linear order have been developed in early days \cite{cmb0,cmb0.5,cmb1,cmb2,cmb3,cmb4,cmb5}, playing
	a critical role in constraining the primordial non-Gaussianity and 
	deriving the lensing potential (see, e.g., \cite{2020A&A...641A...9P,2020A&A...641A...8P}),
	though only the linear-order deviation in the light propagation
	from a straight path is needed for describing the second-order CMB anisotropies. The second-order description of light
	propagation was employed in weak lensing formalism \cite{lens0,lens1,lens2,lens3,lens4,lens5,lens6} and also
	the luminosity distance \cite{ld0,ld1,ld2,ld3,ld4,ld6,ld7,ld8,ld9}. The second-order relativistic description 
	of galaxy clustering and its applications to the bispectrum were recently 
	developed in \cite{YZ,enea,gc0,gc1,gc2,gc3,gc4,gc5,gc6,gc7}, 
	in which the second-order light propagation and
	observations in the rest frame were carefully accounted for.
	
	Despite all the efforts and developments in recent years, there exists little consensus in the second-order
	calculations. Although the extension of the linear-order formalism to the 
	second-order one is conceptually straightforward, it involves remarkably complex expressions and subtleties, making it extremely challenging to track down the
	differences and to make a meaningful comparison. A further complication arises
	from gauge issues. The theory of general relativity is manifestly covariant; 
	however, splitting variables into a background part and perturbations breaks this 
	covariance by introducing a dependence on the choice of coordinates. It is common 
	practice to fix a gauge to work with more manageable expressions. However, by doing 
	so, one loses the ability of checking the gauge-invariance of the final results
	for the cosmological observables, i.e., once gauge freedom is fully fixed, 
	all the resulting expressions are automatically gauge-invariant, not necessarily because 
	the expressions of the cosmological observables are correct, but simply because 
	gauge freedom was removed by hand. In this paper, we work with a general metric 
	representation without fixing a gauge. Only then, it is possible to exploit the geometric
	nature of the defined quantities and check that they gauge-transform correctly 
	under an arbitrary change of coordinates. In the past this consistency test has 
	been performed only for the linear-order formalism.
	
	Drawing on the previous work, here we develop a nonlinear theoretical framework 
	to describe the cosmological observables, present their second-order 
	perturbative expressions, and verify the gauge-transformation properties 
	of the second-order expressions. Adopting a geometric approach to describing 
	the light propagation and observations and using the tetrad formalism \cite{tetrad},
	we describe the observations in our rest frame and physical quantities in the
	source rest frame, and their connection through light propagation was computed 
	in a nonlinear way. Thereby, we can clarify the separation
	between the background quantities and the nonlinear perturbative quantities
	in the theoretical descriptions of the cosmological observables.
	By carefully expanding all the nonlinear expressions along the background light path and around the coordinate-independent reference positions of the observer and the 
	source parameterized by the observables, we derive the second-order expressions 
	of the cosmological observables that contain linear terms in the second-order 
	perturbation variables and quadratic terms of the linear-order perturbation 
	variables. The perturbation contributions at the observer position
	are typically ignored in previous work, but their presence is crucial to ensure the 
	correctness of the expressions \cite{biern1,thepaper,yoolightcone,2020PhRvR...2c3004M,sandra}. In particular, they couple to other perturbation
	variables at second order in perturbations, acquiring the positional dependence.
	Finally, we compute the gauge transformations of all the second-order expressions,
	which are derived directly from the gauge transformation of their metric components
	in the expressions. Utilizing the nonlinear expressions of those, we compute the 
	nonlinear (and second-order) expectations for the gauge transformations and verify 
	the consistency in all of our expressions. Again, we stress that these sanity checks 
	are a necessary step to establish the complete second-order formalism in the era of 
	precision cosmology and they constitute one of the main results in the present work.
	
	The paper is structured as follows. Section $\ref{sec:nonlinear}$ introduces the nonlinear geometric formalism needed to treat all relativistic effects associated with light propagation in observation from a distant source to the observer. In Section \ref{sec:second}, we present the second-order perturbative expressions of the cosmological observables in Section $\ref{sec:nonlinear}$. The main expressions we derive are the observed redshift fluctuations and the physical volume and area occupied by the source. In Section $\ref{sec:gauge}$ we explicitly verify that the perturbative expressions of Section $\ref{sec:second}$ gauge-transform according to the expectation derived from the definitions in Section $\ref{sec:nonlinear}$. Section $\ref{sec:compa}$ compares our expressions with previous work in literature, and in Section $\ref{sec:concl}$, we summarize and discuss our results. Moreover, Appendix $\ref{AppendixA}$ is reserved to further technical calculations related to the observer’s position, while Appendix $\ref{AppendixB}$ displays the second-order expressions in their complete form.

	\section{Nonlinear Expressions for Cosmological Observables}
	\label{sec:nonlinear}
	
	We present nonlinear expressions for cosmological observables obtained by the light propagation from the source to the observer position. In particular, we provide expressions for the observer and source positions, the observed redshift, and the area and volume occupied by the source; these will then be written in terms of nonlinear perturbative and exact background parts. The nonlinear perturbative part can later be expanded order-by-order. This section aims to lay the groundwork necessary to perform rigorous second-order perturbative calculations, which are the main focus of this work and are presented in Section $\ref{sec:second}$. Here we follow the general approach in \cite{YZ,thepaper} with nonlinear generalization in coordinates (see also \cite{tetrad}).

	\subsection{Geometric setup}
	
	We begin by introducing fundamental geometric quantities and stating our notation conventions. We denote with $\bar\mM$ the Friedmann-Lemaître-Robertson-Walker (FLRW) spacetime with metric $\bar g$, it describes a homogeneous isotropic universe and is an excellent approximation to our universe on cosmological scales. We denote with $\mM$ the spacetime describing the real inhomogeneous universe with metric $g$, which deviates perturbatively from the FLRW background metric. The coordinate system we use is that of FLRW, in which the temporal coordinate is the cosmic time $t$, or more commonly the conformal time $\eta$, while the spatial coordinates are the comoving coordinates $x^\al$. We recall that with this choice, the fixed-time sections of $\bar\mM$ are maximally symmetric, and the comoving coordinates stretch according to the Hubble flow.
	
	We adopt the following convention for the components of the background metric:
	\bear
	\bar g_{\eta\eta}(\eta)=-a^2(\eta) \,,\Dquad \bar g_{\al\be}(\eta)= a^2(\eta) \de_{\al\be}\,,
	\enar
	where $\de_{\al\be}$\footnote{We use Greek letters $\al, \be,\cdots$ as indices that can take values from one to three, while $\mu, \nu,\cdots$ are indices that can take values from zero to three.} is the Kronecker delta and $a(\eta)$ is the scale factor, containing the dynamics of the background universe. Note that the metric signature is mostly-plus, and we are restricting to the case of a flat universe with Cartesian comoving coordinates.
	Our convention for the components of the metric of the real inhomogeneous universe is
	\bear\label{metrica}
	g_{\eta\eta}(\eta,\bm x)&=&-a^2(\eta) \left( 1 + 2\AA(\eta,\bm x) \right)~,\Dquad 
	g_{\eta\al}(\eta,\bm x)=-a^2(\eta) \BB_\alpha(\eta,\bm x)~,
	\nnn
	g_{\al\be}(\eta,\bm x)&=&a^2(\eta)\left( \de_{\alpha\beta}+ 2 \CC_{\alpha\beta}(\eta,\bm x)\right)~.
	\enar
	where spacetime fields $\AA,\, \BB_\al,\, \CC_{\al\be}$ are nonlinear perturbations to the background metric, and they are based on the flat 3-metric $\de_{\al\be}$. These perturbations represent the inhomogeneities of the real universe and if they vanish we recover $\bar g_{\mu\nu}$ the components of the background metric.
	
\begin{figure}[H]
		\centering
		\def\svgwidth{0.5\columnwidth}
		\resizebox{0.55\textwidth}{!}{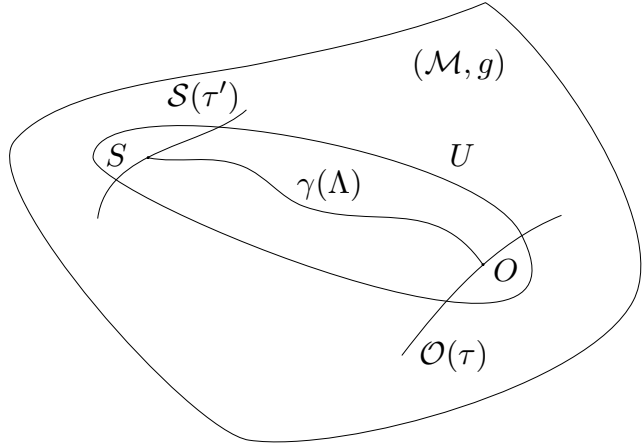}
		\caption{Sketch of the geometric configuration considered in the real inhomogeneous universe. The source follows the time-like world line $\mathcal{S}(\tau')$, and the observer follows the time-like world line $\OO(\tau)$, with $\tau'$ and $\tau$ proper times. Today the observer is at point $O$ and sees the source at light emission point $S$ via the null light path $\ga$. $U$ is the neighborhood in which we give coordinates to the curves with the chart $(U,\psi)$.}
		\label{fig:illustration}
	\end{figure}

	Geometrically, we describe the trajectories of the source and the observer with the curves $\mathcal{S}$ and~$\OO$, respectively. We introduce a curve $\gamma$ to describe the light path intersecting the other two curves at $S$ and $O$. These points denote the position of the source at light emission time and the observer's position today. We affinely parameterize the curves $\mathcal{S}$ and $\OO$ with $\tau'$ and $\tau$, proper times of the source and the observer, such that $\mathcal{S}(\tau')$ is any point on the source world line and~$\mathcal{O}(\tau)$ is any point on the observer world line. To parameterize the curve $\gamma$ we introduce the affine parameter $\Lambda$, which for the moment is arbitrary.
	Moreover, we assume that $\mathcal{S}$ and $\OO$ are time-like curves, while~$\gamma$ is a null geodesic.
	We now consider a neighborhood $U$ of the image of $\gamma$ and assume that there exists~$(U,\psi)$ chart of $\mM$. So, we can express the curves $\OO$, $\mathcal S$ (restricted to $U$), and $\gamma$ in coordinates as
	\bear\label{chart}
	x^\mu(\tau):=(\psi^\mu\circ\OO)(\tau)\,,\qquad x^\mu(\tau'):=(\psi^\mu\circ\mathcal S)(\tau')\,,\qquad x^\mu(\Lambda):=(\psi^\mu\circ\ga)(\Lambda)\,.
	\enar
	Figure $\ref{fig:illustration}$ summarizes what has been introduced so far.
	
	It is convenient to repeat the whole construction at the background level. For this purpose, we consider what we just introduced in the real universe $\mM$ and set the perturbations of the metric~$g$ to zero. Hence, the real position of the observer today reduces to $\bar O$, and the real position of the source at emission time reduces to $\bar S$. Since the geometry of $\bar\mM$ is different from that of $\mM$, the curves introduced will also \textit{differ} in the two manifolds; however, it is still possible to adopt the same parameterizations since these procedures do not depend on the target spaces.
	By assigning coordinates to these background curves in the same way as before, we define the deviations of physical trajectories from the background trajectories as follows
	\bear
	\label{deviations}
	\de x^\mu(\tau):=x^\mu(\tau)-\bar x^\mu(\tau)\,,\,\quad\,
	\de x^\mu(\tau'):=x^\mu(\tau')-\bar x^\mu(\tau')\,,\,\quad\,
	\de x^\mu(\Lambda):=x^\mu(\Lambda)-\bar x^\mu(\Lambda)\,,~
	\enar
	where the overbar denotes the coordinates of the curves defined in the homogeneous and isotropic spacetime $\bar\mM$.
	
	We now narrow our attention to the light path $\ga(\Lambda)$. At each point of this path, through the chart $(U,\psi)$, we have a vector basis $\{\pa_\mu\}$, with which we can represent the components of any tensor. However, the components written in this basis have no physical meaning, as they transform each time we change coordinate system. The coordinate-induced vector basis is not suitable for describing what the observer is measuring. In fact, the measurements made by the observer occur in the privileged system of the laboratory or observatory, where the metric is Minkowski, and the outcome of such measurements does not depend on the coordinate choice of the FLRW manifold. One way to geometrically define the observer rest frame is to define a tetrad basis $\{e_a\}$, with Latin indices\footnote{
		We use Latin letters to denote internal indices, while Greek letters are for spacetime indices. Hence, a quantity with Latin indices is not involved in FLRW coordinate transformation, while a quantity with Greek indices is. Our notation convention is: $\al,\be,\cdots$ and $i,j,\cdots$ run from one to three, while $\mu,\nu,\cdots$ and $a,b,\cdots$ run from zero to three.}
	$a,b,\cdots$ running from zero to three. This frame is fixed in the tangent space of $\mM$, such that tensor components in this basis are independent of the coordinate system. Moreover, the tetrad basis is orthonormal by definition, and the components of the metric tensor in this basis are Minkowski $g(e_a,e_b)=\eta_{ab}$. We conclude that the tetrad basis has all the characteristics needed to describe the observer's measurements; hence we equip the point $O$ with such a frame. We also equip the point $S$ with a tetrad basis to define quantities in the source rest frame. Actually, we will need a tetrad basis at any point of $\ga(\Lambda)$ to describe fictitious observations made by fictitious observers along the light path.\footnote{To be more general one can imagine fictitious observers with tetrad vectors everywhere in the universe and describe the physical events as they could be measured by a fictitious observer in the rest frame.}
	The physical specification of the tetrad basis will be presented in the following subsection because we have not yet introduced all the necessary ingredients.
	
	Let us introduce our notation conventions. We indicate the coordinates of a path, either in the real or background universe, with the affine parameter as a subscript, e.g., $x^\mu_\Lambda$ and~$\bar x^\mu_\Lambda$ stand respectively for $x^\mu(\Lambda)$ and $\bar x^\mu(\Lambda)$. 
	To indicate the coordinates of a specific point along a path, we display as a subscript the lowercase letter corresponding to the point, e.g., $x^\mu_o$ are the coordinates of the observer position $O$, parameterized for example by $\Lambda_o$. These notations also apply to deviations in a path, such that $\de x^\mu_\Lambda$ stands for $\de x^\mu(\Lambda)$ and $\de x^\mu_o$ are the coordinate perturbations at the observer position. For the coordinates of specific points along the background path, we will use roman subscripts to emphasize that these coordinates refer to the homogeneous and isotropic universe, e.g., $\bar x^\mu_{\rm o}$, but note that the point $\bar O$ is still parameterized by $\Lambda_o$. To denote the evaluation of a field, $k^\mu(x)$, at a specific point of a path, we use the same subscript of the coordinates of the point, e.g., $k^\mu_o$ stands for $k^\mu(x_o)$. If the evaluation is on the background path, we additionally put an overbar on the roman subscript to expressly refer to $\bar x^\mu$, e.g., $k^\mu_{\bobs}$ stands for $k^\mu(\bar x_{\rm o})$.
	Finally, we use a prime to indicate partial derivatives with respect to conformal time and a dot to express partial derivatives with respect to cosmic time.

	\subsection{Observer four-velocity and space-like tetrad vectors}
	
	We now introduce a vector field $u$ on $\mM$, which we can express in coordinates via the chart $(U,\psi)$ as
	$u=u^\mu(x)\pa_\mu$. This expression is valid over the whole set $U$, but we are interested in its restriction along the time-like trajectory of the observer $\OO$.
	We want this vector field along the curve to describe the observer's four-velocity, so we impose that the following equations hold at each point of the observer's world line:
	\bear\label{udef}
	u=\frac{d}{d\tau}=\frac{dx^\mu(\tau)}{d\tau}\frac{\pa}{\pa x^\mu}\,,\Dquad u_\mu u^\mu=-1\,.
	\enar
	If we further assume that the observer path is a geodesic, then the observer's acceleration $a^\mu$ vanishes:
	\bear\label{geou}
	a^\mu:=\UU^\nu\nabla_\nu \UU^\mu=\UU^\nu\pa_\nu\UU^\mu+\Ga^\mu_{\rho\si}
	\UU^\rho\UU^\si=0\,,
	\enar
	where $\nabla_\mu$ is the covariant derivative compatible with $g$ and $\Ga^\mu_{\rho\si}$ are the Christoffel symbols specifying the Levi-Civita connection. At the background level, the only non-vanishing component of the observer four-velocity is the temporal one, which is a consequence of the symmetries of the FLRW solution. In the real inhomogeneous universe, we can describe the deviations from the background by introducing the nonlinear perturbations $\de u$ and $\VV^\al$ to the observer four-velocity
	\bear\label{uu}
	u^\mu=:\frac1a\left(1+\de u,\,\VV^\al\right)^\mu\,,
	\enar
	with $\VV^\al$ based on $\de_{\al\be}$ (see \cite{YZ} for notation convention).
	Given this expression, we readily derive
	\bear\label{delubasso}
	u_\eta&=&g_{\eta\mu}u^\mu=-a(1+\de u+2\AA+2\AA\de u+\BB_\al\VV^\al)\,,
	\nnn
	u_\al&=&g_{\al\mu}u^\mu=a(\VV_\al-\BB_\al-\de u\BB_\al+2\CC_{\al\be}\VV^\be)=:a \VL_\alpha\,,
	\enar
	where we defined the nonlinear perturbation $\VL_\al$.
	The normalization condition of the time-like vector~$u_\mu u^\mu=-1$ constrains $\de u$ and $\VV^\al$ to the metric perturbations:
	\bear\label{delu}
	0=\de u+\AA(1+2\de u)+\frac12\de u^2(1+2\AA)+\BB_\al\VV^\al(1+\de u)
	-\frac12\VV_\al\VV^\al(1+2\CC_{\al\be})\,.
	\enar
	Using Eq.\,($\ref{uu}$) the geodesic condition in Eq.\,($\ref{geou}$) can be explicitly written as
	\bear
	a^\eta=0=\de\Ga^{\eta}_{\eta\eta}\left( 1+\dUU \right)^2+\left(2\Ga^{\eta}_{\eta\al}\VV^{\al}+\dUU'\right)\left( 1+\dUU \right)+\Ga^{\eta}_{\al\be}\VV^{\al}\VV^{\be}+\VV^{\al}\dUU_{,\al} \,,
	\enar
	for the temporal component, and
	\bear
	a^\al=0=\Ga^{\al}_{\eta\eta}\left( 1+\dUU \right)^2+\left( 2\Ga^{\al}_{\eta\be}\VV^{\be}+\VV'^{\al}-\HH\VV^{\al} \right)\left( 1+\dUU \right)+\Ga^{\al}_{\be\ga}\VV^{\be}\VV^{\ga}+\VV^{\be}\VV^{\al}{}_{,\be} \,,
	\enar
	for the spatial components.
	Among all the Christoffel symbols in the geodesic condition above, only~$\Ga^\eta_{\eta\eta}$ has a non-vanishing background contribution, such that $a^\mu=0$ at the background. Consequently, the terms like $\Ga^\al_{\be\ga}u^\be u^\ga$ are at least third order in perturbations. With $\de\Ga^\eta_{\eta\eta}$ we denote the nonlinear perturbation $\de\Ga^\eta_{\eta\eta}:=\Ga^\eta_{\eta\eta}-\HH$, where $\HH=a'/a$ is the conformal Hubble parameter.
	
	We are now in a position to resume the discussion about tetrads (see \cite{tetrad} for the tetrad formalism in cosmology). In the previous subsection, we stated that we want to equip the null light path~$\ga$ with tetrad basis vectors at each point; however, there are many possible ways to do this, actually infinite. Indeed, if $\{e_a\}$ is a tetrad basis at a point and we make a change of basis (not a change of coordinates) to arrive at $\{e'_a\}$, then it can be proved that $\{e'_a\}$ is still a tetrad basis only if the transformation made belongs to the Lorentz group. We recall that the finite transformations in the Lorentz group are three independent boosts and three independent spatial rotations. The physical way to fix the freedom in boosts is to align the tetrad frame with the observer trajectory, namely $e_0\equiv u$, so that~$e_0$ is time-like and $\{e_i\}_{i=1,2,3}$ are space-like. We fix the residual rotation freedom by imposing that at the background level the space-like tetrad vectors are aligned parallel to the FLRW spatial directions. Expressing the tetrads in terms of coordinates~$e_a=[e_a]^\mu\pa_\mu$, with the components $[e_a]^\mu$ depending only on the coordinates of the point in which the tetrad is specified, we can fix
	\bear\label{tetrade}
	[e_0]^\mu\equiv u^\mu\,,\Dquad [e_i]^\mu=:\frac1a\left( \de e^\eta_i,\de^\al_i+\de e^\al_i \right)^\mu\,,
	\enar
	where we introduced the nonlinear perturbations $\de e^\eta_i$ and $\de e^\al_i$ \cite{thepaper}.
	
	We can express the metric $g$ both in the coordinate-induced basis and in the tetrad basis; therefore, by relating the components in the two bases, we derive the following equation
	\bear
	g_{\mu\nu}=\eta_{ab}[e^a]_\mu [e^b]_\nu\,,
	\enar
	where $[e^a]_\mu$ is the inverse of $[e_a]^\mu$: $[e^a]_\mu [e_b]^\mu=\de^a_b~$ , $~[e^a]_\mu [e_a]^\nu=\de^\mu_\nu$. Combining these relations we obtain
	\bear\label{orto}
	g_{\mu\nu}[e_a]^\mu[e_b]^\nu=\eta_{ab}\,.
	\enar
	We can use this orthonormality condition to constrain the perturbations in the components $[e_a]^\mu$, but note that this will not constrain $\de e^\al_i$ completely. For this reason, we adopt the following exact splitting
	\bear\label{split}
	\de e^\al_i =:-\de^\al_j(S^j{}_i+A^j{}_i)\,,
	\enar
	with $S_{ij}$ and $A_{ij}$ nonlinear perturbations based on $\de_{ij}$, which are respectively symmetric and antisymmetric under the exchange of spatial indices.
	We stress that only the symmetric part $S_{ij}$ is constrained in terms of metric perturbations via the orthonormality condition \cite{thepaper,tetrad}.

	\subsection{Light wave vector and physical observables}
	
	As in the case of the observer four-velocity, we now define a vector field $k$, which we express in coordinates as $k=k^\mu(x)\pa_\mu$. We consider the restriction of this vector field along the light path $\ga$
	\bear
	k=k^\mu(x_\Lambda)\pa_\mu\,,
	\enar
	and we demand that it describes a null vector at each point:
	\bear
	\label{kp}
	k=\frac{d}{d\Lambda}=\frac{dx^\mu(\Lambda)}{d\Lambda}\frac{\pa}{\pa x^\mu}\,,\Dquad k_\mu k^\mu=0\,.
	\enar
	The requirement that $k$ is not any null vector, but the observed light wave vector, uniquely specifies the affine parameter $\Lambda$ and ensures that the components of $k$ are related to physical observables.
	
	The observer in the rest frame makes cosmological measurements, which we describe geometrically in terms of the tetrad basis at point $O$. The components of the light wave vector in this reference are
	\beeq\label{kobs}
	\begin{dcases}
		k=k^a e_a\,,\\
		k^a\equiv\omega(1,-n^i)^a\,,\Dquad \de_{ij}n^in^j=1\,,
	\end{dcases}
	\eneq
	where we suppressed the subscript $o$. With $n^i$ we denote the angular position of the source in the sky of the observer and the minus sign indicates that it points from the observer to the source. With~$\omega$, or better $\omega_o$, we denote the angular frequency of the light received by the observer; in coordinates it reads $\omega=[e^0]_\mu k^\mu=-u_\mu k^\mu$. To obtain the components of $k$ in coordinates, we use the tetrad components, i.e., $k^\mu=[e_a]^\mu k^a$. Using the results from the previous subsection, we obtain that at the background level, the components of the wave vector are given by
	\bear\label{kbg}
	\bar k^\mu:=\frac{d\bar x^\mu(\Lambda)}{d\Lambda}=\frac{\omega}{a}\left( 1,-n^\al \right)^\mu\,,\Dquad n^\al:=n^i\de^ \al_i\,.
	\enar
	Note that this expression is valid at any point of the background light path. However, only the angular frequency $\omega$ and the scale factor $a(\eta)$ change from point to point, meaning from observer to observer. While we align the internal frames of these fictitious observers such that $n^\al$ is the same at every point of the background light path in the background manifold $\bar\mM$.
	
	In practice, it is more convenient to introduce a conformal wave vector $\hat k$, as it simplifies equations and does not alter physical conclusions. We will discuss the details of conformal transformations in the following subsection.

	\subsection{Conformal light wave vector}
	
	Let us introduce on $\mM$ an additional metric $\hat g$, which is related to the metric $g$ of the real inhomogeneous universe by the conformal transformation	\bear\label{confo}
	g_{\mu\nu}(\eta,\bm x)=a^2(\eta)\hat g_{\mu\nu}(\eta,\bm x)\,.
	\enar
	Notice that the conformal metric $\hat g$ at the background level is the Minkowski metric.
	In the rest of the work we will use $\mM$ as a shorthand for $(\mM,g)$ and $\hat\mM$ for $(\mM,\hat g)$.
	
	The convenience in introducing a conformal transformation is that we can deal with a more straightforward metric, while preserving the causal structure of $\mM$. In fact, if a vector is time-like, space-like, or null with respect to the metric $g$, it will maintain this same property for the metric~$\hat g$. One can say even more, namely, if two Lorentzian manifolds have the same causal structure, then the metrics must be related by a conformal transformation \cite{wald}.
	However, what interests us most is yet another related property of conformal transformations that we now discuss.
	Let $\nabla$ be the affine connection compatible with the metric~$g$, and~$\hat\nabla$ the affine connection compatible with the metric $\hat g$, then we consider a curve $c(\Lambda)$ and a tangent vector $V$. We assume that~$c(\Lambda)$ is an affinely-parameterized geodesic on $\mM$ hence the tangent vector satisfies
	\bear\label{geo}
	\nabla_V V=0\,.
	\enar
	Relating the two covariant derivatives we can see that, in general, $c(\Lambda)$ is not a geodesic in $\hat\mM$. There is an exception in the case of a null geodesic, for which the above equation reduces to \cite{wald}
	\bear\label{nonaffine}
	\hat \nabla_V V=(-2  V^\mu\nabla_\mu \ln a) V\,,
	\enar
	where the scale factor $a$ arises from the conformal transformation in Eq.\,($\ref{confo}$).
	Notice that this equation is of the form
	\bear
	\hat \nabla_V V=f V\,,\Dquad f:=-2  V^\mu\nabla_\mu \ln a=\frac d{d\Lambda}\ln \left(a^{-2}\right)
	\enar
	which describes a path whose tangent vector changes parallelly to itself. This feature defines a geodesic, however the above equation looks different from Eq.\,($\ref{geo}$) due to the parameterization of the curve. If we reparametrize $\Lambda\to \cc$ such that
	\bear
	\frac{d^2\cc}{d\Lambda^2}=f\frac{d\cc}{d\Lambda}\,,
	\enar 
	is satisfied, then we would recover the affinely-parametrized geodesic equation Eq.\,($\ref{geo}$) for $c(\cc)$ with re-scaled tangent vector $(d\Lambda/d\cc)V$. In this sense, null geodesics are conformally invariant.
	
	We now apply all this machinery to the light path $\ga(\Lambda)$ which is a geodesic on $\mM$ with tangent vector $k^\mu$. We then introduce a new parameter $\cc$, such that $\ga(\cc)$ is an affinely parameterized geodesic on $\hat\mM$ with rescaled tangent vector $\hat k^\mu$ defined as
	\beeq\label{khatnohat}
	\begin{dcases}
		k^\mu=\hat k^\mu \frac{d\cc}{d\Lambda}\,,\\
		\hat k^\mu:=\frac{dx^\mu(\cc)}{d\cc}\,,
	\end{dcases}
	\eneq
	where $\hat k^\mu$ satisfies $\hat k^\nu\hat\nabla_\nu \hat k^\mu=0$ and $\hat k_\mu\hat k^\mu=0$.	
	By direct substitution in Eq.\,($\ref{nonaffine}$) we find that $d\cc/d\Lambda$ can be obtained up to a proportionality constant $\NC$ that cannot be constrained by the conformal transformation: \cite{wald}
	\bear
	\label{conforme}
	\frac{d\Lambda}{d\cc}=\NC a^2\,.
	\enar
	Notice that having physically fixed the parameterization $\Lambda$ does not fix the parameterization $\cc$. It is essential to understand that $k^\mu$ is a vector under diffeomorphism in the real inhomogeneous universe~$\mM$ while $\hat k^\mu$ is not, exactly because the Jacobian of the line reparameterization is a function of spacetime coordinates. We will elaborate more on this point in Section $\ref{subsec:gauge}$.
	
	We can now express Eq.\,($\ref{kbg}$) in terms of the conformal wave vector in the background:
	\bear\label{kbarracappuccio}
	\hat{\bar k}^\mu=\NC a\omega\left( 1,-n^\al \right)^\mu\,.
	\enar
	Our choice to fix the constant $\NC$ is to set unity
	\bear\label{NC}
	1\equiv\NC (a\omega)_o\,,
	\enar
	at the observer position.
	The motivation behind this choice is that this combination has the same value at every point of the background light path. We explicitly show this by evaluating the combination~$\NC(a\omega)_{\bobs}$ at the background observer position $\bar O$. First, we normalize the scale factor at the observer position in the background to unity $a(\bar\eta_{\rm o})\equiv1$, and the frequency redshifts to yield $\omega_{\bobs}=(a\omega)_{\bar{\rm p}}$, where $\bar P$ is any point along the background light path. Remember that $\omega$ is the "observed frequency" in the rest frame of a fictitious observer along the light path. Hence, we derived~$\NC(a\omega)_{\bobs}=\NC(a\omega)_{\bar{\rm p}}$. In the presence of perturbations, the product $(a\omega)$ varies from point to point, while $\NC$ remains constant in $\mM$.
	It proves convenient to account for changes in the product~$(a\omega)$ at any point $P$ of the light path by introducing another nonlinear perturbation $\dhnu$, defined as follows:
	\bear\label{dhnu}
	\NC(a\omega)_p=-\NC(au_\mu k^\mu)_p=-(\hat u_\mu\hat k^\mu)_p=:1+\dhnu_p\,.
	\enar
	We introduced the conformal four-velocity components $\hat u^\mu:=au^\mu$, and we used $\omega=-u_\mu k^\mu$ derived in the previous subsection. In these terms, our choice to fix the arbitrary constant $\NC$ at the observer position can be rephrased as $\dhnu_o\equiv 0$, independent of the coordinate system. We stress that the condition ($\ref{NC}$) is evaluated at the observer position $O$ in the real universe, not at $\bar O$ in the background universe. This conformal transformation has been widely used in literature \cite{sasaki, YFZ,YZ, thepaper,jeong,gc0,gc1,gc2}. Another choice is to impose the normalization condition at the source position $\dhnu_s\equiv0$.
	
	We can now discuss the components $\hat k^\mu$ of the conformal wave vector. Combining Eq.\,($\ref{NC}$) and Eq.\,($\ref{kbarracappuccio}$) we introduce the nonlinear parameterization
	\bear\label{hatk}
	\CK^\mu=:\left(1+\dnu~,-n^\al-\dea^\al\right)^{\mu}\,,
	\enar
	defining $\dnu$ and $\dea^\al$ nonlinear perturbations, with $\dea^\al$ based on $\de_{\al\be}$.
	Using the above definition, together with the normalization condition in Eq.\,($\ref{NC}$), and the relation between the two wave vectors in Eq.\,($\ref{khatnohat}$), we obtain the useful nonlinear parameterization for the light wave vector:
	\bear\label{kmu}
	k^\mu=\omega_o\frac{a_o}{a^2}\left( 1+\dnu,-n^\al-\dea^\al \right)^\mu\,.
	\enar
	At the observer position we can derive the perturbations $\dnu_o$ and $\dea^\al_o$ in terms of the tetrad components of Eq.\,($\ref{tetrade}$). From $\hat k^\mu=\NC a^2 e^\mu_a k^a$ we obtain the boundary conditions: \cite{thepaper}
	\bear\label{ko}
	\dnu_o=\dUU_o-(\de e^\eta_i)_on^i~,\qquad
	-\dea^\al_o=\VV^\al_o-n^i(\de e^\al_i)_o
	=\VV^\al_o+\de^\al_j\left(S^j{}_i+A^j{}_i\right)_on^i~,
	\enar
	where not only the symmetric part but also the antisymmetric part contributes. Mind again the notation convention
	\bear
	\dnu_o=\dnu(x_o) \,,\Dquad \dea^\al_o=\dea^\al(x_o)\,.
	\enar
	Using the conformal metric $\CG$, we can compute the exact expressions for $\hat k_\mu=\hat g_{\mu\nu}\hat k^\nu$. By direct substitutions, we derive
	\bear
	\CK_\eta&=&
	-\left(1+2\AA+\dnu-\BBP+2\AA\dnu-\BB_\al\dea^\al\right)~,\nnn
	\CK_\al&=&
	-\left(n_\al+\dea_\al+2\CC_{\al\para}+\BB_\al+\dnu~\BB_\al+
	2\CC_{\al\be}\dea^\be\right)~,
	\enar
	where we introduced the notation that each index contracted with $n^{\al}$ is replaced by $\para$ (parallel to $n^\al$). In turn, the exact null equation is
	\bear\label{null}
	\hat k_\mu\hat k^\mu=0&=&\dnu-\deaP+\AA-\BBP-\CCPP
	-\dnu\BBP-2\CC_{\para\al}\dea^\al+2\AA\dnu-\BB_\al\dea^\al+\frac12\dnu^2
	\nnn
	&&
	-\frac12\dea_\al\dea^\al+\AA\dnu^2-\dnu\BB_\al\dea^\al-\CC_{\al\be}\dea^\al\dea^\be ~.
	\enar
	We conclude by assigning coordinates to the null path parameterized by $\cc$ in the same way we did earlier in Eqs.\,($\ref{chart}$) and ($\ref{deviations}$):
	\bear x^\mu(\cc):=(\psi^\mu\circ\ga)(\cc)\,,\Dquad
	\de x^\mu(\cc):=x^\mu(\cc)-\bar x^\mu(\cc)\,.
	\enar

	\subsection{Light path geodesic equations}
	
	We have motivated that the curve $\ga$ parameterized by $\cc$ is a null geodesic in the conformal manifold$~\hat\mM$. Now we want to express the geodesic equations in coordinates. 
	The temporal and spatial components are 
	\bear\label{kgeo}	0&=&\CK^\mu\hat{\nabla}_{\mu}\CK^\eta{}={d\over d\cc}\dnu+\hat\Ga^\eta_{\mu\nu}\CK^\mu
	\CK^\nu\,,\Dquad
	0=\CK^\mu\hat{\nabla}_{\mu}\CK^\al{}=
	-{d\over d\cc}\dea^\al+\hat{\Ga}^\al_{\mu\nu}\CK^\mu\CK^\nu\,,
	\enar
	where $\hat\nabla_\mu$ is the covariant derivative based on $\hat g$, $\hat\Ga^\mu_{\nu\rho}$ are the Christoffel symbols
	\bear\label{gammahat}
	\hat\Ga^\mu_{\nu\rho}=\frac12 \CG^{\mu\si}\left(\CG_{\si\nu,\rho}+\CG_{\rho\si,\nu}-\CG_{\nu\rho,\si}  \right)\,,
	\enar
	of the corresponding Levi-Civita connection, and the derivative with respect to the affine parameter~$\cc$ is obtained from the parameterization of $\hat k^\mu$ in Eq.\,(\ref{hatk})
	\bear\label{ddlambda}
	\frac{d}{d\cc}=\hat k^\mu \frac{\pa}{\pa x^\mu}=\left(1+\dnu \right)\frac{\pa}{\pa\eta}-\left(n^\al+\dea^\al \right)\frac{\pa}{\pa x^\al}\,.
	\enar
	It is clear from the definition of $n^\al=n^i\de^\al_i$ in Eq.\,($\ref{kbg}$) that any coordinate derivative of $n^\al$ is zero as it does not depend on spacetime coordinates.
	
	Since $\hat g_{\mu\nu}$ in the background is chosen to be the Minkowski metric $\eta_{\mu\nu}$ in Cartesian coordinates, it automatically follows that the Christoffel symbols $\hat\Ga^\mu_{\nu\rho}$ are vanishing in the background. Hence, we can define the following perturbations that source the geodesic equations ($\ref{kgeo}$):
	\bear\label{GGt}
	\GG^\eta:=\hat{\Ga}^\eta_{\mu\nu}\CK^\mu\CK^\nu=\hat\Ga^\eta_{\eta\eta}(1+\dnu)^2
	-2\hat\Ga^\eta_{\al\eta}(1+\dnu)(n^\al+\dea^\al)
	+\hat\Ga^\eta_{\al\be}(n^\al+\dea^\al)(n^\be+\dea^\be)\,,~~~~~~~
	\enar
	\bear\label{GGa}
	\GG^\al:=\hat{\Ga}^\al_{\mu\nu}\CK^\mu\CK^\nu=\hat{\Ga}^\al_{\eta\eta}(1+\dnu)^2-2\hat{\Ga}^\al_{\eta\be}(1+\dnu)
	(n^\be+\dea^\be)
	+\hat{\Ga}^\al_{\be\ga}(n^\be+\dea^\be)(n^\ga+\dea^\ga)\,.~~~~~~~
	\enar
	We emphasize that $\hat{\Ga}^\al_{\be\ga}=0$ in the background given our choice of coordinates, otherwise $\GG^\al$ has extra contributions in proportion to $\hat{\bar{\Ga}}^\al_{\be\ga}$.
	Next, we integrate the geodesic equations to derive a formal solution:
	\bear\label{kint}
	\dnu(x_\cc)-\dnu(x_o)&=&-\int_{0}^{\cc}d\cc'\,\GG^\eta(x_{\cc'})\,,\,\qquad\,
	\dea^\al(x_\cc)-\dea^\al(x_o)=\int_{0}^{\cc}d\cc'\,\GG^\al(x_{\cc'})\,,~~~~~~~
	\enar
	where we set the affine parameter $\cc$ equal to zero at the observer position $\ga(\cc_o)=O$, $\cc_o\equiv0$.

	\subsection{Time lapse and spatial shift at the observer position}
	
	In the real universe, the observer position $x^\mu(\tau)$ drifts away from the background position $\bar x^\mu(\tau)$ due to the metric perturbations and peculiar velocity $\VV^\al$.
	We define the time lapse and spatial shift of the observer today by evaluating the first relation in Eq.\,($\ref{deviations}$) at the observer position $O$ parameterized by $\tau_o$, i.e., $\OO(\tau_o)=O$:
	\bear\label{lapseshift}
	\de t(\tau_o):=t(\tau_o)-\bar t(\tau_o)\,,\Dquad\de x^\al(\tau_o):=x^\al(\tau_o)-\bar x^\al(\tau_o)\,.
	\enar
	We will always work in cosmic time $dt=a d\eta$ when dealing with the time lapse and spatial shift (see \cite{ld9}).
	
	The relation between the vector field $u$ and the associated curve $\OO$ is given by the differential equations for the integral curve, which in coordinates read
	\bear\label{curvint}
	\frac{d x^\mu(\tau)}{d\tau}=u^\mu(x_\tau) \,,\qquad \forall \tau\in[0,\tau_o]\,,\qquad x^\mu(0)=0\,,
	\enar
	which we can formally integrate
	\bear
	x^\mu(\tau_o)=\int_0^{\tau_o}d\tau\,u^\mu(x_\tau)\,,
	\enar
	to obtain the time lapse and spatial shift at the observer
	\bear\label{lapseandshift}
	\de t(\tau_o)=\int_0^{\tau_o}d\tau\,\de u(x_\tau)\,,\Dquad\de x^\al(\tau_o)=\int_0^{\tau_o}d\tau\,\frac1a\VV^\al(x_\tau)\,.
	\enar
	The components of $u^\mu$ in cosmic time are given by
	\bear\label{ucosm}
	u^\mu=
	\left(1+\de u,\,\frac1a\VV^\al\right)^\mu\,,
	\enar
	and we used the solution in the background universe
	\bear\label{dt}
	\bar t_{\rm o}=\int_0^{\tau_o} 
	d\tau=\tau_o=\int_0^\infty{dz\over H(z)(1+z)}\,.
	\enar
	Combining the last equation with the definition of $u$ in Eqs.\,($\ref{udef}$) and ($\ref{ucosm}$), the background path can be parameterized by the cosmic time as
	\bear
	\bar x^\mu_\tau=(\tau=t,0)^\mu=:\bar x^\mu_t\,.
	\enar
	We stress again that Eq.\,($\ref{lapseandshift}$) are formal solutions, as they require the knowledge of the whole trajectory $x^\mu(\tau)$ to obtain $\de x^\mu(\tau_o)$.

	\subsection{Observed redshift and parameterization of the source position}
	
	Using the tetrad basis at the source position $S$, we can define the emitted light frequency $\omega_s$, and compare it with the frequency $\omega_o$ that the observer measures in the rest frame. It is well known that these two frequencies are not equal because the cosmic expansion will stretch the wavelength of the emitted signal as it travels towards the observer. The observed redshift $z$ quantifies the discrepancy between these two frequencies:
	\bear
	1+z:=\frac{\omega_s}{\omega_o}\,.
	\enar
	We can combine this definition with $\dhnu$ in Eq.\,($\ref{dhnu}$) for $P=S$, and with the normalization condition for $\NC$ in Eq.\,($\ref{NC}$) to obtain the nonlinear expression for the observed redshift
	\bear
	1+z=\frac{a_o}{a_s}\left(1+\dhnu_s\right)\,.
	\enar
	We define the nonlinear perturbation $\de z$ in the observed redshift as
	\bear\label{dzdz}
	\frac{1+\de z}{a_s}:=1+z\,.
	\enar
	At the background level, the redshift quantifies how much the universe has expanded since the light left the source at the given scale factor $a_s$ evaluated at the coordinate position $\eta_s$ of the source. While the observed redshift $z$ is well-defined, our choice of splitting the background part and the perturbation $\de z$ is not unique. Note that the coordinate position of the source is different in an inhomogeneous universe from that in the background, and it is also gauge-dependent. Furthermore, we remark that $z$ is a scalar, but $\de z$ does not have a well-defined geometric nature.
	
	Combining the last two equations, we obtain the nonlinear expression for the perturbation $\de z$ in the observed redshift
	\bear\label{dz}
	\de z=a_o\left(1+\dhnu_s\right)-1\,,
	\enar
	and the nonlinear expression for $\dhnu_s$ from its definition in
	Eq.\,($\ref{dhnu}$) is
	\bear\label{dhnuexact}
	\dhnu_s&=&2\AA+\VV_{\para}-\BB_{\para}+(1+2\AA)\left(\dUU+\dnu+\dUU\dnu\right)+\dea^\al\left( \VV_{\al}-\BB_{\al} \right)+\BB^{\al}\VV_{\al}(1+\dnu)
	\nnn
	&&
	-\dUU(\BB_{\para}+\BB^\al\dea_{\al})
	+2\VV^{\al}(\CC_{\al \para}+\CC_{\al\be}\dea^{\al})\,.
	\enar
	We stress again that the right-hand side is evaluated at $x^\mu_s$ but there is also a dependence of $\dhnu_s$ on the reference position $x^\mu_o$ via $\dnu$ and $\dea^\al$; thus, we will use the notation $\dhnu(x_s;x_o)$. This argument also applies to $\dnu(x_s;x_o)$, $\dea^\al(x_s;x_o)$, $\de z(x_s;x_o)$ and $z(x_s;x_o)$.
	
	To illustrate the significance of the observed redshift, we now turn our attention to the relation between $\hat k$ and $\gamma$, given by the differential equations for the integral curve
	\bear\label{sourcepos}
	\frac{d x^\mu(\cc)}{d\cc}=\hat k^\mu(x_\cc) \,,\qquad \forall \cc\in[0,\cc_s]\,,\qquad x^\mu(0)= x^\mu_o\,.
	\enar
	For the moment, we consider the background level, so by integrating the differential equation, we obtain
	\bear
	\bar\eta_{\cc}-\bar\eta_{ \rm o}=\cc\,,\Dquad  \bar x^\al_{\cc}=-\cc n^\al\,,
	\enar
	which shows how the affine parameter is related to the time coordinate of the background light path. However, since coordinates are defined up to diffeomorphisms, there exists ambiguity. We can use the observed redshift $z$ to define an unambiguous time coordinate $\bar\eta_z$, along the light path in the background: \cite{YZ}
	\bear\label{etaz}
	1+z=:\frac{1}{a(\bar\eta_z)}\,,
	\enar
	and this time coordinate is associated to the affine parameter $\cc_z$ defined by
	\bear\label{dr}
	\bar\eta_z-\bar\eta_{ \rm o}=:\cc_z=-\int_0^z\frac{dz'}{H(z')}=-\rbar_z\,,\qquad \bar x^\al_z:=-\cc_zn^\al=\rbar_z n^\al\,,
	\enar
	where $\rbar_z$ is the comoving distance out to redshift $z$ in the
	background. Noting that the observed redshift and the position $\bar x^\mu_z$ are independent of coordinates (or gauge choice), we will use this physical parameterization to denote the reference position of the source in the real universe:
	\beeq\label{xz}
	\eta_s=\bar\eta_{\rm s}+\delta\eta_s=:\bar\eta_z+\Delta\eta_s~,\qquad
	x^\al_s=\bar x^\al_{\rm s}+\delta x^\al_s=:\bar x^\al_z+\DX^\al_s~,
	\qquad \cc_s=:\cc_z+\Dcc_s~,
	\eneq
	where the subscript $z$ indicates evaluation at $\cc_z$, and the subscript $s$ at $\cc_s$. Note that with lowercase delta $\de$ we indicate perturbations around the background source position parameterized by $\cc_s$, while we use uppercase delta $\Delta$ to indicate perturbations around the background source position parameterized physically by the observed redshift. These two types are in general different e.g. $\de\eta_s\neq\DT_s$.
	We want to stress that the parameter $\cc_z$, unlike $\cc_s$, has been introduced to parameterize the light path in the background, where the source position $\bar S$ is identified precisely by the scalar $z$. If we want to use the parameterization $\cc_z$ on $\mM$, we need to account for the perturbation $\Dcc_s$ to identify $S$.
	
	With these definitions, we can obtain useful nonlinear relations. At the source position, we derive
	\bear\label{Dxcc}
	\DX^\mu_s=\de x^\mu_s+\bar x^\mu_{\rm s}-\bar x^\mu_z=\de x_s^\mu+\Dcc_s(1,-n^\al)^\mu\,,
	\enar
	where we used $\bar x^\mu_{\rm s}=\left( \cc_s+\bar\eta_{\rm o},-\cc_sn^\al \right)^\mu$ and $\bar x^\mu_z=\left( \cc_z+\bar\eta_{\rm o},-\cc_zn^\al \right)^\mu$.
	While at the observer position, we obtain the boundary relations
	\bear
	\label{z0}
	\zz(x_o;x_o)=0~,\qquad
	\Dcc_o=0~,\qquad \DT_o=\dT_o\neq0~,\qquad 
	\DX_o^\al=\delta x^\al_o\neq0~.
	\enar
	Using the relation between $\de z$ and $\bar\eta_z$ in Eqs.~\eqref{etaz} and~\eqref{dzdz}, we derive the exact relation
	between $\DT_s$ and~$\dz$:
	\bear
	\label{dzrel}
	1+\dz={a(\bar\eta_z+\DT_s)\over a(\bar\eta_z)}~.
	\enar
	In addition to the observed redshift, we can use the observed angular position $n^i$, parameterized by the angles $\theta,\phi$, to fully specify the real location of the source in terms of quantities defined in the rest frame of the observer. These angles are related to $n^i$ by the following expression in Cartesian coordinates
	\bear
	n^i=(\sin\theta\cos\phi,\sin\theta\sin\phi,\cos\theta)^i\,.
	\enar
	We also define two unit vectors perpendicular to $n^i$ in the rest frame of the observer:
	\bear\label{angoli}
	\ttt^i&:=& \frac{\partial}{\partial\theta}n^i=\left( \cos\ttt\cos\pp,\cos\ttt\sin\pp,-\sin\ttt \right)^i\,,
	\nnn
	\pp^i&:=&\frac1{\sin\theta} \frac{\partial}{\partial\phi} n^i=\left(-\sin\pp,\cos\pp,0\right)^i\,,
	\enar
	which can be defined on $\mM$ in the same way we did for $n^\al$, i.e., $\ttt^\al:=\de^\al_i\ttt^i$ and $\pp^\al:=\de^\al_i\pp^i$.
	The parameterization we adopt for the source position is then \cite{YZ}
	\bear\label{source}
	x^\mu_s=\bar x^\mu_z+\DX^\mu_s\equiv x^\mu(z,\theta,\phi)&=&\bigg(\bar\eta_z+\Delta\eta_s\,,\, \left(\rbar_z+\drr\right)\sin\left( \ttt+\dtt \right)\cos\left( \pp+\dpp \right) ,
	\nnn
	&&
	\left(\rbar_z+\drr\right)\sin\left( \ttt+\dtt \right)\sin\left( \pp+\dpp \right) ,
	\nnn
	&&
	\left(\rbar_z+\drr\right)\cos\left( \ttt+\dtt \right)\bigg)^\mu\,,
	\enar
	with $\DT_s$, $\drr$, $\dtt$, $\dpp$ nonlinear perturbations that are functions of $(z,\theta,\phi)$. For the same reason above, the perturbations $(\de r,\dtt,\dpp)$ should be denoted as $(\Delta r_s,\Delta\ttt_s,\Delta\pp_s)$ but we keep this notation convention as there is no possibility of confusion in this case.
	We stress again that $z,n^i,\ttt^i,\pp^i$ are observables, so their values do not depend on our coordinate-based description, and they are invariant under diffeomorphisms of $\mM$.

	\subsection{Distortion in the source position compared to the observed position}
	
	Let us consider again the relation between the conformal wave vector $\hat k$ and the light path $\gamma$ given in Eq.\,($\ref{sourcepos}$). 
	Isolating the perturbative parts, it reads
	\bear\label{dxder}
	\frac{d}{d\cc}\de x^\mu=(\dnu,-\dea^\al)^\mu\,,
	\enar
	valid at any point along the light path.
	Integrating this equation yields the formal solution
	\bear\label{dx}
	\delta x^\mu_s&=&\left(\dT_s~,~\delta x^\al_s\right)^\mu=\left(\de\eta_o+\int_0^{\cc_s} d\cc~\dnu~,~\de x_o^\al-\int_0^{\cc_s} d\cc~
	\dea^\al\right)^\mu\\
	&=&\left(\dT_o+\lambda_s~\dnu_o-
	\int_0^{\cc_s} d\cc(\cc_s-\cc)\GG^\eta~,
	~\de x^\al_o-\cc_s~\dea^\al_o-\int_0^{\cc_s} d\cc(\cc_s-\cc)\GG^\al\right)^\mu\,,
	\nn
	\enar
	where we used the solution for $\dnu$ and $\dea^\al$ in Eq.\,($\ref{kint}$).
	Combining the last expression with Eq.\,($\ref{Dxcc}$), we derive the following nonlinear expressions for the distortion in the source position with respect to $\bar x^\mu_z$:
	\bear\label{De}
	\DT_s&=&\dT_o+\Dcc_s+\cc_s~\dnu_o-
	\int_0^{\cc_s} d\cc(\cc_s-\cc)\GG^\eta(x_\cc)~,\\
	\label{Dea}
	\DX^\al_s&=&\de x^\al_o-\Dcc_s n^\al-
	\cc_s~\dea^\al_o-\int_0^{\cc_s} d\cc(\cc_s-\cc)\GG^\al(x_\cc)~.
	\enar
	Solving Eq.\,($\ref{De}$) for $\Dcc_s$, and utilizing the solution for $\dea^\al$ in Eq.\,($\ref{kint}$), we derive a more convenient nonlinear expression for the spatial distortion at the source position:
	\bear\label{DXal}
	\DX_s^\al=\de x_o^\al-n^\al(\DT_s-\de\eta_o)-\int_0^{\cc_s}d\cc\left( \dea^\al -n^\al\dnu\right)\,,
	\enar
	where we kept the expression in a compact form by leaving unexpressed the perturbations $\dea^\al$ and$\,\dnu$. Note that $\cc_s$ in the integration limit also has a perturbative part $\Dcc_s$.
	
	\subsection{Physical volume and area occupied by the source}
	\label{subsec:vol}
	So far, we have discussed how the observer can infer the position of a point-like light source via the observed redshift $z$ and the observed angles $\ttt , \pp$. Now we turn our attention to the case of some extended light source subtended by the observed solid angle $\up d\Omega$, over the observed redshift range~$\up dz$. Our interest is to provide expressions for the proper volume and area in terms of the observed quantities in the rest frame of the observer. The former is related to galaxy clustering, and the latter is related to the luminosity distance. Perturbative calculations have been performed in the past \cite{ld0,ld1,ld2,ld3,ld4,ld6,ld7,ld8,ld9,ultimo,gc0,gc1,gc2,gc3,gc4,gc5,gc6,gc7}
	
	Although we are considering an extended source, it can still be geometrically described by the point $S$ on the manifold $\mM$. Indeed, we can define the volume and area in terms of a volume form $\Si$, i.e., a top-form defined at the source position $S$. Having introduced a tetrad basis $\left\{ e_a\right\}$ at the source position, we can define the volume form as the following exterior product:
	\bear\label{forma}
	\Si:=e^0\wedge e^1\wedge e^2\wedge e^3\,,
	\enar
	where $\left\{ e^a\right\}$ is the dual basis of one-forms satisfying $e^a(e_b)=\de^a_b$. We use the tetrad components $[e^a]_\mu$ to recast this volume form at point $S$ in FLRW coordinates, and then using Eq.\,($\ref{source}$), we relate the coordinates of the source position to the observed redshift and angles. Note that the parameterization in Eq.\,($\ref{source}$) is valid at point $S$, when the proper time of the source is $\tau'=\tau_s$. Hence, we derive the volume form in coordinates \cite{Weinberg:1972kfs}
	\bear
	\Si=\sqrt{-g}~\varepsilon_{\rho\mu\nu\si}~u^\rho_s~{\partial x^\mu_s\over
		\partial\zz}{\partial x^\nu_s\over\partial\ttt}
	{\partial x^\si_s\over\partial\pp}~d\tau_s\wedge d\zz\wedge d\ttt \wedge d\pp\,,
	\enar
	where the Levi-Civita symbol is normalized as $\varepsilon_{\eta xyz}=1$, the (square root) of the metric determinant
	\bear\label{detg}
	\sqrt{-g}:=\sqrt{-\up{det}\,g_{\mu\nu}}=:a_s^4(1+\de g)\,,
	\enar
	arises to express the volume form $\Si$ in the FLRW basis $\left\{dx^\mu\right\}$ at the source position,
	$u^\mu_s$ are the components of the source velocity $u_s=d/d\tau_s$, and $d\tau_s,dz,d\ttt,d\pp$ are the one-forms built from the corresponding parameters, and we defined the dimensionless perturbation $\de g$.
	Considering small intervals $\up dz ,\up d\ttt, \up d\pp$ and fixed proper time $\tau_s$, we define the three-dimensional volume element $dV$ as the components of the volume form:
	\bear\label{volume}
	\Si\left({\pa\over\pa\tau_s},\up dz{\pa\over\pa z},\up d\ttt{\pa\over\pa\ttt},\up d\pp{\pa\over\pa\pp}\right)=:\up dV=\sqrt{-g}~\varepsilon_{\rho\mu\nu\si}~u^\rho_s~{\partial x^\mu_s\over
		\partial\zz}{\partial x^\nu_s\over\partial\ttt}
	{\partial x^\si_s\over\partial\pp}~\up dz~\up d\ttt~\up d\pp\,,
	\enar
	where we use the roman $\up d$ to denote intervals, while the italics $d$ denoted one-forms. Note that the expression above is a scalar under diffeomorphisms: $\up dz,\up d\ttt, \up d\pp$ are observables, so by construction, they are not affected by coordinate transformations, then $u_s^\mu$ and $x^\mu_s$ transform as components of four-vectors and their transformation is compensated by that of the scalar density $\sqrt{-g}$. Hence, Eq.\,($\ref{volume}$) indeed describes the physical volume occupied by the source in its rest frame, that appears subtended by $(\up dz,\up d\ttt,\up d\pp)$ in the observer rest frame.
	
	Given this definition of physical volume, there are many ways to introduce an area by considering different decompositions of $\up dV$. Here we define the physical area as the split of $\up dV$ into an element $\up dA$ \textit{perpendicular} to the light propagation in the rest frame of the source. To derive the expression in FLRW coordinates of the light propagation, let us consider the light wave vector $k$ at the source position. Since at point $S$ there are both a tetrad basis $\left\{e_a\right\}$ and a coordinate-induced basis~$\left\{\pa_\mu\right\}$, the following equality holds
	\bear
	k^ae_a=k=k^\mu\pa_\mu\,,
	\enar
	where we suppressed the subscript $s$. Substituting the expression for the internal components of the light wave vector $k^a$ in Eq.\,($\ref{kobs}$), we obtain the propagation direction at the source position
	\bear\label{propag}
	n^\mu_s:=-{k^\mu_s\over\omega_s}+u^\mu_s=\frac1{a_s}\left(-{\hat k^\mu_s\over1+\dhnu_s}+\hat u^\mu_s\right)\,,\qquad (n^\mu n_\mu)_s=1\,, \qquad (n^\mu u_\mu)_s=0\,,
	\enar
	where we additionally made use of the prescription at the source position $[e_0]^\mu\equiv u^\mu$, and via Eq.\,($\ref{dhnu}$) we derived the second equality involving conformal four-vectors. The directional vector is normalized and spatial in the source rest frame. With this expression at hand, we define the physical area from Eq.\,($\ref{volume}$):
	\bear\label{area}
	\up dV&=&:\up dA \left(n_\mu {\pa x^\mu\over\pa z}\right)_s\up dz\,,
	\nnn
	\up dA&=&\sqrt{-g}~\varepsilon_{\rho\mu\nu\si}\UU^\rho_s n^\mu_s
	{\partial x^\nu_s\over\partial\ttt}{\partial x^\si_s\over\partial\pp}~
	\up d\ttt~\up d\pp ~.
	\enar
	The same comment about diffeomorphism invariance applies to this quantity, which indeed corresponds to the physical area occupied by the source in its rest frame. We emphasize again that both covariant expressions ($\ref{volume}$) and ($\ref{area}$) contain derivatives with respect to observed quantities $z, \ttt, \pp$, which affect directly only the observed angles $\ttt^\al, \pp^\al$, while FLRW coordinates are involved via the physical parameterization of the source position $x^\mu_s$ in Eq.\,($\ref{source}$). The physical volume $\up dV$ additionally contains the derivative with respect to the observed redshift $z$, which describes the derivative along the line-of-sight of the observer. Finally, we stress that these are partial derivatives, so they are performed while keeping the other observables fixed.
	
	We are now in the position to consider the perturbations in the physical volume and area due to the light propagation in the real universe $\mM$. We explicitly write $\up dV$ in Eq.\,($\ref{volume}$)
	using Eq.\,($\ref{source}$) to expand the source coordinates, Eq.\,($\ref{uu}$) for the components of $u^\mu_s$, and Eq.\,($\ref{dzdz}$) to express the scale factor at the source position $a_s$ in terms of the observed redshift $z$ and the fluctuation $\de z$:
	\bear\label{dV}
	\up dV&=&{(1+\dz)^3\over(1+\zz)^3}(1+\dg)\bigg\{
	\varepsilon_{\eta\alpha\beta\gamma}(1+\de u)~{\partial(\bar x_z^\alpha+\DX^\al) \over\partial\zz}
	{\partial (\bar x_z^\be+\DX^\be)\over\partial\ttt}{\partial (\bar x_z^\ga+\DX^\ga)\over\partial\pp}
	\nnn
	&&
	+\varepsilon_{\al \mu\nu\rho}~\VV^\al~{\partial (\bar x_z^\mu+\DX^\mu)\over\partial\zz}
	{\partial (\bar x_z^\nu+\DX^\nu)\over\partial\ttt}{\partial (\bar x_z^\rho+\DX^\rho)\over\partial\pp}
	\bigg\}\up dz~\up d\ttt~\up d\pp ~,
	\enar
	where we suppressed the subscript $s$.
	The background contributions to the homogeneous and isotropic universe indeed reduce to the background volume element
	\bear
	\up d\bar V={\rbar_z^2\over H_z(1+\zz)^3}~\up dz ~\up d\Omega\,,
	\enar
	and we define the dimensionless volume fluctuation $\de V$ as
	\bear\label{deV}
	\up dV=:{\rbar_z^2\over H_z(1+\zz)^3}~\up dz ~\up d\Omega~(1+\dV) \,,
	\enar
	with $H_z=H(\bar\eta_z)$ and $\up d\Omega=\sin\ttt\,\up d\ttt\up d\pp$. From this equation we manifestly see that $\de V$ is a scalar under diffeomorphisms. Equations ($\ref{dV}$) and ($\ref{deV}$) provide the exact definition of the gauge-invariant fluctuation $\de V$ in the physical volume.
	
	The relation between the physical area occupied by the source $\up dA$ and the observed solid angle element $\up d\Omega$ defines the angular diameter distance $\dD_A$:
	\bear\label{splitDA}
	\up dA=\dD^2_A\up d\Omega\,,
	\enar
	which we can split into the background and the perturbative parts as
	\bear\label{lumD}
	\dD_A=:\bar\dD_A(1+\de\dD)\,,\Dquad \bar\dD_A=\frac{\rbar_z}{1+z}\,,
	\enar
	valid at any order.
	Solving Eq.\,($\ref{area}$) for the angular diameter distance using Eqs.\,($\ref{splitDA}$) and ($\ref{lumD}$) we obtain
	\bear\label{deA}
	\up \dA^2&=&{\bar\dD_A^2
		\over\rbar_z^2
		\sin\ttt}~ (1+\dg)(1+\dz)^2\bigg[\varepsilon_{\eta\alpha\beta\gamma}(1+\de u)\left({n^\al+\dea^\al\over1+\dhnu}+\VV^\al \right)
	{\partial (\bar x_z^\be+\DX^\be)\over\partial\ttt}{\partial (\bar x_z^\ga+\DX^\ga)\over\partial\pp}
	\nnn
	&&
	\Dquad\Dquad+\varepsilon_{\al \eta\be\ga}~\VV^\al\left(-{\dnu-\dhnu\over1+\dhnu}+\de u \right)
	{\partial (\bar x_z^\be+\DX^\be)\over\partial\ttt}{\partial (\bar x_z^\ga+\DX^\ga)\over\partial\pp}
	\nnn
	&&
	\Dquad\Dquad+\varepsilon_{\al \be\nu\rho}~\VV^\al\left({n^\be+\dea^\be\over1+\dhnu}+\VV^\be \right)
	{\partial (\bar x_z^\nu+\DX^\nu)\over\partial\ttt}{\partial (\bar x_z^\rho+\DX^\rho)\over\partial\pp}
	\bigg]
	\nnn
	&=&\bar\dD_A^2(1+\de\dD)^2=:\bar\dD_A^2(1+\de A)~,
	\enar
	where again we suppressed the subscript $s$, and we substituted the expression for the light propagation direction $n_s^\mu$ in Eq.\,($\ref{propag}$) together with the equations mentioned for the expansion of $\up dV$. Moreover, we defined the perturbation $\de A$ in analogy with the volume fluctuation $\de V$.
	We recall that the angular diameter distance is proportionally related to the luminosity distance $\dD_L$, and it can be decomposed in the same manner:
	\bear
	\dD_L=(1+z)^2\dD_A\,.
	\enar
	This relation suggests that the fluctuations in the angular and diameter distances are identical, and the perturbation in the physical area occupied by the source can be expressed as
	\bear\label{dA}
	\de A=\de\dD^2+2\de\dD\,.
	\enar
	Again, we observe that this fluctuation, just like $\de V$, is a scalar under change of coordinates of $\mM$. Equation ($\ref{deA}$) provides the exact definition of the gauge-invariant fluctuations $\de\dD$ and $\de A$.
	
	For later convenience, we introduce the following decompositions along the line-of-sight and observed angular directions
	\bear
	\VV^\al&=&\VV_\para n^\al+\VV_\theta \theta^\al+\VV_\phi \phi^\al\,,\qquad
	\dea^\al=\dea_\para n^\al+\dea_\theta \theta^\al+\dea_\phi \phi^\al\,,
	\enar
	with $\ttt^\al$ and $\pp^\al$ defined via Eq.\,($\ref{angoli}$).

	\subsection{Gauge transformations of cosmological quantities}
	\label{subsec:gauge}
	The study of gauge transformations of second-order perturbations represents the main effort of this work.
	It will provide a validity check of our second-order expressions presented in Section $\ref{sec:second}$, and it will be carried out explicitly in Section $\ref{sec:gauge}$. Here we present a general introduction to gauge transformations in cosmology, and we establish the notation conventions relevant for the dedicated section (see \cite{bardeen} for a general discussion).
	
	Recall that the theory of general relativity is based on the principle of general covariance, namely we are free to choose any coordinate system to describe physics. One consequence of this freedom is that quantities that have been geometrically defined as tensors must obey well-defined transformation laws under a change of coordinates.
	Let us formalize the problem by considering a chart $(V,\varphi)$ of $\mM$, alternative to $(U,\psi)$, such that $V$ overlaps with $U$ and contains the light path $\ga$. Given a vector field $\zeta$, we can use the transformations that it induces to characterize the change of coordinates at a point $P$ in $U \cap V$:
	\bear\label{xzeta}
	\widetilde{x}^\mu_p(x_p)=e^{\zeta^\nu\pa_\nu}x^\mu\big\rvert_p=x^{\mu}_p+\zeta^{\mu}(x_p)+\frac{1}{2} \zeta^{\nu}(x_p)\pa_\nu\zeta^{\mu}\big\rvert_{x_p}+\cdots\,.
	\enar
	We use a tilde to denote the coordinate system given by the chart $(V,\varphi)$, defined in the same way as in Eq.\,($\ref{chart}$). Note that we can continue to use the parameterization $\cc$, $\tau$, and $\tau'$ for null and time-like curves since the parameterization procedure is chart-independent.
	
	Let us consider a tensor $K$ of arbitrary rank at point $P$. It is well known that the following exact transformation law holds
	\bear\label{tensor}
	\widetilde{K}^{\mu\cdots \nu}_{\rho\cdots\si}(\widetilde x_p)=\left(\frac{\partial \widetilde x^\mu}{\partial x^{\mu_1}}\right)_p\cdots\left(\frac{\partial x^{\rho_1}}{\partial \widetilde x^\rho}\right)_p\cdots\ K^{\mu_1\cdots \nu_1}_{\rho_1\cdots\si_1}(x_p)\,,
	\enar
	which relates the components of $K$ in the new frame, indicated by a tilde, to the components of $K$ in the original frame. We stress that both sides of the equation are evaluated at point $P$, which in the tilde frame has coordinates $\widetilde x^\mu_p$, and in the original frame has coordinates $x^\mu_p$. If the tensor we are considering is a scalar, the above transformation law trivializes:
	\bear\label{scalar}
	\widetilde K(\widetilde x_p)=K(x_p)\,.
	\enar
	
	In cosmology, we do not consider a generic coordinate transformation because we want to exploit the correspondence of the spacetime $(\mM,g)$ with the symmetric background solution given by FLRW $(\bar\mM,\bar g)$. Changing this correspondence between the real universe and the homogeneous isotropic universe is what we call gauge transformation. It is possible to relate a gauge transformation to a change of coordinates by evaluating both sides of Eq.\,($\ref{tensor}$) at the same (numerical) value $x^\mu_p$ in the two frames, which corresponds to two different points of $\mM$. This procedure defines the gauge transformation of the tensor $K$, as generated by the vector field $\zeta$: 
	\bear\label{gheigi}
	\widetilde{K}^{\mu\cdots \nu}_{\rho\cdots\si}(x_p) = e^{-\pounds_\zeta}K^{\mu\cdots \nu}_{\rho\cdots\si}\bigg\rvert_p=K^{\mu\cdots \nu}_{\rho\cdots\si}(x_p)- \left(\pounds_\zeta K^{\mu\cdots \nu}_{\rho\cdots\si}\right)(x_p) +\frac{1}{2} \left(\pounds^2_\zeta K^{\mu\cdots \nu}_{\rho\cdots\si}\right)(x_p)+\cdots \,, \,\,\,\,\,
	\enar
	with $\pounds_\zeta$ being the Lie derivative with respect to $\zeta$. Instead of working with the gauge vector field $\zeta$, we introduce an alternative vector field $\xi$ that simplifies the relation between new and old coordinates given in Eq.\,($\ref{xzeta}$):
	\bear\label{xcsi}
	\widetilde{x}^\mu_p(x)=:x^{\mu}_p+\xi^{\mu}(x_p)\,.
	\enar
	The vector field $\xi$ will be our preferred choice to compute gauge transformations, and it is assumed to be a nonlinear perturbation.
	We adopt the following parameterization for its components
	\bear\label{xicomp}
	\xi^\mu=:(T,\LL^\al)^\mu\,,
	\enar
	with $T$ and $\LL^\al$ nonlinear perturbations.
	The conversion between the two gauge fields $\zeta$ and $\xi$ can be readily derived and is presented at the second perturbative order in Section $\ref{sec:gauge}$.
	
	Note that the whole discussion above applies only to tensors. However, \textit{not} all the quantities that we have introduced so far are tensors under diffeomorphism in $\mM$. The only tensor fields we defined are the metric~$g$, the observer velocity $u$, the light wave vector $k$ and therefore all the tensor products that we can build. In particular, the redshift $z$, the perturbation in the volume occupied by the source $\de V$ and in the luminosity distance $\ddL$ are scalars under diffeomorphisms. Conformally rescaled quantities such as $\hat g$ or $\hat k$ instead, do not transform as tensors. To further illustrate this point let us consider the action of a generic diffeomorphism on the metric $g$:
	\bear
	\widetilde g_{\mu\nu}(\widetilde x)=\frac{\pa x^\rho}{\pa\widetilde x^\mu}\frac{\pa x^\si}{\pa\widetilde x^\nu}g_{\rho\si}(x)\,.
	\enar
	Given the conformal transformation $g=a^2\hat g$, we can derive from the equation above how the conformally rescaled metric $\hat g$ transforms:
	\bear
	\widetilde {\hat g}_{\mu\nu}(\widetilde x)=\frac{a^2(\eta)}{a^2(\widetilde\eta)}\frac{\pa x^\rho}{\pa\widetilde x^\mu}\frac{\pa x^\si}{\pa\widetilde x^\nu}\hat g_{\rho\si}\,,
	\enar
	which is not how a rank-two tensor transforms. The same argument applies to $\hat k$ and in turn affects the perturbations $\dhnu$ and $\de z$.
	Once again, the importance of computing gauge transformations in this work is that they provide constraints that our expressions must satisfy, and thus we use them as a theoretical consistency check. We will continue this discussion in more detail in the dedicated section.

	\section{Second-order Perturbative Calculations}
	\label{sec:second}
	
	Given the definitions and exact relations of the previous section, we now perform perturbative calculations up to \textit{second order}.
	Let us consider some cosmological quantity $Q$ defined in the real universe~$\mM$ and assume the split into background and perturbation as $Q=\bar Q+\delta Q$. We adopt the convention
	\bear\label{exp}
	\de Q =\de Q^{(1)}+\de Q^{(2)}+\cdots\,,
	\enar
	where the superscripts represent the perturbative order.
	For the sake of clarity, we will omit the superscripts for quadratic terms, i.e., terms made by the product of two first-order variables. The linear terms without superscripts, such as $\de Q$, represent first-order and second-order terms.
	
	In the following, we will provide second-order perturbative expressions for the observer and source position, the observed redshift, and the area and volume occupied by the source; these will be obtained perturbing the expressions presented in the previous section. We will begin by computing the perturbations of geometric quantities to finally express all the second-order observables in terms of metric and observer velocity perturbations only. Note that we are not assuming Einstein equations or any other field equations that describe the dynamics of the universe. Hence, metric and observer velocity perturbations cannot be constrained at this stage and are treated as independent spacetime fields.

	\subsection{Metric-related quantities and observer four-velocity}
	
	A comprehensive second-order computation of the metric and the Einstein equations can be found in \cite{hwang}.
	We can compute the second-order expression of the inverse metric components perturbing the relation $g_{\mu\rho}g^{\rho\nu}=\de^\nu_\mu$. We obtain
	\bear\label{met2}
	g^{\eta\eta}&=&{1 \over a^2} \left( - 1 + 2\AA - 4 \AA^2 + \BB_\alpha 
	\BB^\alpha \right)~,\,\,\qquad 
	g^{\eta\al}={1 \over a^2} \left( - \BB^\alpha + 2 \AA \BB^\alpha
	+ 2 \BB_\beta \CC^{\alpha\beta} \right)~,
	\nnn
	g^{\al\be}&=&{1 \over a^2} \left( \de^{\alpha\beta} - 2 \CC^{\alpha\beta}
	-\BB^\alpha \BB^\beta
	+ 4 \CC^\alpha_\gamma \CC^{\beta\gamma} \right)~.
	\enar
	The second-order perturbation in the determinant of the metric in Eq.\,($\ref{detg}$) is\footnote{It can be computed perturbing the following exact equation
		\bear \up{det}\,g=(1/24)\left[(\up{tr}\,g)^4-6\up{tr}(g^2)(\up{tr}\,g)^2+3(\up{tr}(g^2))^2+8\up{tr}(g^3)\up{tr}\,g-6\up{tr}(g^4)\right]\,.
		\nonumber
		\enar}
	\bear
	\label{deg2}
	\dg=\AA+\CC^\alpha_\alpha-{1\over2}\AA^2+\frac12\BB^\alpha\BB_\alpha
	+\AA~\CC^\alpha_\alpha+{1\over2}\CC^\alpha_\alpha\CC^\be_\be-
	\CC^{\alpha\beta}\CC_{\alpha\beta}\,.
	\enar
	From the definition of $\hat\Ga^\rho_{\mu\nu}$ in Eq.\,($\ref{gammahat}$) we derive the second-order Christoffel symbols defining the Levi-Civita connection based on $\hat g$
	\bear
	\hat\Ga^\eta_{\eta\eta}&=&\AA'-2\AA\AA'+\BB^\ga\left(\BB'_\ga-\AA_{,\ga}\right)\,,
	\nnn
	\hat\Ga^\eta_{\al\eta}&=&\AA_{,\al}-2\AA\AA_{,\al}-\BB^\ga\left(\CC'_{\al\ga}-\BB_{[\ga,\al]}\right)\,,
	\nnn
	\hat\Ga^\eta_{\al\be}&=&\CC'_{\al\be}+\BB_{(\al,\be)}-2\AA\left( \CC'_{\al\be}+\BB_{(\al,\be)} \right)+\BB^\ga\left(\CC_{\al\be,\ga}-2\CC_{\ga(\al,\be)}\right)\,,
	\nnn
	\hat\Ga^\al_{\eta\eta}&=&-\BB^{'\al}+\AA^{,\al}+\AA'\BB^\al+2\CC^{\al\ga}(\BB'_\ga-\AA_{,\ga})\,,
	\nnn
	\hat{\Ga}^{\al}_{\be\eta}&=&\CC'^{\al}_{\be}-\BB^{[\al}{}_{,\be]}+\BB^{\al}\AA_{,\be}+2\CC^{\al\ga}(\BB_{[\ga,\be]}-\CC'_{\be\ga})\,,
	\nnn
	\hat{\Ga}^{\al}_{\be\ga}&=&-\CC_{\be\ga}{}^{,\al}+2\CC^{\al}_{(\be,\ga)}+\BB^{\al}\left(\CC'_{\be\ga}+\BB_{(\be,\ga)}\right)
	+2\CC^{\al\de}\left(\CC_{\be\ga,\de}-2\CC_{\de(\be,\ga)}\right)\,.
	\enar
	Consequently, we can obtain $\GG^\eta$ and $\GG^\al$ at second order from Eq.\,($\ref{GGt}$) and Eq.\,($\ref{GGa}$)
	\bear\label{GG2}
	\GG^{\eta}&=&
	\AA'-2\AAP+{\BB}_{\para,\para}+\CCPP'+2\dnu\left(\AA' - \AAP\right)
	- 2\AA\left(\AA'-2\AAP+\BB_{\para,\para}
	+\CC'_{\para\para}\right)
	\nnn
	&&
	-2\dea^\alpha\left(\AA_{,\alpha}
	-\BB_{(\alpha,\para)} -\CC_{\alpha\para}'\right)
	-\BB^\al \left(\AA_{,\alpha}-\BB'_{\alpha}+2\BB_{[\al,\para]} -2\CC'_{\al\para}
	+ 2 \CC_{\alpha \para,\para}
	-\CC_{\para\para,\al}\right)\,,
	\nnn
	\GG^{\al}&=&\AA^{,\alpha}-\BB'^{\alpha}
	-  \BB_{\para}^{\;\;,\alpha} + \BB^\alpha_{\;\;,\para}
	- 2 \CC'^{\alpha}_\para+
	2 \CC^\alpha_{\para,\para} - \CC_{\para\para}^{\;\;\;\;,\alpha}
	+\dnu\left(2\AA^{,\alpha} - 2\BB'^{\alpha}-\BB_\para^{\;\;,\alpha}
	+ \BB^\alpha_{\;\;,\para}- 2 \CC'^{\alpha}_{\para} \right) \nnn
	&&
	-\dea^\ga\left(\BB_\ga^{\;\;,\alpha}
	- \BB^\alpha_{\;\;,\ga}+ 2 \CC'^{\alpha}_\ga 
	-4 \CC^\alpha_{(\para,\gamma)}+2\CC_{\para\gamma}^{\;\;\;\;,\alpha}
	\right)+ \BB^\alpha \left(\AA'+ \BB_{\para,\para} + \CC'_{\para\para}- 2 \AA_{,\para}\right)\nonumber \\
	&&
	- 2\CC^{\alpha\ga}\left(  2 \CC_{\ga\para,\para}
	- \CC_{\para\para,\ga}+\AA_{,\ga} 
	-\BB'_{\ga} -2\BB_{[\para,\gamma]}
	- 2\CC_{\para\gamma}'\right)
	~,
	\enar
	where round brackets around indices represent the symmetrization and square brackets the antisymmetrization with respect to those indices, e.g., $2\BB_{(\al,\be)}=\BB_{\al,\be}+\BB_{\be,\al}\,$ , $\,2\BB_{[\al,\be]}=\BB_{\al,\be}-\BB_{\be,\al}$.
	
	To obtain the perturbed expression of the observer four-velocity we can use the time-like normalization $u_\mu u^\mu=-1$ in Eq.\,($\ref{delu}$). We derive up to second order \cite{YZ}
	\bear\label{du2}
	\dUU=- \AA+\frac32 \AA^2 + {1\over 2}
	\VV^\alpha \VV_\alpha-\VV_{\al}\BB^{\al}\,,
	\enar
	and the covariant components $u_\mu$ by using Eq.\,($\ref{delubasso}$):
	\bear\label{ubasso}
	\UU_\eta = - a \left( 1 + \AA-\frac12\AA^2
	+ {1\over 2} \VV^\alpha \VV_\alpha\right)\,,\qquad
	\UU_\alpha= a \left( \VV_\alpha - \BB_\alpha+\AA\BB_\alpha 
	+ 2 \VV^\beta \CC_{\alpha\beta} \right)=a\VL_\al\,.\,\,\,
	\enar
	As our approach is purely geometric, the observer velocity $\VV^\al$ and the metric perturbations $\AA,\, \BB_\al,\, \CC_{\al\be}$ cannot be constrained.
	All our perturbative expressions presented in this section will ultimately depend on these independent spacetime fields.

	\subsection{Subtleties in second-order computations}
	\label{subsec:sub}
	Second-order perturbative calculations are more challenging than linear order calculations because of the many subtleties that arise.
	Let us take the time lapse at the observer position as an example to clarify this point (see, e.g. \cite{ld9}).
	We begin by considering the differential equation ($\ref{curvint}$) for the integral curve. This equation conveniently relates $\de t$ to $\de u$ so that we can exploit the second-order expression in Eq.\,($\ref{du2}$) to compute the lapse in terms of metric perturbations and observer velocity only.
	To perturb this differential equation we make the following consideration: $\de t(\tau)$ and $\de u(x_\tau)$ are well-defined quantities, and their perturbative expansion can be generically expressed according to Eq.\,($\ref{exp}$):
	\bear
	\de t(\tau)&=& \de t^{(1)}(\tau)+\de t^{(2)}(\tau) +\cdots\,,
	\nnn
	\de u(x_\tau)&=&\de u^{(1)}(x_\tau)+\de u^{(2)}(x_\tau) +\cdots\,.
	\enar
	However, the perturbative expansion of $\de u(x_\tau)$ is somewhat ambiguous, as there exist different ways to define it. This quantity is in fact a functional of the observer world line, so it can be expanded around the background path $\bar x^\mu_\tau$ as
	\bear
	\de u(x_\tau=\bar x_\tau+\de x_\tau)=\de u(\bar x_\tau)+\de x^\mu_\tau\left(\frac{\pa}{\pa x^\mu}\de u\right)\bigg\rvert_{\bar x_\tau}+\cdots\,,
	\enar
	such that the distortion in the path appears explicitly. We now make the following identification between perturbative orders, that we assume to be valid for any field on $\mM$
	\bear\label{2bar}
	\de u^{(1)}(x_\tau)=\de u^{(1)}(\bar x_\tau)
	\,,\Dquad
	\de u^{(2)}(x_\tau)=\de u^{\mu(2)}(\bar x_\tau)+\de x^{\mu(1)}_\tau\left(\frac{\pa}{\pa x^\mu}\de u^{(1)}\right)\bigg\rvert_{\bar x_\tau}\,.
	\enar
	Note that at linear order the deviations from the background path $\bar x^\mu_\tau$ do not come into play as $\de u$ is already a first-order perturbation. We can then write the formal perturbative solutions of Eq.\,($\ref{curvint}$) as
	\bear\label{dt1}
	\de t_o=\int_0^{\bar t_{\rm o}}dt\,\de u(\bar x_t)\,,
	\enar
	at linear order, and
	\bear\label{dt2}
	\de t_o=\int_0^{\bar t_{\rm o}}dt\,\left[\de u(\bar x_t)+\de x^\mu_t(\pa_\mu\de u)\big\rvert_{\bar x_t} \right]\,,
	\enar
	at second order. To emphasize the fact that the integrations are performed along the background path we use the parameterization $t$ instead of $\tau$. The reason why we decide to expand each field around the background is that this allows us to better keep track of the perturbative orders, and in this way, we can actually obtain the solution $\de x^\mu_\tau$ at each perturbative order, which is then fed into a higher order solution. This choice will also prove convenient when we will compute gauge transformations.
	
	Again, we take the time lapse at the observer position to illustrate another subtle point.
	We have repeatedly stated that $\de t(\tau)$ is not a field but rather a (perturbed) time coordinate.
	However, it is also possible to define the time lapse as a field along the observer world line. By specifying a point on the trajectory $x^\mu_\tau$ we are also specifying a value of the field $u^\mu(x_\tau)$ and from Eq.\,($\ref{curvint}$) this also means specifying the value $\de t(\tau)$. This procedure defines a map $\de t(x_\tau)$ from the coordinates of the path to the corresponding time lapse such that $\de t(x_\tau):=\de t(\tau)$ for all $\tau$ in $[0,\tau_o]$.\footnote{In fact, we can think of the time lapse for another trajectory of a fictitious observer. In this way, it can be defined as a field on $\mM$.} Using this field definition of the time lapse, we can replace~$\de u$ in Eq.\,($\ref{dt2}$) by $\dot{\de t}$, as the partial derivatives are now justified. We can now integrate by part, thus obtaining
	\bear\label{dt2nogeo}
	\de t_o=\de x_{o}^\mu (\pa_\mu\de t)\big\rvert_{{\bobs}}+\int_0^{\bar t_{\rm o}}dt\,\left[\de u(\bar x_t)-\dot{\de x}^\mu_t(\pa_\mu\de t)\big\rvert_{\bar x_t} \right]\,,
	\enar
	where the initial conditions in Eq.\,($\ref{curvint}$) have been used. We further remark that the integral in the equation above is indeed the second-order expression for $\de t(\bar x_{\rm o})$, i.e., the time lapse for another observer at $\bar x^\mu_{\rm o}$ in an inhomogeneous universe. Since in the real universe the observer is located at~$O$, and not at $\bar O$, we will need the expression for $\de t_o$ only, and not at all for~$\de t_{\bobs}$. Therefore, $\de t_o$ will not be expanded around $\bar x^\mu_o$ in any of the following equations while the perturbative expression in the right-hand side of Eq.\,($\ref{dt2nogeo}$) is expanded around the background path. The same arguments apply to $\de x^\al_o$, $\dnu_o$, $\dea^\al_o$, $\dnu_s$, $\dea^\al_s$ and all the other quantities we will compute. We conclude by stressing that we will only use the field-definition of path deviations $\de x^\mu$ to derive simpler equations for formal solutions, particularly to perform integration by part. On any other occurrence, they are treated simply as functions of the affine parameter denoting the path.

	\subsection{Time lapse and spatial shift at the observer position}
	
	Resuming the discussion, we now devote ourselves to computing perturbative expressions for the time lapse and spatial shift at the observer position. We will first derive the linear-order expression and then the second-order one. We adopt this structure throughout the whole section.
	\\
	
	\Ietc{Linear-order lapse and shift}
	Combining the expression for $\de u$ in Eq.\,($\ref{du2}$) and the linear-order solution for the time lapse at the observer in Eq.\,($\ref{dt1}$) we obtain at linear order
	\bear\label{dxlin}
	\delta t_o=-\int_0^{\bar t_{\rm o}}dt~\AA(\bar x_t)~,\Dquad
	\delta x^\al_o=\int_0^{\bar t_{\rm o}} dt~\frac1a\VV^\al(\bar x_t)~.
	\enar
	Note that since the integrand is already first order in perturbations, the fields are evaluated at $\bar x^\mu_t$, such that the integration is along the time coordinate $t$ without any spatial shift $\bar x^\al_t=0$. 
	We will use the parameterization $dt$ to emphasize that the integration is performed along the background path.
	
	If the observer experiences no acceleration, the geodesic equations ($\ref{geou}$) provide extra linear-order conditions
	\beeq\label{lindt}
	\AA=\frac{\pa}{\pa t}\left(av\right)~,\Dquad v_\al=0~,
	\eneq
	where we introduced the Helmholtz decomposition $\VL_\al:=v_\al-v_{,\al}$, with $\pa^\al v_\al=0$.
	Using these geodesic conditions, the linear-order time lapse at the observer is further simplified as \cite{biern1,ld9,thepaper}
	\bear\label{dt1geo}
	\de t_o=-v(\bar x_{\rm o})~.
	\enar
	Moreover, we derive the linear-order relations that are valid at any point along the background path:
	\bear\label{dt1nogeo}
	\frac{\pa}{\pa t}{\delta t}=-\AA(\bar x_t)~,\Dquad
	\frac{\pa}{\pa t}{\delta x}^\al={1\over a}\VV^\al(\bar x_t)~.
	\enar
	\\
	
	\Ietc{Second-order lapse and shift}
	Combining the second-order expression for $\de t_o$ in Eq.\,($\ref{dt2nogeo}$) with~$\de u$ in Eq.\,($\ref{du2}$) and Eq.\,($\ref{dt1nogeo}$) we derive 
	\bear\label{dt2met}
	\de t_o=\de x^\mu_o(\pa_\mu\de t)\big\rvert_{\bobs}+\int_0^{\bar t_{\rm o}}dt\left(-\AA+\frac12\AA^2
	+\frac12\VV^\al\VV_\al-\VV^\al\BB_\al-\frac1a\VV^\al\de t_{,\al}\right)\bigg\rvert_{\bar x_t}
	~.
	\enar
	To compute the spatial shift at the observer we can follow the same steps to obtain at the second order
	\bear\label{dx2met}
	\de x^\al_o=\de x_o^\mu(\pa_\mu\de x^\al)\big\rvert_{\bobs}+\int_0^{\bar t_{\rm o}}dt\,\frac1a\left(\VV^\al+\AA\,\VV^\al-\VV^\be\de x^\al{}_{,\be}\right)\bigg\rvert_{\bar x_t}
	\,.
	\enar
	In both expressions, any occurrence of $\de x^\mu$ is intended as the linear-order expression in Eq.\,($\ref{dxlin}$), even if we do not perform the explicit substitution. In this sense the partial derivatives $\pa_\nu\de x^\mu$ are well-defined.
	
	The second-order expression for the time lapse can be further simplified by assuming that the observer follows a geodesic. In particular, the linear-order condition in Eq.\,($\ref{dt1geo}$) at any point of the background path reads $\de t=-av$. Using this equation we can rearrange Eq.\,($\ref{dt2met}$) as
	\bear
	{\de t}_o=\int_0^{\bar t_{\rm o}}dt\left(-\AA+\frac12\AA^2
	+\frac12\VV^\al\VV_\al-\VV^\al\BB_\al+\VV^\al v_{,\al}\right)\bigg\rvert_{\bar x_t}+
	\left(\AA v-\delta x^\al v_{,\al}\right)_{\bobs}~.
	\enar
	Moreover, we derive the second-order geodesic equations from the definition in Eq.\,($\ref{geou}$):
	\bear
	a_\al=0&=&\AA_{,\al} + \frac1a\left[a\left(\VV_\al - \BB_\al\right)\right]^\prime+\AA \left(-2\AA_{,\al}+2\BB'_{\al}+2\HH\BB_{\al}-\VV'_{\al}-\HH\VV_{\al}\right)
	\nnn
	&&
	+ \AA^\prime \BB_\al+2\CC'^{\be}_{\al}\VV_{\be}+2\CC^{\be}_{\al}\left( -\AA_{,\be}+\BB'_{\be}+\HH\BB_{\be} \right)+ \VV^\be \left( \VV_{\al,\be} - 2\BB_{[\al,\be]}\right) \,,
	\nnn
	a^\eta=0&=&\left(\VV^{\al}-\BB^{\al}\right)a_\al\,.
	\enar
	Noting that the linear-order geodesic condition in Eq.\,($\ref{lindt}$) can be recast as $\AA_{,\al}=-(1/a)\left( a\VL_{\al} \right)^\prime$, we can rewrite the previous geodesic condition by using $\VL_\al$ in Eq.\,($\ref{ubasso}$):
	\beeq
	a_\al=0=\AA_{,\al}+\frac1a\left(a\VL_{\al}\right)^\prime\left( 1+\AA \right)+\VV^{\be}\left( \VL_{\al,\be}+\BB_{\be,\al} \right)\,.
	\eneq
	Furthermore, by considering $\VV^{\be}\BB_{\be,\al}=\VV^{\be}\VV_{\be,\al}-\VV^{\be}\VL_{\be,\al}$ we manipulate the equation
	\bear
	&0&=\AA_{,\al}+\frac1a\left(a\VL_{\al}\right)^\prime+2\VV^{\be}\VL_{[\al,\be]}-\AA\AA_{,\al}+\VV^{\be}\VV_{\be,\al}
	=\left(a\VL_{\al}\right)^\cdot+\left[ \AA-\frac12\AA^2+\frac12\VV^{\be}\VV_{\be} \right]_{,\al},\,\,\,\,\,\,\,\,\,
	\quad
	\enar
	which can be readily solved by the Helmholtz decomposition as
	\bear
	\AA-\frac12\AA^2+\frac12\VV^{\be}\VV_{\be}-{\left( av \right)}^\cdot=0\,,\Dquad	v_\al=0\,.
	\enar
	Finally, we manipulate the integrand of Eq.\,($\ref{dt2met}$)
	\beeq
	-\AA+\frac12\AA^2
	+\frac12\VV^\al\VV_\al-\VV^\al\BB_\al-\VV^\al v_{,\al}=-\AA+\frac12\AA^2
	-\frac12\VV^\al\VV_\al
	~,
	\eneq
	such that by combining this equation with the previous one, we derive the second-order time lapse at the observer with geodesic motion
	\beeq\label{dt2geo}
	\de t_o=-\left(v-\AA v+\delta x^\al v_{,\al}\right)_{\bobs}~.
	\eneq
	This result is consistent with the one derived in \cite{ld9}.
	Due to our use of conformal time, it proves more convenient the observer time lapse~$\de\eta_o$ in terms of the conformal time. Given the definition of the cosmic time $dt=a d\eta$, we integrate it to obtain the matching condition for $\eta_o=\bar\eta_o+\de\eta_o$:
	\bear\label{et}
	t_o=\bar t_{\rm o}+\de t_o=\int_0^{\eta_o}d\eta\,a(\eta)&=&\bar t_{\rm o}+\de\eta_o+\frac12\HH(\bar\eta_{\rm o})\de\eta^2_o+\cdots\,,
	\enar
	and this yields the second-order expression for the time lapse in conformal coordinates
	\bear\label{lapseconf}
	\de \eta_o=\de t_o-\frac12\HH_{\bobs}\de t_o^2\,.
	\enar

	\subsection{Distortion in the conformal wave vector}
	Having set up the observer position at second order, we proceed to compute the perturbations in the conformal wave vector $\hat k^\mu$ along the light path using the geodesic equations ($\ref{kint}$).
	\\
	
	\Ietc{Linear-order solution}
	By substituting $\GG^\eta$ and $\GG^\al$ in Eq.\,($\ref{GG2}$) into the formal geodesic solutions in Eq.\,($\ref{kint}$), we obtain at linear order
	\bear\label{dk1s}
	\dnu\big\rvert^s_o&=&\left( \BBP-2\AA \right)\bigg\rvert^z_{\bobs}-\int_0^{\rbar_z} d\rbar\left(\AA-\BBP-\CCPP \right)'~,
	\nnn
	\dea_\al\big\rvert^s_o&=&-\left( \BB_\al+2\CC_{\al\para} \right)\bigg\rvert^z_{\bobs}-\int_0^{\rbar_z} d\rbar\left(\AA-\BBP-\CCPP \right)_{,\al}\,.
	\enar
	Recalling the arguments presented in Subsection ($\ref{subsec:sub}$), the integration is performed along the background path, parameterized by the comoving distance $\rbar$ instead of the affine parameter $\cc$. In fact, from Eq.\,($\ref{dr}$) the following relation holds at the background
	\bear
	\frac{d}{d\cc}=-\frac{d}{d\rbar}={\pa\over\pa\eta}-n^\al{\pa\over\pa x^\al}\,.
	\enar
	
	We can obtain the linear-order expressions for $\hat k_o^\mu$ using the tetrad basis at the observer position.
	From the definition in Eq.\,($\ref{tetrade}$) and the orthonormality condition in Eq.\,($\ref{orto}$) we derive the space-like tetrad components at linear order
	\bear
	\de e_i^\eta=\de^\al_i(\VV_\al-\BB_\al)\,,\Dquad S_{ij}=\de^\al_i\de^\be_j\CC_{\al\be}\,,
	\enar
	where $S_{ij}$ is the symmetric part of the spatial components $\de e_i^\al$ in Eq.\,($\ref{split}$). To derive the antisymmetric part $A^j{}_i$, we cannot rely on the orthonormality condition, but we can find constraints based on gauge transformation properties \cite{thepaper}. The derivation is shown in Appendix $\ref{AppendixA}$, here we state the result:
	\bear\label{asim}
	\de_{\al j}A^j{}_{i}=C_{[\al,\be]}\de^\be_i+\varepsilon_{\al ij}\Om^j\,,
	\enar
	where $C_\al$ is the vector part of $\CC_{\al\be}$ in its scalar-vector-tensor decomposition, $\varepsilon_{ijk}$ is the Levi-Civita symbol with $\varepsilon_{123}=1$, and $\Om^i$ is a gauge-invariant vector describing the misalignment of the space-like tetrad basis against a FLRW coordinate \cite{thepaper}.
	Recall that in Eq.\,($\ref{tetrade}$) we fully specified the tetrad basis only at background level, where we required the components of the space-like tetrads to be fully aligned with FLRW spatial directions. In the real universe $\mM$, we fixed the three degrees of freedom associated with boosts, but we still have three degrees of freedom associated with spatial rotations that are specified by $\Om^i$, but not determined by metric perturbations. Combining these results with the definition of $\hat k^\mu_o$ in Eq.\,($\ref{ko}$) we derive the linear-order distortions at the observer
	\bear\label{dk1o}
	\dnu_{o}=-\left( \AA+\VV_\para-\BBP\right)_{\bobs}\,,\Dquad
	\left(\dea_\al\right)_{o}=-\left(\VV_\al+\CC_{\al\para}+C_{[\al,\para]}+\epsilon_{\al i j}n^i\Om^j\right)_{\bobs}\,.
	\enar
	\\
	
	\Ietc{Second-order solution}
	To obtain the second-order solution we need to expand the formal expression in Eq.\,($\ref{kint}$) around the background. This implies expanding the fields as functionals of the light path, but also expanding the upper limit of integration $\cc_s$ around affine parameter $\cc_z$ via $\cc_s=\cc_z+\Dcc_s$:
	\bear\label{dnu2imp}
	\dnu\big\rvert^s_o=\int_0^{\rbar_z}d\rbar \left[\GG^\eta(\bar x_{\rbar})+\de x^\mu_{\rbar}(\pa_\mu\GG^\eta)\big\rvert_{\bar x_{\rbar}}\right]-\Dcc_s\GG^\eta(\bar x_z)\,.
	\enar
	We can simplify this equation using the geodesic equation ($\ref{kint}$) $\frac{d}{d\cc}\dnu=-\GG^\eta$, and then integrate by part.
	Using the perturbations for $\de\hat k^\mu$ in Eq.\,($\ref{dxder}$) and for $\Delta x_s^\mu$ in Eq.\,($\ref{Dxcc}$) we arrive at
	\bear\label{dnuint}
	\dnu\big\rvert^s_o=\left[\DX^\mu_{\rbar}(\pa_\mu\dnu)\big\rvert_{\bar x_{\rbar}}\right]^z_{\bobs}+\int_0^{\rbar_z}d\rbar \left(\GG^\eta+\dnu\dnu'-\dea^\al\dnu_{,\al}\right)\bigg\rvert_{\bar x_{\rbar}}\,,
	\enar
	where the boundary terms on the right-hand side appear as the difference between the terms inside the square brackets evaluated at $\rbar=\rbar_z$ and $\rbar=\rbar_{\rm o}$.
	We can repeat the same steps for $\dea^\al$ and derive
	\bear\label{deaint}
	\dea^\al\big\rvert^s_o=\left[\DX^\mu_{\rbar}(\pa_\mu\dea^\al)\big\rvert_{\bar x_{\rbar}}\right]^z_{\bobs}-\int_0^{\rbar_z}d\rbar \left(\GG^\al-\dnu\dea'^\al+\dea^\be\dea^\al{}_{,\be}\right)\bigg\rvert_{\bar x_{\rbar}}\,.
	\enar
	To obtain these expressions in terms of perturbations of the metric and velocity, we should substitute the linear-order solutions for $\dnu(\bar x)$ and $\dea^\al(\bar x)$ in Eq.\,($\ref{dk1s}$), then the second-order expressions for~$\GG^\eta(\bar x)$ and $\GG^\al(\bar x)$ in Eq.\,($\ref{GG2}$), as well as the expression for $\DX^\mu_s$ (which we will derive later) and $\de x^\mu_o$ from the previous subsection. The complete expression is presented in Appendix $\ref{AppendixB}$.
	
	Again, we can compute the components of $\hat k^\mu_o$ using the tetrad at the observer in Eqs.\,($\ref{tetrade}$) and ($\ref{split}$). Employing the orthonormality condition ($\ref{orto}$) we derive at second order
	\bear\label{tetrade2}
	\de e^\eta_i&=&\VV_i-\BB_i+\AA(2\BB_i-\VV_i)+\CC_{i\al}(\VV^\al+\BB^\al)+\varepsilon_{i\al j}(\VV^\al-\BB^\al)\Om^j+(\VV^\al-\BB^\al)C_{[i,\al]}\,,
	\nnn
	S_{ij}&=&\CC_{ij}+\frac12\BB_i\BB_j-\frac12\VV_i\VV_j-\frac12\Om_i\Om_j+\frac12\de_{ij}\Om^k\Om_k-\frac32\CC^\al_i\CC_{\al j}-\CC^\al_{(i}\varepsilon_{j)k\al}\Om^k-\frac12C_{\al,(i}\CC^\al_{j)}
	\nnn
	&&
	+\frac12C^\al{}_{,(i}\varepsilon_{j)k\al}\Om^k
	-\frac12C_{(i,\al}\varepsilon_{j)k}{}^\al\Om^k
	+\frac18C^{\al}{}_{,i}C_{\al,j}+\frac18C_{i,}{}^\al C_{j,\al}+\frac12\CC^\al_{(i}C_{j),\al}
	\nnn
	&&
	-\frac14C^\al{}_{,(i}C_{j),\al}\,,
	\enar
	where spacetime indices have been contracted with Kronecker delta involving internal indices. From these relations, we derive the second-order expression for $\dnu_o$ defined in Eq.\,($\ref{ko}$)
	\bear\label{dnu2o}
	\dnu_o&=&\bigg(- \AA-\VV_\para+\BBP+\frac32\AA^2-\CC_{\al\para}\left( \VV^\al+\BB^\al \right)+\AA\left( \VV_\para-2\BBP\right)-\VV^\al\BB_\al
	\nnn
	&&
	+\frac12\VV_\al\VV^\al+\left(\VV^\al-\BB^\al\right)C_{[\al,\para]}+\varepsilon_{\al i j}\left( \VV^\al-\BB^\al \right)n^i\Om^j\bigg)_{\bobs}+\de x^\mu_o\left[\pa_\mu\left(-\AA-\VV_\para+\BB_\para\right)\right]\big\rvert_{\bobs}\,,\,\,\,\,\,\qquad
	\enar
	where the last term is the expansion needed to have the right-hand side evaluated at the background observer position $\bar x^\mu_o$. The second-order expression for $\dea^\al_o$ can also be computed from the definition in Eq.\,($\ref{ko}$), taking into account the above expression for $S_{ij}$
	\bear\label{dea2o}
	-\dea^\al_o&=&\VV^\al_o+\bigg(A^\al{}_i+\CC^\al_{i}+\frac12\BB_i\BB^\al-\frac12\VV_i\VV^\al-\frac12\Om_i\Om^\al+\frac12\de^\al_{i}\Om^k\Om_k-\frac32\CC^\be_i\CC^\al_{\be}-\CC^\be_{(i}\varepsilon^{\al)}{}_{k\be}\Om^k
	\nnn
	&&
	-\frac12C_{\be,(i}\CC^{\al)\be}
	+\frac12C^\be{}_{,(i}\varepsilon^{\al)}{}_{k\be}\Om^k
	-\frac12C_{(i,\be}\varepsilon^{\al)}{}_{k}{}^\be\Om^k
	+\frac18C^{\be}{}_{,i}C_{\be}{}^{,\al}+\frac18C_{i,}{}^\be C^\al{}_{,\be}+\frac12\CC^\be_{(i}C^{\al)}{}_{,\be}
	\nnn
	&&
	-\frac14C^\be{}_{,(i}C^{\al)}{}_{,\be} \bigg)_{\bobs}n^i
	+\de x^\mu_o\left[\pa_\mu\left(\VV^\al+\CC^\al_in^i+C^{[\al}{}_{,i]}n^i+\varepsilon^\al{}_{ij}n^i\Om^j\right)\right]\bigg\rvert_{\bobs}\,.
	\enar
	Note that the second-order expression for the antisymmetric part $A^j{}_i$ is not presented here, as it does not contribute to any of the second-order quantities we explicitly present in Appendix $\ref{AppendixB}$.

	\subsection{Fluctuations in the observed redshift}
	The observed redshift $z$ in Eq.\,($\ref{dzdz}$) is affected by the cosmic expansion set by $a_s$ at the source position and the fluctuations $\de z$ that include the Doppler effect and the Sachs-Wolfe effect. Here we compute the second-order expression for $\de z$ by using Eq.\,($\ref{dz}$).
	Following the steps in Section $\ref{subsec:sub}$, we expand the fields on the right-hand side in Eq.\,($\ref{dz}$) around the background reference positions~$\bar x_z$ and $\bar x_{\rm o}$.
	The scale factor at the observer position is
	\bear
	a(\eta_o=\bar\eta_{\rm o}+\de\eta_o)=1+\HH_{\bobs}\de\eta_o+\frac12\left(\HH^2_{\bobs}+\HH'_{\bobs}\right)\de\eta_o^2\,,
	\enar
	and we expand the perturbation $\dhnu$ around the observer position $\bar x_z$ as
	\bear
	\dhnu_s=\dhnu(x_s=\bar x_z+\DX_s;x_o)=\dhnu(\bar x_z;x_o)+\DX^\mu_s\left(\frac{\pa}{\pa x^\mu}\dhnu\right)\bigg\rvert_{\bar x_z}
	\,.
	\enar
	We choose not to expand the reference position around $\bar x^\mu_o$, keeping it fixed at the real observer position $O$.
	Combining these results, we obtain the second-order expression for the fluctuation in the observed redshift
	\bear\label{dz2}
	\de z(x_s;x_o)&=&\HH_{\bobs}\de\eta_o+\dhnu(\bar x_z;x_o)+\frac12\left(\HH^2_{\bobs}+\HH'_{\bobs}\right)\de\eta_o^2
	+\HH_{\bobs}\de\eta_o\dhnu(\bar x_z;x_o)
	+\DX^\mu_s(\pa_\mu\dhnu)\big\rvert_{\bar x_z}\,.\,\,\,\,\qquad
	\enar
	In order to express $\de z$ in terms of the metric perturbations and the observer velocity, we first need to derive a second-order expression for $\dhnu$. This is readily obtained from the exact relation in Eq.\,($\ref{dhnuexact}$):
	\bear\label{dhnu2}
	\dhnu(\bar x_z;x_o)&=&\dnu+\AA+\VV_\para-\BB_\para+\dnu\AA-\frac12\AA^2+\AA\BB_\para+\dea_\al\left( \VV^\al-\BB^\al \right)
	+\frac12 \VV^\al\VV_\al
	+2\CC_{\al\para}\VV^\al\,,\,\,\,\,\qquad
	\enar
	where both sides are evaluated at $\bar x^\mu_z$, with reference $x^\mu_o$. The full expression for $\de z$ is given in Appendix $\ref{AppendixB}$ (see also \cite{ld9}).

	\subsection{Distortion in the source position compared to the observed position}
	
	In Eq.\,($\ref{xz}$) we introduced two ways to define deviations in the source position, $\de x_s^\mu$ and $\DX^\mu_s$. The first is a deviation compared to the unphysical background position $\bar x^\mu_s$, while the latter is compared to the background observed position $\bar x^\mu_z$. Our interest is indeed in computing the second-order perturbative expression for $\DX^\mu_s$, since it contains the entire chart dependence of the source position $x^\mu_s$.
	\\
	
	\Ietc{Linear-order source position}
	To compute the temporal distortion $\DT_s$, we expand up to first order the exact relationship with the redshift $\de z$ in Eq.\,($\ref{dzrel}$). We obtain
	\bear\label{Deta1}
	\DT_s=\frac{\de z}{\HH_z}=\frac{1}{\HH_z}\left(\HH_{\bobs}\de\eta_o+\dnu+\AA+\VV_\para-\BB_\para\right)\,,
	\enar
	where we substituted the linear-order expression for $\de z$ computed in the previous subsection.
	For the spatial components of the distortion $\DX^\al_s$, from the exact relation in Eq.\,($\ref{DXal}$) we derive instead
	\bear\label{Dx1}
	\DX_s^\al=\de x_o^\al-n^\al(\DT_s-\de\eta_o)+\int_0^{\rbar_z}d\rbar\left( \dea^\al -n^\al\dnu
	\right)\,,
	\enar
	as well as the following linear-order relation valid at any point along the background light path
	\bear\label{dx1}
	\frac{d}{d\rbar}\left( \DX^\al+n^\al \DT \right)=\left(\dea^\al -n^\al\dnu\right)\,.
	\enar
	We stress that the left-hand side of this equation is evaluated at a given comoving distance $\rbar$, while the right-hand side at the background position $\bar x^\mu_{\rbar}$. In analogy with the observer lapse discussed in Subsection $\ref{subsec:sub}$, we can define the coordinate distortion $\DX^\mu(\bar x_{\rbar})$ as a field along the background path, instead of a function of affine parameter.
	
	Given the physical parameterization of the source position in Eq.\,($\ref{source}$), we can use the previous results to compute the radial and transverse distortions:
	\bear\label{deltar}
	\drr&=&n_\alpha  \DX^\alpha_s=n_\al\de x_o^\al-\Dcc_s+\int_0^{\rbar_z}d\rbar\,\dea_\para
	\nnn
	&=&
	\de x_{o \para}+\dT_o-{\dz\over\HH_z}+\int_0^{\rbar_z}d\rbar~\left(
	\AA-\BBP-\CCPP \right)~,
	\enar
	\bear\label{deltat}
	\rbar_z\dtt&=&\ttt_\alpha \DX^\alpha_s=\ttt_\al\de x_o^\al+\int_0^{\rbar_z}d\rbar\, \ttt_\al\dea^\al
	\nnn
	&=&
	-\int_0^{\rbar_z}d\rbar~\left[\theta_\alpha
	\left(\BB^\alpha+2~\CC^\alpha_\para\right)
	+\left({\rbar_z-\rbar\over
		\rbar}\right){\partial\over\partial\ttt}
	\left(\AA-\BBP-\CCPP \right)
	\right]
	\nnn
	&&
	+\ttt_\al\de x_o^\al+\rbar_z\ttt_\alpha\left(\dea^\al+\BB^\alpha+2~\CC^\alpha_\para\right)_{\bobs}\,,
	\enar
	\bear\label{deltap}
	\rbar_z\sin\ttt\dpp&=&\pp_\alpha \DX^\alpha_s=\pp_\al\de x_o^\al+\int_0^{\rbar_z}d\rbar\, \pp_\al\dea^\al
	\nnn
	&=&
	-\int_0^{\rbar_z}d\rbar~\left[\pp_\alpha
	\left(\BB^\alpha+2~\CC^\alpha_\para\right)
	+\left({\rbar_z-\rbar\over
		\rbar\sin\ttt}\right){\partial\over\partial\pp}
	\left(\AA-\BBP-\CCPP \right)
	\right]
	\nnn
	&&
	+\pp_\al\de x_o^\al+\rbar_z\pp_\alpha\left(\dea^\al+\BB^\alpha+2~\CC^\alpha_\para\right)_{\bobs}\,,
	\enar
	where we used $\bar x^\al_z=\rbar_z n^\al(\ttt,\pp)$ to recast the partial derivatives.
	\\
	
	\Ietc{Second-order source position}
	Expanding up to second order the exact relation between $\DT_s$ and~$\de z$ in Eq.\,($\ref{dzrel}$) we obtain
	\bear
	\de z=\HH_z\DT_s+\frac12\left(\HH^2_z+\HH'_z  \right)\DT^2_s\,,
	\enar
	again, we used the fact that $\de z$ has functional dependence only on $x^\mu_s$, not on $x^\mu_o$.
	Substituting the linear-order expression for $\DT_s$ we can solve the above equation for the second-order temporal distortion:
	\bear\label{deta2}
	\DT_s=\frac{\de z}{\HH_z}-\frac12\frac{\HH^2_z+\HH'_z}{\HH_z}\left(\frac{\de z}{\HH_z}\right)^2\,.
	\enar
	For the spatial distortion $\DX^\al_s$ we perturb the exact definition in Eq.\,($\ref{DXal}$) in the same way we did for $\dnu$ and $\dea^\al$, i.e., we expand around the background both the integrand and the integration limit. We obtain the following result
	\bear
	\DX_s^\al&=&\int_0^{\rbar_z}d\rbar\left[ \dea^\al -n^\al\dnu+\de x^\mu_{\rbar}\pa_\mu\left( \dea^\al-n^\al\dnu\right)\big\rvert_{\bar x_{\rbar}}\right]
	\nnn
	&&
	+\de x_o^\al-n^\al(\DT_s-\de\eta_o)-\Dcc_s\left( \dea^\al-n^\al\dnu\right)\big\rvert_{\bar x_z}\,.
	\enar
	We can further simplify this expression using Eq.\,($\ref{dx1}$) to replace $\dea^\al-n^\al\dnu$ evaluated along the background path:
	\bear\label{Dx2}
	\DX_s^\al&=&\int_0^{\rbar_z}d\rbar\left[ \dea^\al -n^\al\dnu+\de x^\mu_{\rbar}\pa_\mu\frac{d}{d\rbar}(\DX^\al+n^\al\DT)
	\right]\bigg\rvert_{\bar x_{\rbar}}
	\nnn
	&&
	+\de x_o^\al-n^\al(\DT_s-\de\eta_o)-\Dcc_s\frac{d}{d\rbar}(\DX^\al+n^\al\DT)\big\rvert_{\bar x_z}\,,
	\enar
	where the field definition of $\de x^\mu$ has been used.
	We can now integrate by part and, making use of Eqs.\,($\ref{Dxcc}$) and ($\ref{dxder}$), we can manipulate the above equation to finally obtain 
	\bear\label{dx2}
	\DX_s^\al&=&\int_0^{\rbar_z}d\rbar\left[ \dea^\al -n^\al\dnu+\dnu\left(\DX'^\al+n^\al\DT'\right)-\dea^\be\left(\DX^\al{}_{,\be}+n^\al\DT_{,\be}\right)
	\right]\bigg\rvert_{\bar x_{\rbar}}
	\nnn
	&&
	+\de x_o^\al-n^\al(\DT_s-\de\eta_o)+\left[\DX_{\rbar}^\mu\,\pa_\mu\left( \DX^\al+n^\al\DT\right)\big\rvert_{\bar x_{\rbar}}\right]^z_{\bobs}\,,
	\enar
	where the partial derivatives of $\DX^\mu$ act on the linear-order expression, and the square brackets represent the difference between the terms evaluated at $\rbar_z$ and $\rbar_{\rm o}$.
	
	We can proceed to compute the radial and transverse distortions as before. By a direct computation we find
	\bear\label{dangoli2}
	\drr&=&n_\al\DX^\al_s+\frac{\rbar_z}{2} \left( \dtt^{2}+\sin^2\ttt~\dpp^{2} \right)\,,
	\nnn\nnn
	\rbar_z\dtt&=&\ttt_\al\DX^\al_s-\drr\dtt +\frac{\rbar_z}{2}\cos\ttt\sin\ttt~\dpp^2\,,
	\nnn\nnn
	\rbar_z\sin\ttt\dpp&=&\pp_\al\DX^\al_s-\sin\ttt~\drr\dpp-\rbar_z\cos\ttt~\dtt\dpp\,,
	\enar
	where again we do not make any substitution, but we derived explicit expressions in terms of metric and observer velocity perturbations for each term appearing.

	\subsection{Distortion in the physical volume and area}
	
	We conclude the section by presenting the second-order expressions for the fluctuations in the volume and area occupied by the source in its rest frame compared to the observed volume and area. The relevant nonlinear equations are Eq.\,($\ref{deV}$) for $\de V$ and Eq.\,($\ref{deA}$) for $\de A$. The distortion in the observed proper area of the source $\de A$ is related to the perturbations $\de\dD$ in the luminosity distance or angular diameter distance via Eq.\,($\ref{dA}$).
	
	For convenience we introduce the distortion matrix $\DD$, usually employed in the standard formalism of gravitational lensing to compare the source angular position $(\ttt+\dtt,\pp+\dpp)$ with the observed angular position $(\ttt,\pp)$ \cite{YZ}:
	\bear
	\DD:=\frac{\partial(\theta+\de\theta,\phi+\de\phi)}{\partial(\theta,\phi)}\,.
	\enar
	Given this definition, we can compute the determinant of the distortion matrix
	\bear\label{dist}
	\up{det}\,\mathbb{D}=\frac{\sin(\ttt+\dtt)}{\sin\ttt}\left( 1+{\pa\over\pa \ttt}\dtt+{\pa\over\pa \pp}\dpp+{\pa\over\pa \ttt}\dtt
	{\pa\over\pa \pp}\dpp-{\pa\over\pa \pp}\dtt{\pa\over\pa \ttt}\dpp \right)=:1+\dDD\,,
	\enar
	where the nonlinear perturbation $\de\DD$ has been introduced. Here we introduced the distortion matrix $\DD$ merely to simplify the second-order expression for $\de V$ and $\de A$, which would be too lengthy otherwise. At linear order, the perturbation $\de\DD$ reads
	\bear
	\de\mathbb{D}=\left(\cot\ttt+{\pa\over\pa \ttt}\right)\dtt+{\pa\over\pa \pp}\dpp=: -2\kappa\,,
	\enar
	with $\kappa$ being the standard gravitational
	lensing convergence describing the linear-order lensing magnification. Already apparent at the linear order, $\de\mathbb{D}$ is gauge-dependent. At second order we obtain
	\bear\label{dDD2}
	\de\mathbb{D}&=&\left(\cot\ttt+{\pa\over\pa \ttt}\right)\dtt+{\pa\over\pa \pp}\dpp+{\pa\over\pa \ttt}\dtt{\pa\over\pa \pp}\dpp-{\pa\over\pa \pp}\dtt{\pa\over\pa \ttt}\dpp
	+\cot\ttt\,\dtt\left({\pa\over\pa \ttt}\dtt+{\pa\over\pa \pp}\dpp\right)
	-\frac12\dtt^2\,.\nnn
	\enar
	To derive the second-order expressions for $\de V$ and $\de A$ we first need to expand up to second-order the physical parameterization of the source position in Eq.\,($\ref{source}$), which together with
	\bear
	{\pa\over\pa z}\bar\eta_z=-\frac1{H_z}\,,\Dquad {\pa\over\pa z}\rbar_z=\frac1{H_z}\,,
	\enar
	obtained from the definition of the comoving distance $\rbar_z$ in Eq.\,($\ref{dr}$), allows us to recast the terms in Eq.\,($\ref{dV}$) for $\up dV$.
	From direct computation we obtain
	\bear
	\varepsilon_{\eta\alpha\beta\gamma}~{\partial x^\alpha\over\partial\zz}
	{\partial x^\beta\over\partial\ttt}{\partial x^\gamma\over\partial\pp}
	&=&{\rbar^2_z\sin\ttt\over H_z}\Bigg\{1+2~{\drr\over\rbar_z}+\de\mathbb{D}
	+H_z{\partial\over\partial\zz}~\drr
	+{\drr^2\over\rbar^2_z}+2~{\drr\over\rbar_z}
	\left(H_z{\partial\over\partial\zz}~\drr-2~\kappa\right)
	\nnn
	&&-2\kappa~H_z{\partial\over\partial\zz}\drr
	-H_z{\partial\over\partial\zz}~\dtt~{\partial\over\partial\ttt}~\drr
	-H_z{\partial\over\partial\zz}~\dpp~{\partial\over\partial\pp}~\drr
	\Bigg\}~,
	\\
	\varepsilon_{\al \mu\nu\rho}~\VV^\al~{\partial x^\mu\over\partial\zz}
	{\partial x^\nu\over\partial\ttt}{\partial x^\rho\over\partial\pp}
	&=&{\rbar^2_z\sin\ttt\over H_z}\Bigg\{\VV_\para\left(1+2~{\drr\over\rbar_z}
	-2~\kappa-H_z{\partial\over\partial\zz}~\DT\right)+\VV_\ttt~\dtt+\VV_\pp\sin\ttt~\dpp
	\nnn
	&&
	-{1\over\rbar_z}\left(\VV_\ttt{\partial\over\partial\ttt}
	+{\VV_\pp\over\sin\ttt}{\partial\over\partial\pp}\right)
	\left(\drr+\DT\right)\Bigg\}~,
	\enar
	where we suppressed the subscript $s$. Substituting these equations in the definition of $\de V$ in Eq.\,($\ref{deV}$) yields
	\bear\label{DV2}
	\dV&=& 3\dz+\dg+2{\drr\over\rbar_z}+\dDD
	+H_z{\pa\over\pa z}\drr+\dUU+\VV_\para+3\,\dz\dg+3~\dz^2
	-\AA\left(2\,{\drr\over\rbar_z}-2\kappa
	+H_z{\pa\over\pa z}\drr\right)
	\nnn
	&&
	+{\drr^2\over\rbar^2_z}+2{\drr\over\rbar_z}
	\left(H_z{\pa\over\pa z}\drr-2\kappa\right)
	-2\kappa H_z{\pa\over\pa z}\drr
	-H_z{\pa\over\pa z}\de\ttt{\pa\over\pa \ttt}\drr
	-H_z{\pa\over\pa z}\dpp{\pa\over\pa \pp}\drr+\VV_\theta\dtt
	\nnn
	&&
	+\VV_\para\left(2{\drr\over\rbar_z}
	-2\kappa-H_z{\pa\over\pa z}\DT\right)
	-{1\over\rbar_z}\left(\VV_\theta{\pa\over\pa \ttt}
	+{\VV_\phi\over\sin\ttt}{\pa\over\pa \pp}\right)
	\left(\drr+\DT\right)+\VV_\phi\sin\ttt\dpp
	\nonumber \\
	&&
	+(\dg+3~\dz)
	\left(2{\drr\over\rbar_z}-2\kappa+H_z{\pa\over\pa z}\drr
	-\AA+\VV_\para\right)
	+\DX^\mu\pa_\mu\left( 3\,\dz+\dg
	-\AA+\VV_\para \right)~,
	\enar
	where the left-hand side is evaluated at the source position $\de V(x_s)$, while the right-hand side is evaluated at the background reference position $\bar x^\mu_z$. The last bracket expresses the expansion of the fields around these coordinates; once again, we point out that $\drr$, $\dtt$, $\dpp$, and $\DT$ are not fields, so they do not contribute to such expansion. To keep the expression as concise as possible, we did not express all the second-order terms and most of the linear-order ones. 
	A further complication in the expression above is that spacetime fields and observed angles mix; both derivatives concerning observables and spacetime coordinates also appear. These aspects will be treated in more detail when we consider gauge transformations in the next section.
	
	We can proceed in the same way to compute the distortion in the observed proper area element of the source. From direct computation, we derive
	\bear
	\varepsilon_{\al\mu\nu\rho}~\VV^\al\left(a\NN^\mu\right)
	{\partial x^\nu\over\partial\ttt}{\partial x^\rho\over\partial\pp}&=&
	\rbar_z^2
	\sin\ttt\Bigg\{-\VV_\para\left(\dhnu-\AA-\dnu\right)-{1\over\rbar_z}
	\left(\VV_\ttt{\partial\over\partial\ttt}+{\VV_\pp\over\sin\ttt}{\partial\over
		\partial\pp}\right)\DT\Bigg\}~,\,\,\,\,\qquad
	\\
	\varepsilon_{\eta\alpha\beta\gamma}\left(a\NN^\alpha_s\right)
	{\partial x^\beta_s\over\partial\ttt}{\partial x^\gamma_s\over\partial\pp}&=&\rbar_z^2
	\sin\ttt\Bigg\{1+
	\VV_\para+\eP-\dhnu+2~{\drr\over\rbar_z}+\de\DD
	+2~{\drr\over\rbar_z}~\left(\VV_\para+\dea_\para-\dhnu\right)
	\nnn
	&&
	-2~\kappa\left(2~{\drr\over\rbar_z}+\VV_\para+\dea_\para-\dhnu\right)
	+{\drr^2\over\rbar^2_z}-\frac12\dtt^2-{1\over2}\left(\sin\ttt~\dpp\right)^2
	\nonumber \\
	&&
	-{1\over\rbar_z}\left[\left(\VV_\ttt+\eTT\right){\partial\over\partial\ttt}
	+{\left(\VV_\pp+\ePP\right)\over\sin\ttt}{\partial\over\partial\pp}\right]\drr
	+\left(\VV_\pp+\ePP\right)\sin\ttt~\dpp
	\nonumber \\
	&&
	+\left(\VV_\ttt+\eTT\right)\dtt+\left(\dtt{\partial\over\partial\ttt}+\dpp{\partial\over\partial\pp}
	\right){\drr\over\rbar_z}+\dhnu\left(\dhnu-\dea_\para\right)\Bigg\}~,\,\,\,\,\,\,\qquad
	\enar
	which can be further simplified using the linear-order null condition in Eq.\,($\ref{null}$) and the linear-order expression for $\dhnu$ in Eq.\,($\ref{dhnu2}$). Substituting these results in Eq.\,($\ref{deA}$) for $\de A$ we obtain
	\bear\label{DA2}
	\de A&=& 2\dz+\dg+2{\drr\over\rbar_z}+\dDD
	+\dea_\para+\dUU+\VV_\para
	-\dhnu-\AA\left(-2\kappa+2\frac{\drr}{\rbar_z}-\CCPP  \right)
	+\VV_\para\left(\BBP-\VV_\para\right)
	\nnn
	&&
	-\frac{1}{\rbar_z}\left( \VV_\ttt{\pa\over\pa \ttt}+\frac{\VV_\pp}{\sin\ttt}{\pa\over\pa \pp} \right)\DT-2\kappa\left(2\frac{\drr}{\rbar_z}-\CCPP\right)-2\frac{\drr}{\rbar_z}\CCPP
	+\frac{\drr^2}{\rbar^2_z}-\frac12\dtt^2-\frac12(\sin\ttt\,\dpp)^2
	\nnn
	&&
	+\dhnu\left(\VV_\para+\CCPP\right)+\dtt\left(\VV_\ttt+\dea_\ttt\right)
	-\frac{1}{\rbar_z}\left[\left(\VV_\ttt+\dea_\ttt\right){\pa\over\pa \ttt}+\frac{1}{\sin\ttt}\left(\VV_\pp+\dea_\pp\right){\pa\over\pa \pp}\right]\drr
	\nnn
	&&
	+\frac{1}{\rbar_z}\dtt{\pa\over\pa \ttt}\drr
	+\sin\ttt\,\dpp\left(\VV_\pp+\dea_\pp\right)+\frac{1}{\rbar_z}\dpp{\pa\over\pa \pp}\drr+\de z^2+2\,\de z\dg
	\nnn
	&&
	+\left( 2\,\de z+\dg\right)\left(  -\AA+2\frac{\drr}{\rbar_z}-2\kappa-\CCPP\right)
	+\DX^\mu\pa_\mu\left( 2\,\de z+\dg-\AA+\VV_\para+\dea_\para-\dhnu \right)\,,\,\,\,\,\qquad
	\enar
	The same remarks made for $\de V$ apply equivalently to $\de A$. Similar expressions are derived in \cite{ld2,ld3,ld8,ultimo,gc0,gc1,gc2}. In Section $\ref{sec:compa}$ we compare our results with previous work.

	\section{Second-order Gauge Transformations}
	\label{sec:gauge}	
	
	To test the validity of the second-order expressions derived in the previous section, we will now turn our attention to the gauge transformation properties of each of those expressions.
	Resuming the discussion of Subsection $\ref{subsec:gauge}$, we can compute the gauge transformation of a tensor using Eq.\,($\ref{gheigi}$), which at second order reads
	\bear\label{ancoragheigi}
	\widetilde{K}^{\mu\cdots \nu}_{\rho\cdots\si}(\widetilde x_p=x_p) = K^{\mu\cdots \nu}_{\rho\cdots\si}(x_p)- \left(\pounds_\zeta K^{\mu\cdots \nu}_{\rho\cdots\si}\right)(x_p) +\frac{1}{2} \left(\pounds^2_\zeta K^{\mu\cdots \nu}_{\rho\cdots\si}\right)(x_p) \,.
	\enar
	Recall that we evaluate both sides of the equation at value $x_p^\mu$ so that the background contributions of the tensor $ \bar K^{\mu\cdots \nu}_{\rho\cdots\si}(t_p)$ are identical in the two frames.
	As anticipated, the gauge field $\zeta(x)$ is treated as a perturbation, and it is expanded according to Eq.\,($\ref{exp}$):
	\bear
	\zeta^\mu(x)=\zeta^{\mu(1)}(x)+\zeta^{\mu(2)}(x)+\cdots\,,
	\enar
	the same kind of expansion holds for the gauge field $\xi^\mu(x)$, which will be our preferred choice since it makes computations simpler. From Eqs.\,($\ref{xzeta}$) and ($\ref{xcsi}$), the second-order relation between the two gauge fields is given by
	\bear\label{zetacsi}
	\zeta^{\mu}=\xi^{\mu}-\frac{1}{2} \xi^{\nu}{\pa\over\pa x^\nu}\xi^{\mu}\,,
	\enar
	which can be expressed in components using Eq.\,($\ref{xicomp}$):
	\bear
	\zeta^\eta=T-\frac{1}{2} T'T  -\frac{1}{2} T_{,\al} \LL^{\al}
	\,,\Dquad
	\zeta^\al=\LL^\al-\frac{1}{2} \LL'^\al T -\frac{1}{2} \LL^\al{}_{,\be} \LL^{\be}\,.
	\enar
	
	With these relations we can readily recast the second-order gauge transformation in Eq.\,($\ref{ancoragheigi}$) in terms of our preferred gauge field $\xi^\mu$. Since we will deal only with scalars, vectors, and symmetric tensors with two covariant indices, the only transformation laws we need to consider are:
	\bear\label{gauge2}
	\widetilde K&=&K-\xi^\mu K_{,\mu}+\xi^\mu\xi^\nu{}_{,\mu}K_{,\nu}+\frac12\xi^\mu\xi^\nu K_{,\mu\nu}
	\nnn
	\widetilde K^\mu&=&K^\mu-\pounds_\xi K^\mu +\frac12\pounds^2_\xi K^\mu+\frac12\xi^\rho\xi^\nu{}_{,\rho}K^\mu{}_{,\nu}-\frac12\left(\xi^\rho{}_{,\nu}\xi^\mu{}_{,\rho}+\xi^\rho\xi^\mu{}_{,\nu\rho}\right)K^\nu
	\nnn
	\widetilde K_{\mu\nu}&=&K_{\mu\nu}-\pounds_\xi K_{\mu\nu}+\frac12\pounds^2_\xi K_{\mu\nu}+\frac12\xi^\si\xi^\rho{}_{,\si}K_{\mu\nu,\rho}+K_{\rho(\mu}\left[ \xi^\si{}_{,\nu)}\xi^\rho{} _{,\si}+ \xi^\si\xi^\rho{}_{,\nu)\si}\right]
	\,,
	\enar
	where the Lie derivative is expressed in terms of $\xi^\mu$ and both sides are evaluated at the same coordinate value $x^\mu_p$. Notice that while a scalar field does not transform under a change of coordinates, it indeed transforms non-trivially under a gauge transformation. In general, we are interested in the gauge transformation of the perturbation variables. Splitting a tensor $K$ into the background and the perturbation $K=\bar K+\de K$ we derive the gauge transformation law for the perturbations:
	\bear
	\widetilde{\de K}&=&\de K-\bar K'T-\xi^\mu\de K_{,\mu}+\bar K'T_{,\mu}\xi^\mu+\frac12\bar K'' T^2
	\nnn
	\widetilde{\de K}{}^\mu&=&\de K^\mu-\bar K'^\mu (T-T_{,\nu}\xi^\nu)+\bar K^\nu(\xi^\mu{}_{,\nu}-\xi^\rho\xi^\mu{}_{,\nu\rho})+ \xi^\mu{}_{,\nu}\de K^\nu-\xi^\nu\de K^\mu{}_{,\nu}-\bar K'^\nu T\xi^\mu{}_{,\nu}+ \frac12\bar K''^\mu T^2
	\nnn
	\widetilde{\de K}_{\mu\nu}&=&\de K_{\mu\nu}-2\bar K_{\rho(\mu}\xi^\rho{}_{,\nu)}-T\bar K_{\mu\nu}-\xi^\rho\de K_{\mu\nu,\rho}-2\de K_{\rho(\mu}\xi^\rho{}_{,\nu)}+2\bar K_{\rho(\mu}\left[\xi^\rho{}_{,\nu)\si}\xi^\si+\xi^\si{}_{,\nu)}\xi^{\rho}{}_{,\si}\right]\nnn
	&&
	+\bar K_{\rho\si}\xi^\rho{}_{,\mu}\xi^\si{}_{,\nu}+2\bar K'_{\rho(\mu}\xi^\rho{}_{,\nu)}T+\frac12 \bar K''_{\mu\nu}T^2+\bar K'_{\mu\nu}T_{,\rho}\xi^\rho\,,
	\enar
	where background quantities are purely time-dependent due to the symmetries of the FLRW solution.
	
	In Eq.\,($\ref{xz}$) we parameterized the source position in terms of the observed position $\bar x^\mu(\cc_z)$ parameterized by the observed redshift and angles
	\bear
	\bar x^\mu(\cc_z)=\left(\bar\eta_z,\rbar_z n^\al\right)^\mu\Dquad
	n^\al=\left(\sin\ttt\cos\pp , \sin\ttt\sin\pp,\cos\pp \right)^\al\,,
	\enar
	and its deviation $\DX^\mu$ around the observed position.
	Here we emphasize why this is more advantageous than the splitting around $\bar x^\mu(\cc_s)$. As we have extensively discussed, the observed redshift $z$ and the observed angles $\ttt,\pp$ are independent of our FLRW coordinate description; hence they are scalars. From Eqs.\,($\ref{etaz}$) and ($\ref{dr}$) we note that these scalars fully determine the background source position $\bar x^\mu_z$. Therefore, the reference coordinates $\bar x^\mu_z$ will not be affected by diffeomorphisms of $\mM$. On the contrary, the coordinates $\bar x^\mu_s$ of the source will be affected by the coordinate transformation, so the parameterization of the source position in terms of deviations around these coordinates is not a convenient choice to deal with gauge transformations. Given the
	boundary relations in Eq.\,($\ref{z0}$), the background observer position $\bar x^\mu(\cc_o)=(\bar\eta_{\rm o},0)^\mu$ also remains unaffected by a change of coordinates. With these remarks, we can readily see that the transformations induced by the gauge field~$\xi$ in Eq.\,($\ref{xcsi}$) will involve only the deviations $\DX^\mu_s$ and $\de x^\mu_o$, such that we can readily derive the second-order gauge transformation
	\bear\label{trasfx2}
	{\widetilde\DX}^\mu_s=\DX^\mu_s+\xi^\mu(\bar x_z)+\DX^\nu_s\left(\frac{\pa}{\pa x^\nu}\xi^\mu\right)\bigg\rvert_{\bar x_z}
	\,,
	\qquad
	{\widetilde{\de x}}^\mu_o=\de x_o^\mu+\xi^\mu(\bar x_{\rm o})+\de x^\nu_o\left(\frac{\pa}{\pa x^\nu}\xi^\mu\right)\bigg\rvert_{\bar x_{\rm o}}\,,\qquad
	\enar
	where the fields are expanded as usual around the background reference points.
	
	In the rest of the section, we will use gauge transformations as a consistency check of the second-order expressions derived in Section $\ref{sec:second}$. First, we compute their \textit{expected} gauge transformations employing the nonlinear relations and exact definitions given in Section $\ref{sec:nonlinear}$. Second, we directly compute the gauge transformations of the second-order expressions in Section $\ref{sec:second}$ in terms of metric perturbations and observer velocity. The consistency of these two independent procedures provides a strong validity test for our second-order expressions in Section $\ref{sec:second}$. For instance, the gauge transformation of $\DX^\mu_s$ in Eq.\,($\ref{trasfx2}$) is the expected transformation from its definition. We can then use its expression in Eqs.\,($\ref{deta2}$) and ($\ref{dx2}$) to check the consistency of the gauge transformation.

	\subsection{Metric and observer velocity perturbations}
	\label{subsec:4.1}
	
	The gauge transformations of the metric perturbations and the observer velocity will play a crucial role in the consistency check because all the second-order expressions in Section $\ref{sec:second}$ finally depend on these quantities. We first apply the second-order gauge transformation equation ($\ref{gauge2}$) to the metric~$g_{\mu\nu}$ in Eq.\,($\ref{metrica}$) and the observer velocity $u^\mu$ in Eq.\,($\ref{uu}$). After straightforward computations, we obtain the second-order gauge transformations of the metric perturbations \cite{YZ}
	\bear
	\label{eq:gtaa}
	{\widetilde\AA}&=& \AA-\left(T^{\prime} +\HH T \right)- \AA^\prime T
	- 2 \AA \left(T'+\HH T\right)+{3\over2}T'^2
	+TT^{\prime\prime} +{1 \over 2}
	\left( 2\HH^2+\HH'\right) T^2  \nnn
	&&
	+ 3\HH T^{\prime}T-\AA_{,\alpha}\LL^\alpha-\BB_\alpha\LL'^{\alpha}
	+T_{,\alpha}\LL'^{\alpha}
	+\LL^\alpha \left(T^{\prime}_{,\alpha}+\HH T_{,\alpha} \right)
	- {1\over 2}\LL'^{\alpha} \LL^\prime_{\alpha}~,
	\enar
	\bear
	\label{eq:gtbb}
	{\widetilde\BB_\alpha}&=&\BB_\alpha -T_{,\alpha}+ \LL^\prime_\alpha
	- 2\AA T_{,\alpha}- \left(\BB^\prime_\alpha + 2 \HH\BB_\alpha \right) T
	- \BB_\alpha T^{\prime}+2 T^{\prime} T_{,\alpha}
	+T \left(T^{\prime}_{,\alpha}+ 2\HH T_{,\alpha} \right)   \nnn
	&&
	-\BB_{\alpha,\beta}\LL^\beta-\BB_\beta
	\LL^\beta_{\;\;,\alpha}+ 2\CC_{\alpha\beta} \LL'^{\beta}
	-\LL^\prime_\alpha T^{\prime}
	+ T_{,\beta}\LL^\beta_{\;\;,\alpha}+\LL^\be T_{,\alpha\be}-T \left( \LL^{\prime\prime}_\alpha+2\HH\LL^\prime_\alpha \right)
	\nonumber \\ 
	&&-2
	\LL_{(\al,\be)}\LL'^{\be}-\LL^\be
	\LL'_{\al,\be}~,
	\enar
	\bear
	\label{eq:gtcc}
	{\widetilde{\CC}_{\alpha\beta}} &=& \CC_{\alpha\beta}-\HH T\de_{\alpha\beta}-\LL_{(\al,\beta)}
	+\BB_{(\alpha} T_{,\beta)}-\left(\CC^\prime_{\alpha\beta}+
	2 \HH\CC_{\alpha\beta} \right)T
	+ {1 \over 2} \left(
	2\HH^2+\HH'\right)  T^2 \de_{\alpha\beta}  -{1\over 2}T_{,\alpha}T_{,\beta}
	\nnn
	&&
	+ \HH TT^{\prime}\de_{\alpha\beta} -\CC_{\alpha\beta,\gamma} \LL^\gamma
	- 2\CC_{\gamma(\alpha} \LL^\gamma_{\;\;,\beta)} + \LL^\prime_{(\alpha}
	T_{,\beta)}+\HH\LL^\gamma T_{,\gamma}\de_{\alpha\beta} + 2\HH
	\LL_{(\al,\beta)}T+ 
	\LL'_{(\al,\beta)} T  \nonumber \\
	&&
	+\LL^\ga{}_{,(\al} \LL_{\be),\ga}+\LL^\ga
	\LL_{(\al,\beta)\ga}+\frac12\LL^\ga{}_{,\al}\LL_{\ga,\be}\,,
	\enar
	where we used Eq.\,($\ref{xicomp}$) to express the components of the gauge field $\xi^\mu$. Remember that we chose rectangular coordinates to simplify the calculations. In the same way, we obtain the second-order gauge transformation of the observer velocity
	\bear\label{deutilde}
	\widetilde{\de u}&=&\de u+\HH T+T'+\left(\mathcal{A}'  -  \HH\mathcal{A} -T''\right)T+ \tfrac{1}{2} \left(\mathcal{H}^2  -  \mathcal{H}' \right)T^2 -  \mathcal{A}T'   + \mathcal{U}^{\be} T_{,\beta}
	\nnn
	&&
	+ \LL^\be\left( \mathcal{A}_{,\be} -  \mathcal{H} T_{,\beta} -  T'_{,\beta}\right) \,,
	\enar
	\bear
	{\widetilde{\VV}}^\al&=&{\VV}^\al+ \LL'^\al+\left(\HH T-\AA\right) \LL'^\al  +\left(\HH\VV^\al-
	\LL''^\al-\mathcal{U}'^{\al} \right)T   + \mathcal{U}^{\be} \LL^\al{}_{,\beta} 
	\nnn
	&&
	-  \LL^\beta\left( \mathcal{U}^{\al}{}_{,\beta}+\LL'^\al{}_{,\beta}\right)\,.
	\enar
	We can readily check that Eq.\,($\ref{deutilde}$) is consistent with the transformation obtained by directly substituting the gauge transformations of the metric perturbations derived above in the second-order expression for $\de u$ in Eq.\,($\ref{du2}$).
	
	Note that both sides of each equation presented in this subsection are evaluated at the same coordinate position $x^\mu$. For later convenience, we will assume these coordinates to be either on the background observer path $\bar x^\mu_t$ or on the background light path $\bar x^\mu_{\rbar}$.

	\subsection{Time lapse and spatial shift at the observer position}
	
	We begin the consistency checks by computing the second-order transformations of the time lapse~$\de t_o$ and spatial shift $\de x^\al_o$ at the observer position. On the one hand, we compute the transformations from the definitions of the lapse and shift in Eq.\,($\ref{lapseshift}$). On the other hand, we compute the transformations of the second-order expressions in Eqs.\,($\ref{dt2met}$) and ($\ref{dx2met}$) in terms of metric and observer velocity perturbations. The former procedure will be denoted \textit{second-order expectation}, while the latter \textit{second-order direct computation}. We employ this terminology throughout the whole section. By checking that the two procedures give the same result, we ensure the sanity of our second-order expressions. We will only show the second-order gauge-transformation, from which the linear-order gauge transformation can be trivially read off.
	\\
	
	\Ietc{Second-order expectation}
	Let us write in components the transformation of the time lapse and spatial shift in Eq.\,($\ref{trasfx2}$) in the observer position:
	\bear\label{detao}
	{\widetilde{\de\eta}}_o=\de\eta_o+T(\bar x_{\rm o})+\de x^\mu_o\left(\frac{\pa}{\pa x^\mu} T\right)\bigg\rvert_{\bar x_{\rm o}}\,,
	\qquad\quad
	{\widetilde{\de x}}^\al_o=\de x_o^\al+\LL^\al(\bar x_{\rm o})+\de x^\mu_o\left(\frac{\pa}{\pa x^\mu}\LL^\al\right)\bigg\rvert_{\bar x_{\rm o}}\,.\qquad
	\enar
	To derive the expectation for the time lapse in proper time, let us use the relation between $\de t_o$ and $\de\eta_o$ in Eq.\,($\ref{lapseconf}$). Using the gauge transformation of $\de\eta_o$ above, we derive the expectation
	\bear
	{\widetilde{\de t}}_o&=&{\widetilde{\de\eta}}_o+\frac12\HH_{\bobs}\widetilde{\de\eta}_o^2=\de t_o+T_{\bobs}+\de x^\mu_o\left[{\pa\over\pa x^\mu} (aT)\right]\bigg\rvert_{\bobs}+\frac12\HH_{\bobs}T^2_{\bobs}\,.
	\enar
	\\
	
	\Ietc{Second-order direct computation}
	We start the direct computations from the spatial shift at the observer position in Eq.\,($\ref{dx2met}$). Directly substituting the gauge transformations of the metric perturbations and the observer velocity presented in Subsection $\ref{subsec:4.1}$, and using the linear-order transformations of the coordinate perturbations in Eq.\,($\ref{trasfx2}$), we obtain
	\bear
	{\widetilde{\delta x}}^\al_o&=& \widetilde{\delta x}{}_o^\mu\left(
	\pa_\mu\widetilde{\delta x}{}^\al\right)\bigg\rvert_{\bobs}+\int_0^{\bar t_{\rm o}}dt \frac1a
	\left({\widetilde\VV}^\al+\widetilde\AA\,\widetilde\VV^\al-\widetilde\VV^\be\widetilde{\delta x}{}^\al{}_{,\be}\right)
	\nnn
	&=&{\delta x}^\al_o+\de x^\mu_o(\pa_\mu\LL^\al)\big\rvert_{\bobs}+\xi^\mu_{\bobs}\left[\pa_\mu\left( \de x^\al+\LL^\al \right)\right]\big\rvert_{\bobs}
	\nnn
	&&
	+\int_0^{\bar t_{\rm o}}dt\frac1a\bigg[\frac{\pa}{\pa\eta}\bigg(\LL^\al-T\VV^\al-T\LL'^\al
	-\LL^\be\LL^\al{}_{,\be}\bigg)-\LL^\be\VV^\al{}_{,\be}-\LL'^\be\de x^\al{}_{,\be}\bigg]\bigg\rvert_{\bar x_t}\,,
	\enar
	where the boundary terms in the first line come exclusively from the gauge transformation of $\de x_o^\mu(\pa_\mu \de x^\al)\big\rvert_{\bobs}$.
	From the linear-order condition in Eq.\,($\ref{dt1nogeo}$), valid at any point of the background observer path $\bar x^\mu_t$, we can rewrite the last two terms in the above integral as
	\bear
	-\LL^\be\VV^\al{}_{,\be}-\LL'^\be\de x^\al{}_{,\be}=-\frac{\pa}{\pa\eta}\left(\LL^\be\de x^\al{}_{,\be}  \right)\,.
	\enar
	With this result and the relation $dt=a\,d\eta$, we notice that the integrand above is a partial derivative with respect to cosmic time $t$, and the integration yields boundary terms evaluated at the background observer position. Most of the terms canceled in the resulting expression to yield
	\bear
	{\widetilde{\delta x}}^\al_o={\delta x}^\al_o+\LL^\al_{\bobs}+\de x_o^\mu\left({\pa\over\pa x^\mu}\LL^\al\right)\bigg\rvert_{\bobs}\,,
	\enar
	which indeed coincides with the second-order expectation.
	
	We can proceed in the same way for the time lapse at the observer position. We have derived two expressions for the second-order lapse, the general expression in Eq.\,($\ref{dt2nogeo}$) and the simplified expression in Eq.\,($\ref{dt2geo}$) under the assumption that the observer motion is geodesic. We will compute the direct transformation of the general expression:
	\bear
	{\widetilde{\delta t}}_o&=& \widetilde{\delta x}{}^\mu_o(\pa_\mu
	\widetilde{\delta t})\big\rvert_{\bobs}+\int_0^{\bar t_{\rm o}}dt
	\left(-{\widetilde\AA}+\frac12\widetilde\AA^2+\frac12\widetilde\VV^\al\widetilde\VV_\al
	-\widetilde\VV^\al\widetilde\BB_\al-\frac1a\widetilde\VV^\al\widetilde{\delta t}_{,\al}\right)
	~ \nnn
	&=&{\delta t}_o+\de x^\mu_o[\pa_\mu(aT)]\big\rvert_{\bobs}+\xi^\mu_{\bobs}[\pa_\mu\big(\de t+aT\big)]\big\rvert_{\bobs}
	\nnn
	&&
	+\int_0^{\bar t_{\rm o}}dt\,\bigg[\frac{1}{a}\frac{\pa}{\pa\eta}\bigg( aT+a\AA T-a\LL^\al T_{,\al}
	-aTT'-\frac12 a\HH T^2 \bigg)+\LL^\al\AA_{,\al}
	-\frac1a\LL'^\al\de t_{,\al} \bigg]\,.\quad
	\qquad
	\enar
	Again, we can use the linear-order condition in Eq.\,($\ref{dt1nogeo}$), valid on the observer background path, to rearrange the last two terms in the integral
	\bear
	\LL^\al\AA_{,\al}
	-\frac1a\LL'^\al\de t_{,\al}=-\frac1a\frac{\pa}{\pa\eta}\left( \LL^\al\de t_{,\al}  \right)\,,
	\enar
	such that the integration yields boundary terms at the observer position. Simplifying the resulting expression, we obtain
	\bear
	{\widetilde{\de t}}_o=\de t_o+T_{\bobs}+\de x^\mu_o \left[{\pa\over\pa x^\mu}(aT)\right]\bigg\rvert_{\bobs}+\frac12\HH_{\bobs}T^2_{\bobs}\,,
	\enar
	in agreement with the expected result (see \cite{ld9}).

	\subsection{Distortion in the conformal wave vector}
	
	Here we want to compute the second-order gauge transformation of the components of the conformal light wave vector $\hat k^\mu$ defined in Eq.\,($\ref{hatk}$). In the real universe $\mM$, we have argued that the conformal light wave vector is \textit{not} an actual vector, and its components $\dnu$ and $\dea^\al$ depend on the position on the light path $x^\mu_\cc$ and the reference position $x^\mu_o$. As far as gauge transformations are concerned, we are considering an active change of coordinates only for the position on the light path, while the reference is fixed at the physical point $O$ in both frames.\footnote{Note that if we also expanded $\widetilde x^\mu_o$ around $x^\mu_o$, this position would not correspond to the point $O$ in the frame denoted by a tilde.} For this reason, in this section, we will consider the perturbations as a function of the light path only: $\dnu=\dnu(x_\cc)\,,\, \dea^\al=\dea^\al(x_\cc)$.
	\\
	
	\Ietc{Second-order expectation}
	To compute the gauge transformation of $\hat k^\mu$, we first compute the gauge transformation of $k^\mu$ parameterized as in Eq.\,($\ref{kmu}$), since the light wave vector $k^\mu$ is a four-vector in $\mM$ while $\hat k^\mu$ is not. Mind the following subtle point: the left-hand side of Eq.\,($\ref{kmu}$) is evaluated at a point $x^\mu_\Lambda$ in the real universe, while the right-hand side depends on $x^\mu_\cc$ and $x^\mu_o$. Consequently, a change of coordinates in a point of the real universe implies a change of coordinates in two points of the conformal universe.
	
	To compute the second-order expectation of $\widetilde\dnu$ and $\widetilde\dea{}^\al$ we employ the transformation law for the components of a tensor evaluated at a point $P$, given in Eq.\,($\ref{tensor}$). This equation provides the expression for a tensor of arbitrary rank; in the case of a four-vector, it reads
	\bear\label{ktrasf}
	\widetilde k^\mu(\widetilde x_p)=\left(\frac{\partial \widetilde x^\mu}{\partial x^\nu}\right)_pk^\nu(x_p)\,.
	\enar
	In principle, we could also apply Eq.\,($\ref{gauge2}$) directly to derive the sought gauge transformations, but with this procedure we will derive intermediate relations that will be crucial for the consistency check. Substituting in the above equation the parameterization of $k^\mu$ in Eq.\,($\ref{kmu}$) we derive
	\bear\label{imp}
	1+\widetilde\dnu(\widetilde x)&=&\frac{a^2(\widetilde\eta)}{a^2(\eta)}\frac{a(\eta_o)}{a(\widetilde\eta_o)}\frac{\partial \widetilde \eta}{\partial x^\nu}\bigg(1+\dnu(x)\,,\,-n^\al-\dea^\al(x)\bigg)^\nu
	\nnn
	&=&\frac{a^2(\widetilde\eta)}{a^2(\eta)}\frac{a(\eta_o)}{a(\widetilde\eta_o)}\Big[\left(1+T'\right)\left(1+\dnu\right)-T_{,\al}\left(n^\al+\dea^\al \right)\Big](x)\,,
	\enar
	for the zero-th component $\mu=\eta$, and
	\bear\label{imp2}
	-n^\al-\widetilde\dea{}^\al(\widetilde x)&=&\frac{a^2(\widetilde\eta)}{a^2(\eta)}\frac{a(\eta_o)}{a(\widetilde\eta_o)}\frac{\partial \widetilde x^\al}{\partial x^\nu}\bigg(1+\dnu(x)\,,\,-n^\al-\dea^\al(x)\bigg)^\nu
	\nnn
	&=&\frac{a^2(\widetilde\eta)}{a^2(\eta)}\frac{a(\eta_o)}{a(\widetilde\eta_o)}\Big[\LL'^\al\left(1+\dnu\right)-\left(\de^\al_\be+\LL^\al{}_{,\be}\right)\left(n^\be+\dea^\be \right)\Big](x)\,,
	\enar
	for the spatial components $\mu=\al$. In both equations, we suppressed the subscript $p$, and we used $\widetilde\omega_o=\omega_o$, as the observed frequency does not depend on our coordinate description. Note that these equations are valid at any perturbative order at this stage and $n^\al$ is the observed angle in the rest frame of the observer, not a function of position $x^\mu$. By expanding $\widetilde x^\mu$ in Eq.\,($\ref{xcsi}$) around $x^\mu$ in the right-hand side, we derive at the second order:
	\bear\label{dnuPtilde}
	{\widetilde\dnu}(\widetilde x) &=&\dnu -\frac{d}{d\rbar}T +\left(2\HH T-\HH_oT_o\right)\left(1+\dnu -\frac{d}{d\rbar}T\right) + \left(2 \mathcal{H}^2 +  \mathcal{H}'\right)T^2 
	\nnn
	&&
	+T'\dnu- T_{,\al}\dea^\al-2 \mathcal{H} \mathcal{H}_{o} T T_{o} +\tfrac12 \left( \mathcal{H}_o^2 - \mathcal{H}'_o\right)T_o^2 
	\,,
	\enar
	\bear\label{deaPtilde}
	{\widetilde{\dea}}{}^\al(\widetilde x)&=&\dea^\al+\frac{d}{d\rbar}\LL^\al+\left(2\HH T-\HH_oT_o\right)\left(n^\al+\dea^\alpha +\frac{d}{d\rbar}\LL^\al \right) -\LL'^\al\dnu+\LL^\al{}_{,\be}\dea^\be
	\nnn
	&&
	+ \left(2 \mathcal{H}^2 +  \mathcal{H}' \right) T^2n^\alpha
	+\tfrac12 \left( \mathcal{H}_o^2 -  \mathcal{H}'_o\right)T_o^2n^\al  - 2 \mathcal{H} \mathcal{H}_{o}  T T_{o} n^\alpha  \,,
	\enar
	which will be used to check the validity of our expressions. Note that the left-hand side is evaluated at $\widetilde x^\mu$, while the right-hand side at $x^\mu$. To obtain the gauge-transformation relation at the same value, we first expand the left-hand side around $x^\mu$, move the quadratic terms to the right-hand side, and finally use the linear-order relation:
	\bear\label{dnuP}
	{\widetilde\dnu} &=&\dnu -\frac{d}{d\rbar}T + T\frac{d}{d\rbar}T' +\left(2\HH T-\HH_oT_o\right)\left(1+\dnu -\frac{d}{d\rbar}T\right) 
	\nnn
	&&
	-  \mathcal{L}^{\al} \left(\dnu_{,\al}+2 \mathcal{H}T_{,\al}-\frac{d}{d\rbar}T_{,\al}  \right)+ \left(2 \mathcal{H}^2 -  \mathcal{H}'\right)T^2 -  2\mathcal{H}TT'
	\nnn
	&&
	+T'\dnu-T_{,\al}\dea^\al-T\dnu'-2 \mathcal{H} \mathcal{H}_{o} T T_{o} +\tfrac12 \left( \mathcal{H}_o^2 - \mathcal{H}'_o\right)T_o^2 
	\,,
	\enar
	\bear\label{deaP}
	{\widetilde{\dea}}{}^\al&=&\dea^\al+\frac{d}{d\rbar}\LL^\al-T\frac{d}{d\rbar}\LL'^\al+\left(2\HH T-\HH_oT_o\right)\left(n^\al+\dea^\alpha +\frac{d}{d\rbar}\LL^\al \right)  
	\nnn
	&&
	-  \LL^\beta\left( \dea^\al{}_{,\be}+2 \mathcal{H}  T_{,\beta}n^\alpha  +\frac{d}{d\rbar}\LL^\al{}_{,\be} \right)-2\HH TT'n^\al+ \left(2 \mathcal{H}^2 -  \mathcal{H}' \right) T^2n^\alpha
	\nnn
	&&
	-\LL'^\al\dnu+\LL^\al{}_{,\be}\dea^\be-T\dea'^\al+\tfrac12 \left( \mathcal{H}_o^2 -  \mathcal{H}'_o\right)T_o^2n^\al  - 2 \mathcal{H} \mathcal{H}_{o}  T T_{o} n^\alpha  \,.
	\enar
	As a concluding remark we stress again that the coordinates of the reference position have not been expanded as they refer to the physical point $O$. In practice, this means that in the left-hand side the equations read $\widetilde\dnu(\widetilde x=x;\widetilde x_o)\,,\,\ \widetilde\dea^\al(\widetilde x=x;\widetilde x_o)$, while the right-hand sides contains only fields evaluated at $x^\mu$ and $x^\mu_o$.
	\\
	
	\Ietc{Second-order direct computation}
	Our second-order expressions for $\dnu$ and $\dea^\al$ are given in Eq.\,($\ref{dnuint}$) and
	Eq.\,($\ref{deaint}$) respectively. We first check the consistency of the expression for $\dnu$ by computing the gauge transformation
	\bear\label{dnutilde1}
	{\widetilde\dnu}(\widetilde{x}_s)-{\widetilde\dnu}(\widetilde{x}_o)&=&\left[\widetilde\DX{}^\mu_{\rbar}(\pa_\mu\widetilde\dnu)\big\rvert_{\bar x_{\rbar}}\right]^z_{\bobs}+\int_0^{\rbar_z}d\rbar \left({\widetilde\GG}{}^\eta+\widetilde\dnu\widetilde\dnu'-\widetilde\dea{}^\al\widetilde\dnu_{,\al}\right)\bigg\rvert_{\bar x_{\rbar}}
	\nnn
	&=&
	\bigg\{
	\DX^\mu\pa_\mu\left(\dnu+ 2\HH T-\frac{d}{d\rbar}T \right)+T\dnu'+\LL^\al\dnu_{,\al}-T\frac{d}{d\rbar}T'
	\nnn
	&&
	-\LL^\al\frac{d}{d\rbar}T_{,\al}
	+2\HH'T^2+2\HH TT'+2\HH\LL^\al T_{,\al}\bigg\}\bigg\rvert^z_{\bobs}
	\nnn
	&&
	+\int_0^{\rbar_z}d\rbar \left({\widetilde\GG}{}^\eta+\widetilde\dnu\widetilde\dnu'-\widetilde\dea{}^\al\widetilde\dnu_{,\al}\right)\bigg\rvert_{\bar x_{\rbar}}\,,
	\enar
	where only the boundary terms have been expanded, for which we used the gauge transformation of~$\DX^\mu$ in Eq.\,($\ref{trasfx2}$) and the linear-order expression of $\dnu$ in Eq.\,($\ref{dk1s}$). We present the gauge transformation of the integral separately to avoid lengthy expressions:
	\bear\label{dnutilde2}
	&&\int_0^{\rbar_z} d\rbar\,\left( {\widetilde\GG}{}^\eta+\widetilde\dnu\widetilde\dnu'-\widetilde\dea{}^\al\widetilde\dnu_{,\al} \right)\bigg\rvert_{\bar x_{\rbar}}=
	\nnn&&
	\Dquad=\int_0^{\rbar_z} d\rbar\,\left( \GG^\eta+\dnu\dnu'-\dea^\al\dnu_{,\al} \right)\bigg\rvert_{\bar x_{\rbar}}
	+\bigg\{-\frac{d}{d\rbar}T+T\frac{d}{d\rbar}T'+2\HH T\left( 1+\dnu-\frac{d}{d\rbar}T \right)
	\nnn
	&&\Dquad\quad
	-2\HH TT'-\LL^\al\left( \dnu_{,\al}+2\HH T_{,\al}-\frac{d}{d\rbar}T_{,\al} \right)
	-\HH_oT_o\left( \dnu-\frac{d}{d\rbar}T \right)+\left( 2\HH^2-\HH' \right)T^2
	\nnn
	&&\Dquad\quad
	-T\dnu'+T'\dnu-T_{,\al}\dea^\al-2\HH\HH_oTT_o\bigg\}\bigg\rvert^z_{\bobs}\,,
	\enar
	where we used $\GG^\eta$ in Eq.\,($\ref{GG2}$), $\dnu$ and $\dea^\al$ in Eq.\,($\ref{dk1s}$), and we stress that all the fields inside the curly brackets are evaluated along the background light path $\bar x^\mu_{\rbar}$. Combining the two previous equations ($\ref{dnutilde1}$) and ($\ref{dnutilde2}$), we obtain after straightforward simplifications:
	\bear\label{direct}
	{\widetilde\dnu}(\widetilde{x}_s)-{\widetilde\dnu}(\widetilde{x}_o)&=&{\dnu}(x_s)-{\dnu}(x_o)+\bigg\{-\frac{d}{d\rbar}T+2\HH T+\DX^\mu\pa_\mu\left( 2\HH T-\frac{d}{d\rbar}T \right)
	\nnn
	&&
	+T'\dnu-T_{,\al}\dea^\al+\left(2\HH T
	-\HH_oT_o\right)\left( \dnu-\frac{d}{d\rbar}T \right)+\left(2\HH^2+\HH'\right)T^2
	\nnn
	&&
	-2\HH\HH_oTT_o\bigg\}\bigg\rvert^z_{\bobs}\,.
	\enar
	The second-order expectation for this quantity is given by evaluating Eq.\,($\ref{dnuPtilde}$) at the source position $P=S$ and the observer position $P=O$. The difference of these two contributions yields
	\bear\label{expect}
	{\widetilde\dnu}(\widetilde{x_s})-{\widetilde\dnu}(\widetilde{x_o})&=&{\dnu}(x_s)-{\dnu}(x_o)+\bigg\{-\frac{d}{d\rbar}T+2\HH T
	+T'\dnu-T_{,\al}\dea^\al
	\nnn
	&&
	+\left(2\HH T-\HH_oT_o\right)\left( \dnu-\frac{d}{d\rbar}T \right)+\left(2\HH^2+\HH'\right)T^2
	-2\HH\HH_oTT_o\bigg\}\bigg\rvert^s_o\,.
	\qquad
	\enar
	There seems to be a discrepancy between the two equations but note the different evaluation positions. The fields in the curly brackets in Eq.\,($\ref{direct}$) are evaluated at coordinates $\bar x^\mu_z$ and $\bar x^\mu_o$ along the background paths $\bar x^\mu_{\rbar}$. On the other hand, the fields in the curly brackets of the expectation in Eq.\,($\ref{expect}$) are evaluated at coordinates $x^\mu_s$ and $x^\mu_o$ along the real path $x^\mu_\cc$.
	When we discussed subtleties at second order in Subsection $\ref{subsec:sub}$, we argued that evaluating a field either on the real path or on the background path makes a difference at second order, and the discrepancy is given in Eq.\,($\ref{2bar}$). Hence, we need the following second-order relations to compare the two equations properly:
	\bear
	\frac{d}{d\rbar}T\bigg\rvert_{x}=\frac{d}{d\rbar}T\bigg\rvert_{\bar x}+\DX^\mu\left[\pa_\mu\left(\frac{d}{d\rbar} T\right)\right]\bigg\rvert_{\bar x}\,,\qquad
	(\HH T)(x)=\HH(\bar\eta)T(\bar x)+\DX^\mu\pa_\mu\left(\HH T\right)\bigg\rvert_{\bar x}\,,\quad
	\enar
	where we do not show any subscript as these equations are valid both for source and observer positions. Moreover, with $\DX^\mu$ we refer both to $\DX^\mu_s$ and $\de x^\mu_o$.
	Substituting these relations in Eq.\,($\ref{expect}$), we readily check that the expectation and the direct computation agree and indeed the apparent discrepancy was just a consequence of different evaluation positions. 
	
	We can repeat the same procedure for the spatial perturbation $\dea^\al$ given in Eq.\,($\ref{deaint}$). The transformation of the boundary terms is given by
	\bear
	\left[\widetilde\DX{}^\mu_{\rbar}(\pa_\mu\widetilde\dea{}^\al)\big\rvert_{\bar x_{\rbar}}\right]^z_{\bobs}&=&
	\bigg\{
	\DX^\mu\pa_\mu\left(\dea^\al+ 2\HH Tn^\al+\frac{d}{d\rbar}\LL^\al \right)+T\dea'^\al+\LL^\be\dea^\al{}_{,\be}+T\frac{d}{d\rbar}\LL'^\al
	\nnn
	&&
	+\LL^\be\frac{d}{d\rbar}\LL^\al{}_{,\be}
	+2\HH'T^2n^\al+2\HH TT'n^\al+2\HH\LL^\be T_{,\be}n^\al\bigg\}\bigg\rvert^z_{\bobs}\,,
	\enar
	while for the integral we derive
	\bear
	&&\int_0^{\rbar_z} d\rbar\,\left( {\widetilde\GG}{}^\al-\widetilde\dnu\widetilde\dea{}'^\al+\widetilde\dea{}^\be\widetilde\dea{}^\al{}_{,\be} \right)\bigg\rvert_{\bar x_{\rbar}}=
	\nnn&&\Dquad
	=\int_0^{\rbar_z} d\rbar\,\left( \GG^\al-\dnu\dea'^\al+\dea^\be\dea^\al{}_{,\be} \right)\bigg\rvert_{\bar x_{\rbar}}
	-\bigg\{\frac{d}{d\rbar}\LL^\al+2\HH T\left( n^\al+\dea^\al+\frac{d}{d\rbar}\LL^\al \right)
	\nnn
	&&\Dquad\quad
	-T\frac{d}{d\rbar}\LL'^\al
	-2\HH TT'n^\al+\left( 2\HH^2-\HH' \right)T^2n^\al
	-\LL^\be\left( \dea^\al{}_{,\be}+2\HH T_{,\be}n^\al+\frac{d}{d\rbar}\LL^\al{}_{,\be} \right)
	\nnn
	&&\Dquad\quad
	-\HH_oT_o\left( \dea^\al+\frac{d}{d\rbar}\LL^\al \right)-T\dea'^\al
	-\LL'^\al\dnu+\LL^\al{}_{,\be}\dea^\al-2\HH\HH_oTT_on^\al\bigg\}\bigg\rvert^z_{\bobs}\,.
	\enar
	The substitutions employed to obtain these equations are the same mentioned in the previous case of~$\dnu$, and we do not repeat them here.
	The sum of these two contributions leads, after straightforward simplifications, to
	\bear
	{\widetilde\dea}{}^\al(\widetilde{x_s})-{\widetilde\dea}{}^\al(\widetilde{x_o})&=&{\dea}{}^\al(x_s)-{\dea}{}^\al(x_o)+\bigg\{\frac{d}{d\rbar}\LL^\al+2\HH Tn^\al+\DX^\mu\pa_\mu\left( 2\HH Tn^\al+\frac{d}{d\rbar}\LL^\al \right)
	\nnn
	&&
	-\LL'^\al\dnu+\LL^\al{}_{,\be}\dea^\be+\left(2\HH T
	-\HH_oT_o\right)\left( \dea^\al+\frac{d}{d\rbar}\LL^\al \right)-2\HH\HH_oTT_on^\al
	\nnn
	&&
	+\left(2\HH^2+\HH'\right)T^2n^\al\bigg\}\bigg\rvert^z_{\bobs}\,.
	\enar
	As before, the expectation for this equation can be read from Eq.\,($\ref{dnuPtilde}$), and we find no discrepancy once the following second-order relations are taken into account
	\bear
	\frac{d}{d\rbar}\LL^\al\bigg\rvert_{x}=\frac{d}{d\rbar}\LL^\al\bigg\rvert_{\bar x}+\DX^\mu\left[\pa_\mu\left(\frac{d}{d\rbar} \LL^\al\right)\right]\bigg\rvert_{\bar x}\,,\qquad
	(\HH T)(x)=\HH(\bar\eta)T(\bar x)+\DX^\mu[\pa_\mu\left(\HH T\right)]\bigg\rvert_{\bar x}\,.\qquad
	\enar
	\\

	\subsection{Fluctuations in the observed redshift}
	
	The observed redshift $z(x_s;x_o)$ is a cosmological observable, and as such, does not depend on the choice of FLRW coordinates we use to describe it. Geometrically, this means that $z$ is a scalar under diffeomorphisms of $\mM$.
	We will make use of this fact to compute the gauge transformation of the perturbation $\de z(x_s;x_o)$, which again is not a tensor. As discussed in the previous subsection, we will suppress the dependence of $z$ and $\de z$ from the reference position $x^\mu_o$, since the evaluation at the observer position $O$ does not participate in gauge transformations.
	\\
	
	\Ietc{Second-order expectation}
	We recall the transformation law of a scalar under change of coordinates in Eq.\,($\ref{scalar}$), therefore the observed redshift satisfies
	\beeq
	\widetilde z ( \widetilde x_s)=z(x_s)\,,
	\eneq
	where the reference coordinates in the left-hand side are $\widetilde x^\mu_o$, and in the right-hand side $x^\mu_o$. Using our definition for the splitting in Eq.\,($\ref{dzdz}$)
	we can recast the previous equation in terms of $\de z$ as
	\bear\label{dzimp}
	\widetilde{\de z}( \widetilde x)=\frac{a(\widetilde\eta)}{a(\eta)}\left(1+\de z(x)\right)-1\,,
	\enar
	where we suppressed the subscript $s$. This expression is valid at any perturbative order, and we can obtain a second-order expression expanding $\widetilde x^\mu$ in Eq.\,($\ref{xcsi}$) around $x^\mu$. The expansion of the scale factor is
	\bear
	a(\widetilde\eta=\eta+T)=a(\eta)\left\{1+\HH T+\frac12(\HH^2+\HH')T^2+\cdots\right\}.
	\enar
	Substituting this result in Eq.\,($\ref{dzimp}$) we obtain at second order
	\bear\label{dztildetilde}
	\widetilde{\de z}(\widetilde x)=\de z+\HH T(1+\de z) +\frac12\left(\HH^2+\HH'  \right)T^2\,,
	\enar
	which is the expression that we will need for the consistency check. Combining the three equations above, we derive the second-order gauge transformation of $\de z$:
	\bear\label{dezGT}
	{\widetilde{\de z}}&=&{\de z}+\HH T(1+\de z) -\HH TT'-\HH\LL^\al T_{,\al}+\frac12\left(\HH^2-\HH'  \right)T^2
	-T\de z'-\LL^\al\de z_{,\al}\,,
	\enar
	where all the quantities are evaluated at the coordinate value of the source position $x^\mu_s$, and the dependence on the observer reference position is only implicit.
	\\
	
	\Ietc{Second-order direct computation}
	To compute the gauge transformation of our second-order expression of the distortion in the observed redshift given in Eq.\,($\ref{dz2}$), we need to compute the gauge transformations of the perturbations $\de\eta_o$ and $\dhnu$. The perturbation $\de\eta_o$ is related to the time lapse at the observer position via Eq.\,($\ref{lapseconf}$), and we have already checked the consistency of our second-order expression for the time lapse at the observer position. We only need to check the gauge transformation of $\dhnu$ directly from Eq.\,($\ref{dhnu2}$). Note that this expression contains perturbations of the conformal wave vector and perturbations of the metric. Since we have already checked the consistency of perturbations in the conformal wave vector, we can use the results of the previous subsection, together with the results in Subsection $\ref{subsec:4.1}$, to derive the second-order gauge transformation
	\bear
	{\widetilde{\dhnu}}=&&\dhnu+\left(\mathcal{H} T -\mathcal{H}_{o}T_{o}\right)(1+\dhnu)+ \tfrac{1}{2}\left(\mathcal{H}^2  -  \mathcal{H}' \right) T^2+ \tfrac{1}{2} \left(\mathcal{H}_{o}^2 -  \mathcal{H}_{o}'\right) T_{o}^2- \dhnu'T
	\nnn
	&&
	-  \mathcal{L}^{\al} \left(\dhnu_{,\al} +  \mathcal{H}  T_{,\al}\right) -  \mathcal{H} \mathcal{H}_{o} TT_{o}  -\HH TT' \,.
	\enar
	This expression is evaluated at $\bar x^\mu_z$ as in Eq.\,($\ref{dhnu2}$) with the reference observer position $O$. For consistency, we expand the second-order fields around the background observer position~$\bar x^\mu_{\rm o}$ via Eq.\,($\ref{2bar}$):
	\bear\label{HT}
	(\HH T)(x_o)=\HH_{\bobs}T_{\bobs}+\de x^\mu_o\left[\frac{\pa}{\pa x^\mu}(\HH T)\right]\bigg\rvert_{\bobs}.
	\enar
	Substituting the previous two equations in the expression for the redshift perturbation, together with the transformation of $\de\eta_o$ in Eq.\,($\ref{detao}$), we derive the second-order
	\bear
	\widetilde\de z(\widetilde x)=\de z(x)+\HH_zT_z(1+\de z) +\frac12\left(\HH^2+\HH' \right)T^2+\DX^\mu_s\left[{\pa\over\pa x^\mu}\left( \HH T \right)\right]\bigg\rvert_{\bar x_z}\,,
	\enar
	where the subscript $z$ indicates the evaluation of the second-order fields at $\bar x^\mu_z$.
	Again, the fields in the expectation in Eq.\,($\ref{dztildetilde}$) are evaluated at the source position $x^\mu_s$, while the expression above is evaluated at $\bar x^\mu_z$. Once we expand $(\HH T)(x_s)$ around $\bar x^\mu_z$ as in Eq.\,($\ref{HT}$) we recover the consistency between the expectation and direct computation \cite{ld9}.
	\\

	\subsection{Distortion in the source position compared to the observed position}
	
	Here we will compute the transformation of the distortion in the source position around the background $\bar x^\mu_z$ parameterized by the observed redshift and angles. The computations will be performed both in rectangular coordinates $\DX^\mu_s$ and in spherical coordinates $\drr\,,\,\dtt\,,\,\dpp$. The expression in spherical coordinates around observable quantities will be needed to compute the gauge transformations in the next subsection.
	\\
	
	\Ietc{Second-order expectation}
	The expectation for the second-order transformation of $\DX_s^\mu$ has already been derived in Eq.\,($\ref{trasfx2}$), here we write the components explicitly:
	\bear\label{gtdx2}
	{\widetilde\DT}_s&=&\DT_s+T(\bar x_z)+\DT_sT'\big\rvert_{\bar x_z}+\DX^\al_sT{}_{,\al}\big\rvert_{\bar x_z}\,,
	\nnn
	{\widetilde\DX}^\al_s&=&\DX^\al_s+\LL^\al(\bar x_z)+\DT_s\LL'^\al\big\rvert_{\bar x_z}+\DX^\be_s\LL^\al{}_{,\be}\big\rvert_{\bar x_z}\,.
	\enar
	With the transformation of the spatial distortion $\DX_s^\al$, we can derive the transformation laws satisfied by the deviations in spherical coordinates. By direct substitution, we obtain
	\bear
	{\widetilde\drr}&=&n_\al{\widetilde\DX}{}^\al_s+\frac{\rbar_z}{2} \left( \widetilde\dtt^{2}+\sin^2\ttt~\widetilde\dpp^{2} \right)
	\nnn
	&=&{\drr}+\LL_\para+\DX^\mu_s\pa_\mu\LL_{\para}+\frac{1}{2\rbar_z}\left(\LL^2_\ttt+\LL^2_\pp\right)+\LL_\ttt\,\dtt+\LL_\pp\sin\ttt\,\dpp
	\,,
	\nnn
	\rbar_z{\widetilde\dtt}&=&\ttt_\al{\widetilde\DX}{}^\al_s-\widetilde\drr\,\widetilde\dtt +\frac{\rbar_z}{2}\cos\ttt\sin\ttt~\widetilde\dpp^2
	\nnn
	&=&\rbar_z{\dtt}+\LL_\ttt+\DX^\mu_s\pa_\mu\LL_{\ttt}-\frac{\drr}{\rbar_z}\LL_\ttt-\LL_\para\left(\dtt+\frac{\LL_\ttt}{\rbar_z}\right)+\frac{1}{2\rbar_z}\LL^2_\pp\cot{\ttt}
	+\LL_\pp\cos\ttt\,\dpp
	\,,
	\nnn
	\rbar_z\sin\ttt\,{\widetilde\dpp}&=&\pp_\al{\widetilde\DX}{}^\al_s-\sin\ttt~\widetilde\drr\,\widetilde\dpp-\rbar_z\cos\ttt~\widetilde\dtt\,\widetilde\dpp
	\nnn
	&=&\rbar_z\sin\ttt\,{\dpp}+\LL_\pp+\DX^\mu_s\pa_\mu\LL_{\pp}-\frac{\drr}{\rbar_z}\LL_\pp-\frac{\LL_\para}{\rbar_z}\left( \rbar_z\sin\ttt\,\dpp+\LL_\pp  \right)
	-\LL_\pp\cot\ttt\,\dtt
	\nnn
	&&
	-{\LL_\ttt\over\rbar_z}\left( \rbar_z\cos\ttt\,\dpp-\cot\ttt \LL_\pp \right)
	\,,
	\enar
	where we defined $\LL_\para:=\LL^\al n_\al\,$, $\LL_\ttt:=\LL^\al \ttt_\al\,$, $\LL_\pp:=\LL^\al \pp_\al\,$ and expressed each field in the right-hand side at the reference position $\bar x^\mu_z$.
	Note that, once we have verified the consistency of our second-order expressions for $\DT_s$ and $\DX_s^\al$ in Eqs.\,($\ref{deta2}$) and ($\ref{dx2}$) with the expectations in Eq.\,($\ref{gtdx2}$), then the consistency of our second-order expressions for the distortion in spherical coordinates in Eq.\,($\ref{dangoli2}$) is automatically recovered because $\drr,\dtt,\dpp$ are simply reparameterizations of $\DX^\al_s$.
	\\
	
	\Ietc{Second-order direct computation}
	We start by checking the consistency of $\DT_s$ given in Eq.\,($\ref{deta2}$). Our second-order expression depends only on $\de z$, of which we verified the consistency in the previous subsection. We can therefore substitute Eq.\,($\ref{dztildetilde}$) for the gauge transformation of $\de z$. After straightforward manipulations, we recover the expected result
	\bear
	{\widetilde\DT}_s&=&\DT_s+T(\bar x_z)+\DT_sT'\big\rvert_{\bar x_z}+\DX^\al_sT{}_{,\al}\big\rvert_{\bar x_z}\,.
	\enar
	
	We now turn our attention to the spatial components $\DX_s^\al$ of the distortion in the source position. From Eq.\,($\ref{Dx2}$), we need to compute the gauge transformations of boundary terms and line of sight integration. To avoid cumbersome equations, we present the terms separately. For the boundary terms, we derive
	\bear
	{\widetilde{\de x}}{}^\al_o&=&\de x^\al_o+\LL^\al_{\bobs}+\de x^\mu_o(\pa_\mu\LL^\al)\big\rvert_{\bobs}\,,
	\\
	\nnn
	-n^\al({\widetilde\DT}_s-{\widetilde{\de\eta}}_o)&=&-n^\al\bigg({\DT}_s-{{\de\eta}}_o+ T_z-T_{\bobs}+\DX^\mu_s(\pa_\mu T)\big\rvert_{z}
	-\de x^\mu_o(\pa_\mu T)\big\rvert_{\bobs}\bigg)\,,\quad
	\\
	\nnn
	\widetilde\DX^\mu\pa_\mu\left( \widetilde\DX{}^\al+n^\al\widetilde\DT\right)&=&\DX^\mu\pa_\mu\left( \DX{}^\al+n^\al\DT+\LL^\al+n^\al T\right)
	+T\big( \DX'^\al+n^\al\DT'\nnn
	&&
	+\LL'^\al+n^\al T' \big)+\LL^\be\big( \DX^\al{}_{,\be}+n^\al\DT_{,\be}+\LL^\al{}_{,\be}+n^\al T_{,\be} \big)\,.\qquad\,\,\,
	\enar
	To compute the transformations of the distortion in the observer and source position, $\de x^\mu_o$ and $\DT_s$, we used Eq.\,($\ref{trasfx2}$) as we already verified the consistency of these expressions. From the results of previous subsections, we can readily see that also the transformation of the linear-order perturbation~$\DX_s^\al$ in Eq.\,($\ref{Dx1}$) satisfies the linear-order expectation in Eq.\,($\ref{trasfx2}$).
	The gauge transformation of the integral yields
	\bear
	&&\int_0^{\rbar_z}d\rbar\left[ {\widetilde\dea}{}^\al -n^\al{\widetilde\dnu}+\widetilde\dnu\left(\widetilde\DX{}'^\al+n^\al\widetilde\DT{}'\right)-\widetilde\dea{}^\be\left(\widetilde\DX{}^\al{}_{,\be}+n^\al\widetilde\DT_{,\be}\right)
	\right]\bigg\rvert_{\bar x_{\rbar}}=
	\nnn
	&&
	\int_0^{\rbar_z}d\rbar\bigg\{ {\dea}{}^\al -n^\al{\dnu}+\dnu\left(\DX{}'^\al+n^\al\DT{}'\right)-\dea^\be\left(\DX{}^\al{}_{,\be}+n^\al\DT_{,\be}\right)
	+\frac{d}{d\rbar}\bigg[\LL^\al+n^\al T
	\nnn
	&&
	-T\big( \DX'^\al+n^\al\DT'+\LL'^\al+n^\al T' \big)-\LL^\be\big( \DX^\al{}_{,\be}+n^\al\DT_{,\be}+\LL^\al{}_{,\be}+n^\al T_{,\be} \big)\bigg]\bigg\}\bigg\rvert_{\bar x_{\rbar}}\,,\quad
	\enar
	where for the gauge transformations of the components of the conformal wave vector we used Eqs.\,($\ref{dnuP}$) and ($\ref{deaP}$) since we have already checked the consistency of our second-order expressions.
	Summing the contributions of the boundary terms and the integral, most of the terms cancel, and we recover
	\bear
	{\widetilde\DX}{}^\al_s&=&\DX^\al_s+\LL^\al(\bar x_z)+\DT_s\LL'^\al\big\rvert_{\bar x_z}+\DX^\be_s\LL^\al{}_{,\be}\big\rvert_{\bar x_z}\,,
	\enar
	in agreement with the expectation.

	\subsection{Distortion in the physical volume and area}
	
	\Ietc{Second-order expectation}
	In Subsection $\ref{subsec:vol}$ we showed that the fluctuations $\de V$ and $\de A$ in Eqs.\,($\ref{deV}$) and ($\ref{deA}$) are invariant under diffeomorphisms of $\mM$, which is consistent with the requirement that they are observed quantities. Consequently, the following relations hold under a change of coordinates
	\bear\label{legge}
	\widetilde{\de V}(\widetilde x_s)=\de V(x_s)\,,\Dquad \widetilde{\de A}(\widetilde x_s)=\de A(x_s)\,,
	\enar
	i.e. invariant under diffeomorphism \cite{2017JCAP...09..016Y,tetrad}.
	These are valid at any perturbative order, while the second-order gauge transformations can be read from Eq.\,($\ref{scalar}$):
	\bear
	\widetilde{\de V}(\bar x_z)=\de V(\bar x_z)-T\de V'\big\rvert_{\bar x _z}-\LL^\al\de V_{,\al}\big\rvert_{\bar x _z}
	\,,\quad
	\widetilde{\de A}(\bar x_z)=\de A(\bar x_z)-T\de A'\big\rvert_{\bar x _z}-\LL^\al\de A_{,\al}\big\rvert_{\bar x _z}\,,\,\,\,\,\quad
	\enar
	\\
	
	\Ietc{Second-order direct computation}
	Given the intricacy of our second-order expressions for volume and area fluctuations in Eqs.\,($\ref{DV2}$) and ($\ref{DA2}$), the direct computation of the gauge transformation of the perturbations $\de V$ and $\de A$ is a very laborious task. Beyond having to make many substitutions, the main complication is that we need to relate partial derivatives with respect to spacetime coordinates $x^\mu$ to derivatives with respect to the observed quantities $z,\ttt,\pp$. Instead of explicitly showing the consistency between the expectation and the direct computation, which would involve many equations, we will provide the critical relations used to prove the expected consistency so that the interested reader can reproduce the calculations.
	\begin{itemize}
		\item Relations between partial derivatives:
		\bear
		\frac{\pa}{\pa z}=-\frac1{H_z}{\pa\over\pa\eta}+\frac1{H_z}n^\al{\pa\over\pa x^\al}\equiv \frac1{H_z} \frac{d}{d\rbar}\,,\qquad
		\frac{\pa}{\pa\ttt}=\rbar_z\ttt^\al\frac{\pa}{\pa x^\al}\,,\qquad
		\frac{\pa}{\pa\pp}=\rbar_z\sin\ttt\,\pp^\al\frac{\pa}{\pa x^\al}\,.\qquad
		\enar
		\\
		
		\item Identities with unit vectors
		\bear
		\left(\cot\ttt+	\frac{\pa}{\pa\ttt}\right)\ttt^\al+\frac{1}{\sin\ttt}\frac{\pa}{\pa\pp}\pp^\al &=& -2n^\al\,,
		\qquad
		\frac{\pa}{\pa\pp}\ttt^\al= \pp^\al\cos\ttt\,,\qquad
		\frac{\pa}{\pa\ttt}\ttt^\al= -n^\al\,,
		\nnn
		n^\al\sin\ttt+\ttt^\al\cos\ttt&=&-\frac{\pa}{\pa\pp}\pp^\al\,.
		\enar
		\\
		
		\item Relations involving partial derivatives with respect to Cartesian coordinates and spherical coordinates
		\bear
		\LL^\al&=&\LL_\para n^\al+\LL_\ttt\ttt^\al+\LL_\pp\pp^\al\,,
		\\
		\nnn
		\LL^\al{}_{,\al}=\LL_{\para,\para}+\LL_{\ttt,\al}\ttt^\al+\LL_{\pp,\al}\pp^\al&=&\LL_{\para,\para}+\frac1\rbar_z\ttt^\al{\pa\over\pa \ttt}\LL_{\al}
		+\frac{1}{\rbar_z\sin\ttt}\pp^\al{\pa\over\pa \pp}\LL_{\al}\,,
		\\
		\nnn
		\frac12\LL^\be{}_{,\al}\LL^\al{}_{,\be}-\frac12\LL^\al{}_{,\al}\LL^\be{}_{,\be}&=&\LL_{\ttt,\para}\LL_{\para,\al}\ttt^\al+\LL_{\pp,\para}\LL_{\para,\al}\pp^\al+\LL_{\ttt,\al}\pp^\al\LL_{\pp,\be}\ttt^\be-\LL_{\para,\para}\LL_{\ttt,\al}\ttt^\al
		\nnn
		&&
		-\LL_{\para,\para}\LL_{\pp,\al}\pp^\al-\LL_{\ttt,\al}\ttt^\al\LL_{\pp,\be}\pp^\be\,,
		\enar
		which also apply to $\VV^\al$, or any other field.
		\\
		
		\item Relations involving derivatives of spatial deviations
		\bear
		\DX^\al&=&n^\al\drr+\ttt^\al\rbar_z\,\dtt+\pp^\al\rbar_z\sin\ttt\,\dpp\,,
		\\
		\nnn
		{\pa\over\pa z}\DX^\al&=&n^\al{\pa\over\pa z}\drr+\frac1{H_z}\ttt^\al\dtt+\ttt^\al\rbar_z{\pa\over\pa z}\dtt+\frac1{H_z}\pp^\al\sin\ttt\,\dpp+\pp^\al\rbar_z\sin\ttt{\pa\over\pa z}\dpp\,,
		\\
		\nnn
		{\pa\over\pa \ttt}\DX^\al&=&n^\al{\pa\over\pa \ttt}\drr+\ttt^\al\drr+\rbar_z\dtt{\pa\over\pa \ttt}\ttt^\al+\ttt^\al\rbar_z{\pa\over\pa \ttt}\dtt+\pp^\al\rbar_z\cos\ttt\,\dpp+\pp^\al\rbar_z\sin\ttt{\pa\over\pa \ttt}\dpp\,,\qquad\,\,\,
		\\
		\nnn
		{\pa\over\pa \pp}\DX^\al&=&n^\al{\pa\over\pa \pp}\drr+\pp^\al\sin\ttt\,\drr+\rbar_z\dtt{\pa\over\pa \pp}\ttt^\al+\ttt^\al\rbar_z{\pa\over\pa \pp}\dtt+\rbar_z\sin\ttt\,\dpp{\pa\over\pa\pp}\pp^\al
		\nnn&&
		+\pp^\al\rbar_z\sin\ttt{\pa\over\pa \pp}\dpp\,.
		\enar
	\end{itemize}
	Employing these relations, and directly substituting in Eqs.\,($\ref{DV2}$) and ($\ref{DA2}$) the gauge-transformation laws derived previously in the section we obtain
	\bear
	{\widetilde{\de V}}(\widetilde x_s)={\de V}(\bar x_z)+\DX^\mu_s\left({\pa\over\pa x^\mu}\de V\right)\bigg\rvert_{\bar x_z}\,,
	\qquad
	{\widetilde{\de A}}(\widetilde x_s)={\de A}(\bar x_z)+\DX^\mu_s\left({\pa\over\pa x^\mu}\de A\right)\bigg\rvert_{\bar x_z}\,,\quad
	\enar
	in agreement with the expectation in Eq.\,($\ref{legge}$) once the right-hand side is expanded up to second order around $\bar x^\mu_z$.

	\section{Comparison with Previous Works}
	\label{sec:compa}	
	
	In this section, we compare our main results with previous work in the literature. The physical volume and area, as well as the observed redshift and source position, have already been computed at the second order in perturbations, and many applications have been proposed \cite{cmb0,cmb0.5,cmb1,cmb2,cmb3,cmb4,cmb5,lens0,lens1,lens2,lens3,lens4,lens5,lens6,ld0,ld1,ld2,ld3,ld4,ld6,ld7,ld8,ld9,YZ,enea,gc0,gc1,gc2,gc3,gc4,gc5,gc6,gc7}.
	However, in none of these previous works have the gauge transformation properties of the main results been explicitly verified. Given the complex nature of the second-order relativistic expressions of the cosmological observables, complete verification of the gauge-transformation properties in our work provides strong support for the validity of our results that are likely to have less (or no) flaws. Due to the different notations and intricacy of the
	expressions, it is not straightforward to make a detailed comparison between our results and those of other groups. Hence, we can only conduct a qualitative comparison at this stage, and any quantitative comparison is reserved for future works.
	Therefore, we will list previous efforts in literature, highlighting only the main differences with our results.

	\begin{itemize}
		\item Yoo, J., \& Zaldarriaga, M. (2014). Beyond the linear-order relativistic effect in galaxy clustering: second-order gauge-invariant formalism \cite{YZ}.
		
		This article can be regarded as a prior work of what we have set out here. In fact, except for a few minor details and notation conventions, the approach and the methodology coincide but here we focus more on the exact nonlinear relations and definitions, separating perturbative expressions. In \cite{YZ}, the general relativistic description of galaxy clustering was extended to the second order in metric perturbations without assuming any gauge conditions or adopting any restrictions on vector and tensor perturbations. Moreover, second-order gauge-invariant formalism was discussed, computing second-order gauge-invariant variables.
		However, no explicit check of gauge transformation was made, over which we improve here. Furthermore, with the recent development of the tetrad formalism \cite{thepaper}, we have clarified the relation of the physical observables to coordinate-dependent quantities.

		\item Di Dio, E., Durrer, R., Marozzi, G., \& Montanari, F. (2014). Galaxy number counts to second order and their bispectrum \cite{enea}.
		
		The authors present the galaxy number counts to the second order in perturbations
		and compute the angular bispectrum. They identify the dominant contributions 
		to the angular bispectrum of the galaxy number counts,
		which originate from the density fluctuation, the redshift space distortion, and 
		gravitational lensing. Using the geodesic light-cone coordinate \cite{REF0,REF1}, the authors
		derive the nonlinear expressions for the cosmological observables without solving
		for the Einstein equation (see, e.g., \cite{REF2}). Then they transform their expressions to 
		the second-order Poisson gauge, neglecting vector and tensor contributions 
		to the metric perturbations but without assuming vanishing anisotropic stress. 
		Their expression in Eq.~(3.23) shows the relation of perturbation variables
		to the observed redshift:
		\bear
		1+z_s=\frac{a(\eta_o)}{a(\eta_s)}\left[1+\de z^{(1)}+\de z^{(2)} \right]\,,
		\enar
		which is different from ours given in Eq.~(\ref{dzdz}), but we stress that the
		difference arises only from a different choice of splitting the background and the
		perturbation. However, their explicit expression in Eq.~(3.23) in terms of metric 
		perturbations contains only the contributions evaluated at the source position and 
		along the-line-of-sight direction, while no contributions at the observer position
		are present. This equation can be compared to Eq.~(\ref{B1}) in our work, where
		there exist not only the contributions at the observer position, but also
		the quadratic contributions that involve one at the observer position and the
		other at the source position. Similar differences are present in their expressions
		for the angular position in Eq.~(3.15), which can be compared to Eq.~(\ref{dx2}) in our
		work. These differences persist in their subsequent expressions based on the source
		position such as the fluctuation in the physical volume and the area.

		\item D. Bertacca, R. Maartens, and C. Clarkson. Observed galaxy number counts on the light cone up
		to second order: II. Derivation \cite{gc1}. 
		
		This paper aims to develop the second-order framework for studying
		the relativistic effects beyond the linear order, in the same spirit
		of our current work. The final results are presented both in the Poisson gauge and
		in a general metric representation without assuming a gauge condition,
		but the vector and tensor perturbations at the linear order are ignored.
		Due to the different notation convention, it is not easy to conduct
		a detailed comparison, but the calculational steps are quite similar to ours.
		In \cite{gc1} (BMC hence forth) they use the comoving distance as an affine parameter~$\chi$, which corresponds
		to our affine parameter~$-\lambda$ with negative sign and gives rise to
		the sign difference in the conformal wave vector. The perturbative deviation of their affine parameter $\chi$ around the background comoving distance $\bar\chi$ is equivalent to ours in Eq.~($\ref{xz}$).
		The main difference lies in their notation convention for the conformal wave vector, which is given in their Eq.\,(11):
		\bear
		k^\mu(\bar\chi)=\frac{dx^\mu}{d\bar\chi}(\bar\chi)=\left( -1+\dnu,\,n^i+\dea^i\right)^\mu(\bar\chi)\,.
		\enar
		The corresponding quantity $\hat k^\mu$ in our notation is defined in Eq.\,($\ref{khatnohat}$) and it their notation is
		\bear
		k^\mu(\chi)=\frac{dx^\mu}{d\chi}(\chi)=\left( -1-\dnu,\,n^i+\dea^i\right)^\mu(\chi)\,.
		\enar
		Therefore, what we call $\dnu$ and $\dea^i$ in our papers are different quantities, where the difference amounts to another choice of the parameterization of the light path.
		For instance, the difference in what we call $\dnu$ is given by
		\bear
		\dnu_{MY}=-\dnu_{BMC}-\left( 1-\dnu_{BMC}\right)\frac{d}{d\chi}\de\chi\,.
		\enar
		Notice also that their geodesic equation for the conformal light wave vector above Eq.\,(86) looks different from what we use to derive Eqs\,($\ref{dnuint}$) and ($\ref{deaint}$), namely
		\bear
		\frac{d}{d\cc}\hat k^\mu+{\hat\Ga}^\mu_{\nu\rho}\big\rvert_{\bar x_\cc}\hat k^\nu\hat k^\rho+\de x^\si\pa_\si{\hat\Ga}^\mu_{\nu\rho}\big\rvert_{\bar x_\cc}\hat k^\nu\hat k^\rho=0\,.
		\enar
		However, further expanding this equation around $\bar x(\cc_z)$ and $\cc_z$, and using the linear-order relation
		\bear
		\Delta x^\mu=\de x^\mu+\Dcc_z \hat {\bar k}^\mu\,,
		\enar
		we find consistency with their expression.
		The use of $\bar\chi$ instead of $\chi$ yields their definition of~$\de x^{0}$ and $\de x^i$ in Eqs.\,(16) and (17), which amounts to
		adopting the Born approximation. However, the authors do include post-Born terms (see for instance their Eq.\,(113)) which make
		a difference at the second order in which we have to account for the evolution
		of perturbations on a perturbed light path. The observed redshift is defined in their Eq.~(27), and the normalization condition for the conformal wave vector in their Eq.~(28) corresponds to our
		condition in Eq.~($\ref{NC}$). However, adopting~$a(\eta_o)=1$, they ignored
		the contributions at the observer position. Indeed, as apparent in our Eq.~($\ref{lapseandshift}$)
		the observer position in an inhomogeneous universe is not uniquely set, but stochastically
		set in a given realization of the inhomogeneous universe. With such normalization, all
		the background equations with~$a(\bar\eta_{\rm o})$, which are mostly not shown explicitly,
		result in extra perturbative quantities, because of the normalization condition.
		Similarly, the spatial shift of the observer position~$\delta x^\alpha_o$ in our Eq.~($\ref{dxlin}$)
		is ignored, which matters starting at the second order in perturbations.
		Finally, the antisymmetric part of the tetrad vectors in our Eq.~($\ref{asim}$) is not
		considered in BMC, which again matters in the second-order expressions for the
		cosmological observations.
		No verification of the gauge-transformation properties of their expressions are made.

		\item O. Umeh, C. Clarkson, and Roy Maartens. Nonlinear relativistic corrections to cosmological
		distances, redshift and gravitational lensing magnification: II. Derivation \cite{ld3}.

		This paper studies light propagation in a generic spacetime to derive a nonlinear expression for the area distance. This is computed up to second order in perturbations in the Poisson gauge, and their results are expressed using the observed redshift as parameterization.
		It is not straightforward to directly compare our results for the area distance with those obtained by the authors. These expressions are lengthy, the notation conventions are different, and the authors fix a gauge. For these reasons, we will compare the calculations in the intermediate steps rather than the final but lengthy result (see also \cite{ld6,ultimo} for similar
		derivations based on the geodesic light cone gauge that include the contributions
		at the observer position).
		The authors consider two spacetimes: quantities defined on the physical spacetime are denoted by a hat, while quantities defined on a conformally related manifold have no hat. They specify the conformal transformation in their Eq. (21), where the only difference with our derivation is that they fix the constant $\NC=1$. A consequence of this choice is that quantities like the light wave vector or the perturbation in the observed redshift are different from our expressions. However, this should not impact the final expressions for the cosmological observables. Another difference between the two approaches is that the authors fix $\de k^\mu_o=0$ in the light wave vector  at the observer position, both at first and second order. In our case, such quantity satisfies the boundary conditions in Eq.\,($\ref{ko}$) involving tetrad vectors at the observer. Such contributions at the observer position appear in our final result for the area distance but are neglected in the paper we are discussing. Moreover, in their definition of the redshift perturbation in Eq.\,(69), they adopt $a(\eta_o)=1$, ignoring contributions at the observer position.
		Notice that the authors obtain their expression for the angular diameter distance by solving the differential equation
		\bear
		\frac{d^2\de^nD_A}{d\cc^2}=\de^nS(\cc)\,,
		\enar
		with boundary conditions
		\bear
		\de^nD_A(\cc_o)=0\,,\Dquad\frac{d\de^nD_A}{d\cc}\bigg\rvert_o=-\de^nE_o\,,
		\enar
		where $\de^n S$ contains the area distance up to order $n-1$ and, in our notation, $E_0=-(\hat u^\mu\hat k_\mu)_o=1+\dhnu_o$. Given our choice of the constant $\NC$ we fixed $\dhnu_o=0$ while for the authors $E_o$ contain also a perturbative part. As a consequence, the boundary condition in Eq.\,(11) are different in the two approaches, hence also the solution for the angular diameter distance will be different.
		There is another difference present for the relation between the affine parameter and the observed redshift. In their Eq.\,(126)
		\bear
		\cc=\nu+\de\cc+\frac12\de^2\cc\,,
		\enar
		where the definition of $\cc$ coincides with our $\cc_s$, while $\nu$ coincides with our $\bar\eta_z$. In Eqs.\,(132) and (133) they derive their expressions for $\de\cc$ and $\de^2\cc$. Such expressions coincide with what we derived for $\DT_s$ in Eq.\,($\ref{deta2}$). So in our notation we could rewrite Eq.\,(12) as
		\bear
		\cc_s=\bar\eta_z+\DT_s\,,
		\enar
		which actually differs from our expression in Eq.\,($\ref{xz}$) of the terms $-\bar\eta_{\rm o}-\de\eta_s$ and would then impact all the expansions in the mapping from affine parameter to observed redshift.
		
	\end{itemize}

	\section{Summary and Discussion}
	\label{sec:concl}
	In this paper we have developed the nonlinear gauge-invariant formalism for describing the cosmological observables and presented complete second-order expressions associated with light propagation and observations without choosing a gauge condition. Our second-order expressions in general-metric representations allow us to explicitly check the validity by directly computing their gauge transformations and by comparing to the expectations. Our main results are summarized as follows:
	\begin{itemize}
		\item Adopting a geometric approach to describing the light propagation and observations, we have developed the nonlinear gauge-invariant formalism that is fundamental to computing all the general relativistic effects in cosmological observables. In particular, we have clarified the separation between the background and the perturbation quantities in a fully nonlinear way. Drawing on the previous work, we have introduced
		tetrad vectors at the observer and the source positions to describe the observations recorded in the rest frame of the observer and to define the physical quantities in the source rest frame that are measured by the observer. The observable quantities are the redshift and the angular position of the source and their spectroscopic information; the physical quantities are the physical size and shape of the source and their spectrum in the rest
		frame. These quantities are connected by light propagation and observations, and expressed in terms of intermediate quantities such as the coordinate position of the source or the distortion in the observed redshift. We have presented fully nonlinear expressions for the physical volume and area occupied by the source that can be measured in galaxy clustering and the luminosity distance. For future work we have left a nonlinear computation of the distortion of the source shape that can be measured in weak gravitational lensing.
		\item By expanding the nonlinear expressions, we have computed the second-order expressions for the cosmological observables and their intermediate quantities in terms of metric perturbations and the velocities of the source and the observer. Any nonlinear quantities can be perturbatively expanded and truncated in many different ways, though the resulting expressions should be all consistent, if the perturbation order is defined in the same way. Paying particular attention to these subtleties, we have expanded the second-order quantities along the
		background photon path and around the coordinate-independent reference points of the observer and the source parameterized by the observables. The reason behind this expansion is that first we can keep perturbative orders under control by expanding each quantity around background paths. More importantly, by having employed coordinate-independent parameterization for physical positions, the entire coordinate-dependence is contained in the deviations $\de x^\mu_o$ and $\DX^\mu_s$. The main results are the second-order gauge-invariant expressions of the physical volume and area for galaxy clustering and the luminosity distance.
		\item Utilizing the geometric nature and nonlinear relations, we have first derived the expected second-order gauge-transformation properties of our theoretical descriptions of the observable quantities and the intermediate quantities. With the theoretical descriptions in a general metric representation, we have next performed the second-order gauge-transformation of the metric perturbation variables in those theoretical descriptions. By comparing the direct gauge-transformation properties to the expectations from the nonlinear relations, we have {\it explicitly verified} the consistency of our theoretical expressions, among which only the expressions for the physical volume and area occupied by the source are gauge-invariant \cite{2017JCAP...09..016Y,tetrad}.
		\item We have compared our results to those in the previous work and discussed the physical ingredients that are missed or ignored in the previous work.\footnote{As the current work was near completion, a work by \cite{loro} appeared on arXiv in which they present a nonlinear framework for computing the cosmological observables, but no second-order expressions or verification of gauge invariance are presented.} Among many differences, we have highlighted that the contributions at the observer position are typically neglected in literature and as a consequence all the quadratic terms that couple one contribution at the observer position and another contribution at the second order are also ignored. With any contributions missing, the second-order expressions in literature are incomplete and gauge-dependent.
	\end{itemize}
	Given the technical difficulties and theoretical subtleties in the second-order relativistic perturbation calculations of the cosmological observables, the second-order expressions in the previous work (including one of our own)
	are generally considered {\it incomplete} with a few or several missing relativistic contributions. The explicit check of second-order gauge-invariance of the theoretical descriptions that we have performed {\it for the first time} constitutes a major step forward in establishing the {\it complete} second-order theoretical descriptions of the cosmological observables. With full degrees of freedom intact, our second-order expressions can be used to describe the cosmological observables in any theories of metric gravity. In the future, we will apply our second-order gauge-invariant formalism to compute the three-point correlation function of cosmological observables such as galaxy clustering, the luminosity distance, weak gravitational lensing, and cosmic microwave background anisotropies. Given the interest in the primordial non-Gaussianity, our complete second-order relativistic expressions will play a crucial role in providing accurate predictions for the higher-order statistics.

	\section*{Acknowledgment}
	We acknowledge support by a Consolidator Grant of the European Research Council 
	(ERC-2015-CoG grant 680886).

	\appendix
	
	\section{Gauge transformation of the light wave vector at the observer position}
	\label{AppendixA}
	Here we present the consistency check for the components of the light wave vector $\hat k^\mu_o$ at the observer position.
	We start by computing the second-order expectation for the gauge transformations of $\dnu_o$ and $\dea^\al_o$. Taking the limit for $P\to O$ of Eq.\,($\ref{dnuP}$) and ($\ref{deaP}$) we obtain
	\bear\label{dnuO}
	{\widetilde\dnu}_o:={\widetilde\dnu}(\widetilde x=x_o) &=&\dnu_o+\bigg[ -\frac{d}{d\rbar}T + T\frac{d}{d\rbar}T' +\HH T\left(1+\dnu -\frac{d}{d\rbar}T\right) 
	\nnn
	&&
	-  \mathcal{L}^{\al} \left(\dnu_{,\al}+2 \mathcal{H}T_{,\al}-\frac{d}{d\rbar}T_{,\al}  \right)+ \frac12\left(\mathcal{H}^2 - 3\mathcal{H}'\right)T^2 -  2\mathcal{H}TT'
	\nnn
	&&
	+T'\dnu-T_{,\al}\dea^\al-T\dnu' \bigg]_o
	\,,
	\enar
	\bear\label{deao}
	{\widetilde{\dea}}{}^\al_o:={\widetilde{\dea}}{}^\al(\widetilde x=x_o)&=&\dea^\al_o+\bigg[\frac{d}{d\rbar}\LL^\al-T\frac{d}{d\rbar}\LL'^\al+\HH T\left(n^\al+\dea^\alpha +\frac{d}{d\rbar}\LL^\al \right)  
	\nnn
	&&
	-  \LL^\beta\left( \dea^\al{}_{,\be}+2 \mathcal{H}  T_{,\beta}n^\alpha  +\frac{d}{d\rbar}\LL^\al{}_{,\be} \right)+ \frac12\left( \mathcal{H}^2 - 3 \mathcal{H}' \right) T^2n^\alpha
	\nnn
	&&
	-2\HH TT'n^\al-\LL'^\al\dnu+\LL^\al{}_{,\be}\dea^\be-T\dea'^\al\bigg]_o\,.
	\enar
	We emphasize that both sides of the equations are evaluated at the coordinate value $x^\mu_o$, while the coordinates of the reference position $O$ are different in the two charts.
	An independent way to derive the gauge transformation of $\dnu_o$ can be obtained considering the perturbation $\dhnu$. Let us rewrite the nonlinear expression in Eq.\,($\ref{dhnuexact}$) in the new coordinate system denoted by a tilde:
	\bear
	\widetilde\dhnu&=&2\widetilde\AA+\widetilde\VV_{\para}-\widetilde\BB_{\para}+(1+2\widetilde\AA)\left(\widetilde\dUU+\widetilde\dnu+\widetilde\dUU\,\widetilde\dnu\right)+\widetilde\dea{}^\al\left( \widetilde\VV_{\al}-\widetilde\BB_{\al} \right)+\widetilde\BB{}^{\al}\widetilde\VV_{\al}(1+\widetilde\dnu)
	\nnn
	&&
	-\widetilde\dUU(\widetilde\BB_{\para}+\widetilde\BB{}^\al\widetilde\dea_{\al})
	+2\widetilde\VV{}^{\al}(\widetilde\CC_{\al \para}+\widetilde\CC_{\al\be}\widetilde\dea{}^{\al})\,,
	\enar
	where both sides are evaluated at the coordinates $\widetilde x_p^\mu$ of a generic point $P$ on the light path. Given our choice for the normalization constant in Eq.\,($\ref{NC}$) and the relation ($\ref{dhnu}$), we readily see that the evaluation at $P=O$ implies $\widetilde\dhnu(\widetilde x_o)=0$. Hence, we can solve the equation above for the perturbation $\widetilde{\dnu}(\widetilde x_o$):
	\bear
	\widetilde\dnu(\widetilde x_o)&=&-\bigg[2\widetilde\AA+\widetilde\VV_{\para}-\widetilde\BB_{\para}+\widetilde\dUU+\widetilde\dUU\,\widetilde\dnu+2\widetilde\AA\left(\widetilde\dUU+\widetilde\dnu+\widetilde\dUU\,\widetilde\dnu\right)+\widetilde\dea{}^\al\left( \widetilde\VV_{\al}-\widetilde\BB_{\al} \right)
	\nnn
	&&
	+\widetilde\BB{}^{\al}\widetilde\VV_{\al}(1+\widetilde\dnu)
	-\widetilde\dUU(\widetilde\BB_{\para}+\widetilde\BB{}^\al\widetilde\dea_{\al})
	+2\widetilde\VV{}^{\al}(\widetilde\CC_{\al \para}+\widetilde\CC_{\al\be}\widetilde\dea{}^{\al})\bigg](\widetilde x_o)\,.
	\enar
	We are interested in the evaluation at $x^\mu_o$ since these coordinates correspond to the observer position in the original chart. Therefore, we expand the coordinates $\widetilde x^\mu_o$ around the original coordinates $x^\mu_o$ according to the coordinate transformation $\widetilde x^\mu(x)$ in Eq.\,($\ref{xcsi}$). At second order we obtain
	\bear
	{\widetilde\dnu}(\widetilde x=x_o)&=&-
	\bigg[\xi^\mu\pa_\mu\left(\dnu+\HH T+\AA+\VV_\para-\BBP \right)+{\widetilde\AA}+{\widetilde\VV}_\para-{\widetilde\BBP}+\widetilde\dnu\widetilde\AA-\frac12\widetilde\AA^2
	\nnn
	&&
	+\widetilde\AA\,\widetilde\BB_\para+\widetilde\dea_\al\big( \widetilde\VV^\al-\widetilde\BB^\al \big)
	+\frac12 \widetilde\VV^\al\widetilde\VV_\al
	+2\widetilde\CC_{\al\para}\widetilde\VV^\al\bigg](\widetilde x= x_o)   \,,
	\enar
	where we used the linear-order gauge transformation of $\dnu$. Employing the linear-order definition of~$\dhnu$ from Eq.\,($\ref{dhnuexact}$), we can see that the above equation yields
	\bear
	{\widetilde\dnu}(x_o)&=&-
	\bigg[\xi^\mu\pa_\mu\left(\dhnu+\HH T \right)+{\widetilde\AA}+{\widetilde\VV}_\para-{\widetilde\BBP}+\widetilde\dnu\widetilde\AA-\frac12\widetilde\AA^2
	\nnn
	&&
	+\widetilde\AA\,\widetilde\BB_\para+\widetilde\dea_\al\big( \widetilde\VV^\al-\widetilde\BB^\al \big)
	+\frac12 \widetilde\VV^\al\widetilde\VV_\al
	+2\widetilde\CC_{\al\para}\widetilde\VV^\al\bigg](x_o)   \,,
	\enar
	Replacing all the gauge transformations with the results of Section $\ref{sec:gauge}$ we finally derive
	\bear
	{\widetilde\dnu}_o&=&\dnu_o+\bigg[ -\frac{d}{d\rbar}T + T\frac{d}{d\rbar}T' +\HH T\left(1+\dnu -\frac{d}{d\rbar}T\right) 
	-  \mathcal{L}^{\al} \left(\dnu_{,\al}+2 \mathcal{H}T_{,\al}-\frac{d}{d\rbar}T_{,\al}  \right)
	\nnn
	&&
	+ \frac12\left(\mathcal{H}^2 - 3\mathcal{H}'\right)T^2 -  2\mathcal{H}TT'
	+T'\dnu-T_{,\al}\dea^\al-T\dnu' \bigg]_o
	\,,
	\enar
	which agrees with the previous computation in Eq.\,($\ref{dnuO}$).
	
	We now want to check that our second-order expression for $\hat k^\mu_o$ gauge transforms consistently. Note that we have a second-order expression for $\dnu_o$ in Eq.\,($\ref{dnu2o}$), but we do not have a complete second-order expression for $\dea^\al_o$ (see Eq.\,($\ref{dea2o}$)). However, its relevant contribution to second-order is given by $S_{ij}$ in Eq.\,($\ref{tetrade2}$), so we only need to prove the consistency of such expression.
	
	To derive the second-order expectations we consider that at any spacetime point $x^\mu$ the tetrad components $[e_a]^\mu$ transform as a four-vector under diffeomorphisms of $\mM$. The corresponding gauge transformation can be read from Eq.\,($\ref{gauge2}$). Using the definition in Eq.\,($\ref{tetrade}$) for the space-like tetrad components, we obtain at second order:
	\bear\label{A1}
	\widetilde{[e_i]}{}^\alpha &=&[e_i]^\alpha+\frac{1}{a} \bigg\{ 
	\delta^\alpha_i (\mathcal{H} T) + \delta^\beta_i 
	\LL^\alpha{}_{,\be} +
	\delta^\beta_i  \LL'^{\al} (\VV_\beta - \BB_\beta)
	- \LL^{ \alpha}{}_{,\be}
	\left[\delta^\gamma_i (\CC^\be_\ga+C^{[\be}{}_{,\ga]}) + \epsilon^{\beta}{}_{ik} \Omega^k  \right]
	\nnn
	&&
	+ \big( T\partial_\eta + \LL^{\beta}\partial_\beta - \mathcal{H} T \big) 
	\left[ \delta^\gamma_i (\CC^\al_\ga+C^{[\al}{}_{,\ga]}) + \epsilon^{\alpha}{}_{ik} \Omega^k\right]
	-\mathcal{H} T T'\delta^\alpha_i -  \mathcal{H} \LL^{\beta} T_{,\be} \delta^\alpha_i
	\nnn
	&&
	- \frac12(\mathcal{H}' - \mathcal{H}^2) T^2 \delta^\alpha_i -T \LL'^{\alpha}{}_{,\be} \delta^\be_i
	-\LL^{ \ga}\LL^{ \alpha}{}_{,\be\ga} \delta^\be_i + \mathcal{H}T \LL^{\alpha}{}_{,\be} \delta^\be_i   \bigg\}\,,
	\enar
	where both sides are evaluated at the same coordinate value $x_o^\mu$. Taking the symmetric and antisymmetric part of the last equation and contracting the Kronecker deltas, we derive
	\bear\label{Sij}
	{\widetilde S}_{ij}-S_{ij}&=&-\de_{ij}\HH T-\LL_{(i,j)}+\BB_{(j}\LL'_{i)}-\CC'_{ij}T+\HH T\CC_{ij}-\frac12\delta_{ij}\HH^2T^2+\frac12\de_{ij}\HH'T^2
	\nnn
	&&
	+\de_{ij}\HH TT'
	+\LL_{(i,\al}\varepsilon_{j)k}{}^{\al}\Om^k
	-\LL'_{(i}\VV_{j)}-\HH T\LL_{(i,j)}+T\LL'_{(i,j)}-\LL^\al\CC_{ij,\al}
	\nnn
	&&
	+\frac12C^\al{}_{,(i}\LL_{j),\al}+\de_{ij}\HH\LL^\al T_{,\al}+\LL^\al\LL_{(i,j)\al}-\frac12\LL_{(i,\al}C_{j)}{}^{,\al}+\CC_{\al(i}\LL_{j)}{}^{,\al}
	\,,
	\\
	\nnn
	\widetilde{A}_{ij}-A_{ij}&=&-\LL_{[i,j]}+\LL'_{[i}\left(\BB_{j]}-\VV_{j]}\right)-T\left( C'_{[i,j]}-\LL'_{[i,j]}+\varepsilon_{ijk}\Om'^k \right)
	\nnn
	&&
	+\HH T \left( C_{[i,j]}-\LL_{[i,j]}+ \varepsilon_{ijk}\Om^k\right) -\LL^\al\left(C_{[i,j]\al}-\LL_{[i,j]\al}+ \varepsilon_{ijl}\Om^l{}_{,\al}\right)
	\nnn
	&&
	+\frac12\LL_{[i}{}^{,\al}\left(2\CC_{j]\al}+C_{\al,j]}-C_{j],\al}-2\varepsilon_{j]\al l}\Om^l\right)
	\,,
	\enar
	according to the definition of the split in Eq.\,($\ref{split}$). From the gauge transformation $\widetilde A_{ij}$ we can infer the linear-order expression of $A_{ij}$ in Eq.\,($\ref{asim}$) which indeed transforms accordingly considering that $\Om^i$ is gauge-invariant and $C_\al$ at linear order gauge transforms as the vector part of $\CC_{\al\be}$:
	\bear
	\widetilde C_\al=C_\al-L_\al\,.
	\enar
	Finally, directly substituting all the gauge transformations in our second-order expression for $S_{ij}$ in Eq.\,($\ref{tetrade2}$), we recover Eq.\,($\ref{Sij}$), proving the validity of our expression.

	\section{Explicit expressions of the most relevant second-order quantities}
	\label{AppendixB}
	
	In Section $\ref{sec:second}$ we derived all the second-order expressions related to the light propagation from a luminous source to the observer position. In Table $\ref{tab1}$ we provide the references to such equations.
	\renewcommand{\arraystretch}{2}
	\begin{center}	\captionof{table}{References to the second-order equations given in Section $\ref{sec:second}$.} 
		
		\begin{tabular}{|c|c||c|c|}
			\hline
			$\de g$ & Eq.\,($\ref{deg2}$) &  $\de z$  & Eq.\,($\ref{dz2}$) \\ \hline % \hhline{|=|=|=|=|=|} 
			$\GG^\eta$ , $\GG^\al$ & Eq.\,($\ref{GG2}$) & $\dhnu$ &  Eq.\,($\ref{dhnu2}$) \\ \hline  
			$\de u$ &  Eq.\,($\ref{du2}$) & $\de\eta_o$ &  Eq.\,($\ref{lapseconf}$) \\ \hline
			$\de t_o$ &  Eq.\,($\ref{dt2met}$) & $\DT_s$&  Eq.\,($\ref{deta2}$) \\ \hline
			$\de x^\al_o$ &  Eq.\,($\ref{dx2met}$) & $\DX^\al_s$ & Eq.\,($\ref{Dx2}$)\\ \hline
			$\dnu_s$ &  Eq.\,($\ref{dnuint}$) & $\drr$ , $\dtt$ , $\dpp$ &  Eq.\,($\ref{dangoli2}$)\\ \hline
			$\dea^\al_s$ &  Eq.\,($\ref{deaint}$) & $\dDD$ &  Eq.\,($\ref{dDD2}$)\\ \hline
			$\dnu_o$ &  Eq.\,($\ref{dnu2o}$)& $\de V$ &  Eq.\,($\ref{DV2}$) \\ \hline
			$\dea^\al_o$ &  Eq.\,($\ref{dea2o}$)& $\de A$ &  Eq.\,($\ref{DA2}$)\\ \hline
		\end{tabular}
		\label{tab1}
	\end{center}
	Of all these quantities, the most relevant are those related to cosmological observables, i.e., the redshift fluctuation $\de z$, the distortion in the source position $\drr,\dtt,\dpp$, and the perturbation in the physical volume $\de V$ and area $\de A$.
	
	We now present the second-order expressions of these quantities in terms of metric perturbations and observer velocity. For clarity we will keep some of those quantities in the table at the linear order (i.e., they will not be expressed in terms of the metric), but we provide the reference to their expressions.
	Note that the second-order antisymmetric part $A^j{}_i$ of the space-like tetrad perturbation $\de e^\al_i$ never appears.

	Distortion in the source position $(\drr, \dtt, \dpp)$ in Eq.\,($\ref{dangoli2}$)
	\bear
	\drr&=&
	-\rbar_z\Bigg[\VV_\para+n^in_j\de^j_\al\bigg(A^\al{}_i+\CC^\al_{i}+\frac12\BB_i\BB^\al-\frac12\VV_i\VV^\al-\frac12\Om_i\Om^\al+\frac12\de^\al_{i}\Om^k\Om_k-\frac32\CC^\be_i\CC^\al_{\be}
	\nnn
	&&
	-\CC^\be_{(i}\varepsilon^{\al)}{}_{k\be}\Om^k-\frac12C_{\be,(i}\CC^{\al)\be}
	+\frac12C^\be{}_{,(i}\varepsilon^{\al)}{}_{k\be}\Om^k
	-\frac12C_{(i,\be}\varepsilon^{\al)}{}_{k}{}^\be\Om^k
	+\frac18C^{\be}{}_{,i}C_{\be}{}^{,\al}
	\nnn
	&&
	+\frac18C_{i,}{}^\be C^\al{}_{,\be}+\frac12\CC^\be_{(i}C^{\al)}{}_{,\be}
	-\frac14C^\be{}_{,(i}C^{\al)}{}_{,\be} \bigg) - \AA-\VV_\para+\BBP+\frac32\AA^2
	\nnn
	&&
	-\CC_{\al\para}\left( \VV^\al+\BB^\al \right)+\AA\left( \VV_\para-2\BBP\right)-\VV^\al\BB_\al+\frac12\VV_\al\VV^\al+\left(\VV^\al-\BB^\al\right)C_{[\al,\para]}
	\nnn
	&&
	+\varepsilon_{\al i j}\left( \VV^\al-\BB^\al \right)n^i\Om^j\Bigg]\Bigg\rvert_{\bobs}+n_\al\de x_o^\mu(\pa_\mu\de x^\al)\big\rvert_{\bobs}+n_\al\int_0^{\bar t_{\rm o}}dt\,\frac1a\left(\VV^\al+\AA\,\VV^\al-\VV^\be\de x^\al{}_{,\be}\right)\bigg\rvert_{\bar x_t}
	\nnn
	&&
	-\frac{1}{\HH_z}\Bigg[\bigg(- \AA-\VV_\para+\BBP+\frac32\AA^2-\CC_{\al\para}\left( \VV^\al+\BB^\al \right)+\AA\left( \VV_\para-2\BBP\right)-\VV^\al\BB_\al
	\nnn
	&&
	+\frac12\VV_\al\VV^\al+\left(\VV^\al-\BB^\al\right)C_{[\al,\para]}+\varepsilon_{\al i j}\left( \VV^\al-\BB^\al \right)n^i\Om^j\bigg)_{\bobs}+\int_0^{\rbar_z}d\rbar \bigg[\dnu\dnu'-\dea^\al\dnu_{,\al}
	\nnn
	&&
	+\AA'-2\AAP+{\BB}_{\para,\para}+\CCPP'+2\dnu\left(\AA' - \AAP\right)
	- 2\AA\left(\AA'-2\AAP+\BB_{\para,\para}
	+\CC'_{\para\para}\right)
	\nnn
	&&
	-2\dea^\alpha\left(\AA_{,\alpha}
	-\BB_{(\alpha,\para)} -\CC_{\alpha\para}'\right)
	-\BB^\al \left(\AA_{,\alpha}-\BB'_{\alpha}+2\BB_{[\al,\para]} -2\CC'_{\al\para}
	+ 2 \CC_{\alpha \para,\para}
	-\CC_{\para\para,\al}\right)
	\bigg]\bigg\rvert_{\bar x_{\rbar}}
	\nnn
	&&
	+\AA+\VV_\para-\BB_\para+\dnu\AA-\frac12\AA^2+\AA\BB_\para+\dea_\al\left( \VV^\al-\BB^\al \right)
	+\frac12 \VV^\al\VV_\al
	+2\CC_{\al\para}\VV^\al
	\nnn
	&&
	+\frac12\left(\HH^2_{\bobs}+\HH'_{\bobs}\right)\de\eta_o^2
	+\HH_{\bobs}\de\eta_o\dhnu
	+\DX^\mu_s(\pa_\mu\dhnu)\big\rvert_{\bar x_z} \Bigg]
	+\frac12\frac{\HH^2_z+\HH'_z}{\HH_z}\left(\frac{\de z}{\HH_z}\right)^2
	\nnn
	&&
	-\Bigg(\frac{\HH_{\bobs}}{\HH_z}-1\Bigg)\Bigg[\de x^\mu_o(\pa_\mu\de t)\big\rvert_{\bobs}+\int_0^{\bar t_{\rm o}}dt\left(-\AA+\frac12\AA^2
	+\frac12\VV^\al\VV_\al-\VV^\al\BB_\al-\frac1a\VV^\al\de t_{,\al}\right)\bigg\rvert_{\bar x_t} \Bigg]
	\nnn
	&&
	+\frac12\HH_{\bobs}\de t_o^2+n_\al\left[\DX_{\rbar}^\mu\,\pa_\mu\left( \DX^\al+n^\al\DT\right)\big\rvert_{\bar x_{\rbar}}\right]^z_{\bobs}
	+n_\al\int_0^{\rbar_z}d\rbar\Bigg[ -\dea^\be\left(\DX^\al{}_{,\be}+n^\al\DT_{,\be}\right)
	\nnn
	&&
	+\dnu\left(\DX'^\al+n^\al\DT'\right)
	-\left(\rbar_z-\rbar\right)\bigg(-\dnu\dea'^\al+\dea^\be\dea^\al{}_{,\be}+\AA^{,\alpha}-\BB'^{\alpha}
	-  \BB_{\para}^{\;\;,\alpha} + \BB^\alpha_{\;\;,\para}
	\nnn
	&&
	- 2 \CC'^{\alpha}_\para+
	2 \CC^\alpha_{\para,\para} - \CC_{\para\para}^{\;\;\;\;,\alpha}
	+\dnu\left(2\AA^{,\alpha} - 2\BB'^{\alpha}-\BB_\para^{\;\;,\alpha}
	+ \BB^\alpha_{\;\;,\para}- 2 \CC'^{\alpha}_{\para} \right) \nnn
	&&
	-\dea^\ga\left(\BB_\ga^{\;\;,\alpha}
	- \BB^\alpha_{\;\;,\ga}+ 2 \CC'^{\alpha}_\ga 
	-4 \CC^\alpha_{(\para,\gamma)}+2\CC_{\para\gamma}^{\;\;\;\;,\alpha}
	\right)+ \BB^\alpha \left(\AA'+ \BB_{\para,\para} + \CC'_{\para\para}- 2 \AA_{,\para}\right)\nonumber \\
	&&
	- 2\CC^{\alpha\ga}\left(  2 \CC_{\ga\para,\para}
	- \CC_{\para\para,\ga}+\AA_{,\ga} 
	-\BB'_{\ga} -2\BB_{[\para,\gamma]}
	- 2\CC_{\para\gamma}'\right)+n^\al\bigg(\dnu\dnu'-\dea^\al\dnu_{,\al}+\AA'-2\AAP+{\BB}_{\para,\para}
	\nnn
	&&
	+\CCPP'+2\dnu\left(\AA' - \AAP\right)
	- 2\AA\left(\AA'-2\AAP+\BB_{\para,\para}
	+\CC'_{\para\para}\right)-2\dea^\alpha\left(\AA_{,\alpha}
	-\BB_{(\alpha,\para)} -\CC_{\alpha\para}'\right)
	\nnn
	&&
	-\BB^\al \left(\AA_{,\alpha}-\BB'_{\alpha}+2\BB_{[\al,\para]} -2\CC'_{\al\para}
	+ 2 \CC_{\alpha \para,\para}
	-\CC_{\para\para,\al}\right)\bigg)
	\Bigg]\Bigg\rvert_{\bar x_{\rbar}}
	+\frac{\rbar_z}{2} \left( \dtt^{2}+\sin^2\ttt~\dpp^{2} \right)\,,
	\enar

	\bear
	\dtt&=&\frac1\rbar_z\ttt_\al\Bigg\{
	-\rbar_z\Bigg[\VV^\al+n^i\bigg(A^\al{}_i+\CC^\al_{i}+\frac12\BB_i\BB^\al-\frac12\VV_i\VV^\al-\frac12\Om_i\Om^\al+\frac12\de^\al_{i}\Om^k\Om_k-\frac32\CC^\be_i\CC^\al_{\be}
	\nnn
	&&
	-\CC^\be_{(i}\varepsilon^{\al)}{}_{k\be}\Om^k-\frac12C_{\be,(i}\CC^{\al)\be}
	+\frac12C^\be{}_{,(i}\varepsilon^{\al)}{}_{k\be}\Om^k
	-\frac12C_{(i,\be}\varepsilon^{\al)}{}_{k}{}^\be\Om^k
	+\frac18C^{\be}{}_{,i}C_{\be}{}^{,\al}
	\nnn
	&&
	+\frac18C_{i,}{}^\be C^\al{}_{,\be}+\frac12\CC^\be_{(i}C^{\al)}{}_{,\be}
	-\frac14C^\be{}_{,(i}C^{\al)}{}_{,\be} \bigg)\Bigg]\Bigg\rvert_{\bobs}+\de x_o^\mu(\pa_\mu\de x^\al)\big\rvert_{\bobs}+\left[\DX_{\rbar}^\mu\,\pa_\mu\left( \DX^\al+n^\al\DT\right)\big\rvert_{\bar x_{\rbar}}\right]^z_{\bobs}
	\nnn
	&&
	+\int_0^{\bar t_{\rm o}}dt\,\frac1a\left(\VV^\al+\AA\,\VV^\al-\VV^\be\de x^\al{}_{,\be}\right)\bigg\rvert_{\bar x_t}
	+\int_0^{\rbar_z}d\rbar\Bigg[ -\dea^\be\left(\DX^\al{}_{,\be}+n^\al\DT_{,\be}\right)
	\nnn
	&&
	+\dnu\left(\DX'^\al+n^\al\DT'\right)
	-\left(\rbar_z-\rbar\right)\bigg(-\dnu\dea'^\al+\dea^\be\dea^\al{}_{,\be}+\AA^{,\alpha}-\BB'^{\alpha}
	-  \BB_{\para}^{\;\;,\alpha} + \BB^\alpha_{\;\;,\para}
	\nnn
	&&
	- 2 \CC'^{\alpha}_\para+
	2 \CC^\alpha_{\para,\para} - \CC_{\para\para}^{\;\;\;\;,\alpha}
	+\dnu\left(2\AA^{,\alpha} - 2\BB'^{\alpha}-\BB_\para^{\;\;,\alpha}
	+ \BB^\alpha_{\;\;,\para}- 2 \CC'^{\alpha}_{\para} \right) \nnn
	&&
	-\dea^\ga\left(\BB_\ga^{\;\;,\alpha}
	- \BB^\alpha_{\;\;,\ga}+ 2 \CC'^{\alpha}_\ga 
	-4 \CC^\alpha_{(\para,\gamma)}+2\CC_{\para\gamma}^{\;\;\;\;,\alpha}
	\right)+ \BB^\alpha \left(\AA'+ \BB_{\para,\para} + \CC'_{\para\para}- 2 \AA_{,\para}\right)\nonumber \\
	&&
	- 2\CC^{\alpha\ga}\left(  2 \CC_{\ga\para,\para}
	- \CC_{\para\para,\ga}+\AA_{,\ga} 
	-\BB'_{\ga} -2\BB_{[\para,\gamma]}
	- 2\CC_{\para\gamma}'\right)
	\Bigg]\Bigg\rvert_{\bar x_{\rbar}}
	\Bigg\}
	-\frac{\drr}{\rbar_z}\dtt +\frac{1}{2}\cos\ttt\sin\ttt~\dpp^2\,.
	\enar
	
	The explicit expressions of the linear-order terms are given in Eqs.\,($\ref{dxlin}$), ($\ref{lapseconf}$), ($\ref{dk1s}$), ($\ref{dk1o}$), ($\ref{dhnu2}$), ($\ref{Deta1}$), ($\ref{Dx1}$), ($\ref{dz2}$), ($\ref{deltar}$), ($\ref{deltat}$), ($\ref{deltap}$). According to Eq.\,($\ref{dangoli2}$) the expression for $\dpp$ is analogous to that of $\dtt$ replacing $\ttt_\al$ with $\pp_\al/\sin\ttt$ and $-\frac{\drr}{\rbar_z}\dtt +\frac{1}{2}\cos\ttt\sin\ttt~\dpp^2$ with $-\frac{\drr}{\rbar_z}\dpp-\rbar_z\cot\ttt~\dtt\dpp$.

	Fluctuation in the observed redshift $\de z$ in Eq.\,($\ref{dz2}$)
	\bear\label{B1}
	\de z&=&\HH_{\bobs}\Bigg(\de x^\mu_o(\pa_\mu\de t)\big\rvert_{\bobs}-\frac12\HH_{\bobs}\de t_o^2+\int_0^{\bar t_{\rm o}}dt\left(-\AA+\frac12\AA^2
	+\frac12\VV^\al\VV_\al-\VV^\al\BB_\al-\frac1a\VV^\al\de t_{,\al}\right)\bigg\rvert_{\bar x_t} \Bigg)
	\nnn
	&&
	+\bigg(- \AA-\VV_\para+\BBP+\frac32\AA^2-\CC_{\al\para}\left( \VV^\al+\BB^\al \right)+\AA\left( \VV_\para-2\BBP\right)-\VV^\al\BB_\al
	\nnn
	&&
	+\frac12\VV_\al\VV^\al+\left(\VV^\al-\BB^\al\right)C_{[\al,\para]}+\varepsilon_{\al i j}\left( \VV^\al-\BB^\al \right)n^i\Om^j\bigg)_{\bobs}+\int_0^{\rbar_z}d\rbar \bigg[\dnu\dnu'-\dea^\al\dnu_{,\al}
	\nnn
	&&
	+\AA'-2\AAP+{\BB}_{\para,\para}+\CCPP'+2\dnu\left(\AA' - \AAP\right)
	- 2\AA\left(\AA'-2\AAP+\BB_{\para,\para}
	+\CC'_{\para\para}\right)
	\nnn
	&&
	-2\dea^\alpha\left(\AA_{,\alpha}
	-\BB_{(\alpha,\para)} -\CC_{\alpha\para}'\right)
	-\BB^\al \left(\AA_{,\alpha}-\BB'_{\alpha}+2\BB_{[\al,\para]} -2\CC'_{\al\para}
	+ 2 \CC_{\alpha \para,\para}
	-\CC_{\para\para,\al}\right)
	\bigg]\bigg\rvert_{\bar x_{\rbar}}
	\nnn
	&&
	+\AA+\VV_\para-\BB_\para+\dnu\AA-\frac12\AA^2+\AA\BB_\para+\dea_\al\left( \VV^\al-\BB^\al \right)
	+\frac12 \VV^\al\VV_\al
	+2\CC_{\al\para}\VV^\al
	\nnn
	&&
	+\frac12\left(\HH^2_{\bobs}+\HH'_{\bobs}\right)\de\eta_o^2
	+\HH_{\bobs}\de\eta_o\dhnu
	+\DX^\mu_s(\pa_\mu\dhnu)\big\rvert_{\bar x_z}\,.
	\enar
	The explicit expressions of the linear-order terms are given in Eqs.\,($\ref{dxlin}$), ($\ref{lapseconf}$), ($\ref{dk1s}$), ($\ref{dk1o}$), ($\ref{dhnu2}$), ($\ref{Deta1}$), ($\ref{Dx1}$).

	Perturbation in the physical volume $\de V$ occupied by the source in Eq.\,($\ref{DV2}$)
	\bear
	\dV&=& 3\Bigg[ \HH_{\bobs}\Bigg(\de x^\mu_o(\pa_\mu\de t)\big\rvert_{\bobs}-\frac12\HH_{\bobs}\de t_o^2+\int_0^{\bar t_{\rm o}}dt\left(-\AA+\frac12\AA^2
	+\frac12\VV^\al\VV_\al-\VV^\al\BB_\al-\frac1a\VV^\al\de t_{,\al}\right)\bigg\rvert_{\bar x_t} \Bigg)
	\nnn
	&&
	+\bigg(- \AA-\VV_\para+\BBP+\frac32\AA^2-\CC_{\al\para}\left( \VV^\al+\BB^\al \right)+\AA\left( \VV_\para-2\BBP\right)-\VV^\al\BB_\al
	\nnn
	&&
	+\frac12\VV_\al\VV^\al+\left(\VV^\al-\BB^\al\right)C_{[\al,\para]}+\varepsilon_{\al i j}\left( \VV^\al-\BB^\al \right)n^i\Om^j\bigg)_{\bobs}+\int_0^{\rbar_z}d\rbar \bigg[\dnu\dnu'-\dea^\al\dnu_{,\al}
	\nnn
	&&
	+\AA'-2\AAP+{\BB}_{\para,\para}+\CCPP'+2\dnu\left(\AA' - \AAP\right)
	- 2\AA\left(\AA'-2\AAP+\BB_{\para,\para}
	+\CC'_{\para\para}\right)
	\nnn
	&&
	-2\dea^\alpha\left(\AA_{,\alpha}
	-\BB_{(\alpha,\para)} -\CC_{\alpha\para}'\right)
	-\BB^\al \left(\AA_{,\alpha}-\BB'_{\alpha}+2\BB_{[\al,\para]} -2\CC'_{\al\para}
	+ 2 \CC_{\alpha \para,\para}
	-\CC_{\para\para,\al}\right)
	\bigg]\bigg\rvert_{\bar x_{\rbar}}
	\nnn
	&&
	+\AA+\VV_\para-\BB_\para+\dnu\AA-\frac12\AA^2+\AA\BB_\para+\dea_\al\left( \VV^\al-\BB^\al \right)
	+\frac12 \VV^\al\VV_\al
	+2\CC_{\al\para}\VV^\al
	\nnn
	&&
	+\frac12\left(\HH^2_{\bobs}+\HH'_{\bobs}\right)\de\eta_o^2
	+\HH_{\bobs}\de\eta_o\dhnu
	+\DX^\mu_s(\pa_\mu\dhnu)\big\rvert_{\bar x_z} \Bigg]+\AA+\CC^\alpha_\alpha-{1\over2}\AA^2+\frac12\BB^\alpha\BB_\alpha
	+\AA~\CC^\alpha_\alpha
	\nnn
	&&
	+{1\over2}\CC^\alpha_\alpha\CC^\be_\be-
	\CC^{\alpha\beta}\CC_{\alpha\beta}+H_z\frac{\rbar_z}{2} {\pa\over\pa z}\left( \dtt^{2}+\sin^2\ttt~\dpp^{2} \right)+\frac{3}{2} \left( \dtt^{2}+\sin^2\ttt~\dpp^{2} \right)
	\nnn
	&&
	+\left({2\over\rbar_z}+H_z{\pa\over\pa z}\right)\Bigg\{
	-\rbar_z\Bigg[\VV_\para+n^in_j\de^j_\al\bigg(A^\al{}_i+\CC^\al_{i}+\frac12\BB_i\BB^\al-\frac12\VV_i\VV^\al-\frac12\Om_i\Om^\al
	\nnn
	&&
	+\frac12\de^\al_{i}\Om^k\Om_k-\frac32\CC^\be_i\CC^\al_{\be}
	-\CC^\be_{(i}\varepsilon^{\al)}{}_{k\be}\Om^k-\frac12C_{\be,(i}\CC^{\al)\be}
	+\frac12C^\be{}_{,(i}\varepsilon^{\al)}{}_{k\be}\Om^k
	-\frac12C_{(i,\be}\varepsilon^{\al)}{}_{k}{}^\be\Om^k
	\nnn
	&&
	+\frac18C^{\be}{}_{,i}C_{\be}{}^{,\al}
	+\frac18C_{i,}{}^\be C^\al{}_{,\be}+\frac12\CC^\be_{(i}C^{\al)}{}_{,\be}
	-\frac14C^\be{}_{,(i}C^{\al)}{}_{,\be} \bigg) - \AA-\VV_\para+\BBP+\frac32\AA^2
	\nnn
	&&
	-\CC_{\al\para}\left( \VV^\al+\BB^\al \right)+\AA\left( \VV_\para-2\BBP\right)-\VV^\al\BB_\al+\frac12\VV_\al\VV^\al+\left(\VV^\al-\BB^\al\right)C_{[\al,\para]}
	\nnn
	&&
	+\varepsilon_{\al i j}\left( \VV^\al-\BB^\al \right)n^i\Om^j\Bigg]\Bigg\rvert_{\bobs}+n_\al\de x_o^\mu(\pa_\mu\de x^\al)\big\rvert_{\bobs}+n_\al\int_0^{\bar t_{\rm o}}dt\,\frac1a\left(\VV^\al+\AA\,\VV^\al-\VV^\be\de x^\al{}_{,\be}\right)\bigg\rvert_{\bar x_t}
	\nnn
	&&
	-\frac{1}{\HH_z}\Bigg[\bigg(- \AA-\VV_\para+\BBP+\frac32\AA^2-\CC_{\al\para}\left( \VV^\al+\BB^\al \right)+\AA\left( \VV_\para-2\BBP\right)-\VV^\al\BB_\al
	\nnn
	&&
	+\frac12\VV_\al\VV^\al+\left(\VV^\al-\BB^\al\right)C_{[\al,\para]}+\varepsilon_{\al i j}\left( \VV^\al-\BB^\al \right)n^i\Om^j\bigg)_{\bobs}+\int_0^{\rbar_z}d\rbar \bigg(\dnu\dnu'-\dea^\al\dnu_{,\al}
	\nnn
	&&
	+\AA'-2\AAP+{\BB}_{\para,\para}+\CCPP'+2\dnu\left(\AA' - \AAP\right)
	- 2\AA\left(\AA'-2\AAP+\BB_{\para,\para}
	+\CC'_{\para\para}\right)
	\nnn
	&&
	-2\dea^\alpha\left(\AA_{,\alpha}
	-\BB_{(\alpha,\para)} -\CC_{\alpha\para}'\right)
	-\BB^\al \left(\AA_{,\alpha}-\BB'_{\alpha}+2\BB_{[\al,\para]} -2\CC'_{\al\para}
	+ 2 \CC_{\alpha \para,\para}
	-\CC_{\para\para,\al}\right)
	\bigg)\bigg\rvert_{\bar x_{\rbar}}
	\nnn
	&&
	+\AA+\VV_\para-\BB_\para+\dnu\AA-\frac12\AA^2+\AA\BB_\para+\dea_\al\left( \VV^\al-\BB^\al \right)
	+\frac12 \VV^\al\VV_\al
	+2\CC_{\al\para}\VV^\al
	\nnn
	&&
	+\frac12\left(\HH^2_{\bobs}+\HH'_{\bobs}\right)\de\eta_o^2
	+\HH_{\bobs}\de\eta_o\dhnu
	+\DX^\mu_s(\pa_\mu\dhnu)\big\rvert_{\bar x_z} \Bigg]
	+\frac12\frac{\HH^2_z+\HH'_z}{\HH_z}\left(\frac{\de z}{\HH_z}\right)^2
	\nnn
	&&
	-\Bigg(\frac{\HH_{\bobs}}{\HH_z}-1\Bigg)\Bigg[\de x^\mu_o(\pa_\mu\de t)\big\rvert_{\bobs}+\int_0^{\bar t_{\rm o}}dt\left(-\AA+\frac12\AA^2
	+\frac12\VV^\al\VV_\al-\VV^\al\BB_\al-\frac1a\VV^\al\de t_{,\al}\right)\bigg\rvert_{\bar x_t} \Bigg]
	\nnn
	&&
	+\frac12\HH_{\bobs}\de t_o^2+n_\al\left[\DX_{\rbar}^\mu\,\pa_\mu\left( \DX^\al+n^\al\DT\right)\big\rvert_{\bar x_{\rbar}}\right]^z_{\bobs}
	+n_\al\int_0^{\rbar_z}d\rbar\Bigg[ -\dea^\be\left(\DX^\al{}_{,\be}+n^\al\DT_{,\be}\right)\nonumber
	\nonumber
	\enar
	\bear
	\quad&&
	+\dnu\left(\DX'^\al+n^\al\DT'\right)
	-\left(\rbar_z-\rbar\right)\bigg(-\dnu\dea'^\al+\dea^\be\dea^\al{}_{,\be}+\AA^{,\alpha}-\BB'^{\alpha}
	-  \BB_{\para}^{\;\;,\alpha} + \BB^\alpha_{\;\;,\para}
	\nnn
	&&
	- 2 \CC'^{\alpha}_\para+
	2 \CC^\alpha_{\para,\para} - \CC_{\para\para}^{\;\;\;\;,\alpha}
	+\dnu\left(2\AA^{,\alpha} - 2\BB'^{\alpha}-\BB_\para^{\;\;,\alpha}
	+ \BB^\alpha_{\;\;,\para}- 2 \CC'^{\alpha}_{\para} \right) \nnn
	&&
	-\dea^\ga\left(\BB_\ga^{\;\;,\alpha}
	- \BB^\alpha_{\;\;,\ga}+ 2 \CC'^{\alpha}_\ga 
	-4 \CC^\alpha_{(\para,\gamma)}+2\CC_{\para\gamma}^{\;\;\;\;,\alpha}
	\right)+ \BB^\alpha \left(\AA'+ \BB_{\para,\para} + \CC'_{\para\para}- 2 \AA_{,\para}\right)\nonumber \\
	&&
	- 2\CC^{\alpha\ga}\left(  2 \CC_{\ga\para,\para}
	- \CC_{\para\para,\ga}+\AA_{,\ga} 
	-\BB'_{\ga} -2\BB_{[\para,\gamma]}
	- 2\CC_{\para\gamma}'\right)+n^\al\bigg(\dnu\dnu'-\dea^\al\dnu_{,\al}
	\nnn
	&&
	+\AA'-2\AAP+{\BB}_{\para,\para}+\CCPP'+2\dnu\left(\AA' - \AAP\right)
	- 2\AA\left(\AA'-2\AAP+\BB_{\para,\para}
	+\CC'_{\para\para}\right)
	\nnn
	&&
	-2\dea^\alpha\left(\AA_{,\alpha}
	-\BB_{(\alpha,\para)} -\CC_{\alpha\para}'\right)
	-\BB^\al \left(\AA_{,\alpha}-\BB'_{\alpha}+2\BB_{[\al,\para]} -2\CC'_{\al\para}
	+ 2 \CC_{\alpha \para,\para}
	-\CC_{\para\para,\al}\right)\bigg)
	\Bigg]\Bigg\rvert_{\bar x_{\rbar}}
	\Bigg\}
	\nnn
	&&
	+\left(\cot\ttt+{\pa\over\pa \ttt}\right)\Biggr(
	\frac1\rbar_z\ttt_\al\Bigg\{
	-\rbar_z\Bigg[\VV^\al+n^i\bigg(A^\al{}_i+\CC^\al_{i}+\frac12\BB_i\BB^\al-\frac12\VV_i\VV^\al-\frac12\Om_i\Om^\al+\frac12\de^\al_{i}\Om^k\Om_k
	\nnn
	&&
	-\frac32\CC^\be_i\CC^\al_{\be}
	-\CC^\be_{(i}\varepsilon^{\al)}{}_{k\be}\Om^k-\frac12C_{\be,(i}\CC^{\al)\be}
	+\frac12C^\be{}_{,(i}\varepsilon^{\al)}{}_{k\be}\Om^k
	-\frac12C_{(i,\be}\varepsilon^{\al)}{}_{k}{}^\be\Om^k
	+\frac18C^{\be}{}_{,i}C_{\be}{}^{,\al}
	\nnn
	&&
	+\frac18C_{i,}{}^\be C^\al{}_{,\be}+\frac12\CC^\be_{(i}C^{\al)}{}_{,\be}
	-\frac14C^\be{}_{,(i}C^{\al)}{}_{,\be} \bigg)\Bigg]\Bigg\rvert_{\bobs}+\de x_o^\mu(\pa_\mu\de x^\al)\big\rvert_{\bobs}+\left[\DX_{\rbar}^\mu\,\pa_\mu\left( \DX^\al+n^\al\DT\right)\big\rvert_{\bar x_{\rbar}}\right]^z_{\bobs}
	\nnn
	&&
	+\int_0^{\bar t_{\rm o}}dt\,\frac1a\left(\VV^\al+\AA\,\VV^\al-\VV^\be\de x^\al{}_{,\be}\right)\bigg\rvert_{\bar x_t}
	+\int_0^{\rbar_z}d\rbar\Bigg[ -\dea^\be\left(\DX^\al{}_{,\be}+n^\al\DT_{,\be}\right)
	\nnn
	&&
	+\dnu\left(\DX'^\al+n^\al\DT'\right)
	-\left(\rbar_z-\rbar\right)\bigg(-\dnu\dea'^\al+\dea^\be\dea^\al{}_{,\be}+\AA^{,\alpha}-\BB'^{\alpha}
	-  \BB_{\para}^{\;\;,\alpha} + \BB^\alpha_{\;\;,\para}
	\nnn
	&&
	- 2 \CC'^{\alpha}_\para+
	2 \CC^\alpha_{\para,\para} - \CC_{\para\para}^{\;\;\;\;,\alpha}
	+\dnu\left(2\AA^{,\alpha} - 2\BB'^{\alpha}-\BB_\para^{\;\;,\alpha}
	+ \BB^\alpha_{\;\;,\para}- 2 \CC'^{\alpha}_{\para} \right) \nnn
	&&
	-\dea^\ga\left(\BB_\ga^{\;\;,\alpha}
	- \BB^\alpha_{\;\;,\ga}+ 2 \CC'^{\alpha}_\ga 
	-4 \CC^\alpha_{(\para,\gamma)}+2\CC_{\para\gamma}^{\;\;\;\;,\alpha}
	\right)+ \BB^\alpha \left(\AA'+ \BB_{\para,\para} + \CC'_{\para\para}- 2 \AA_{,\para}\right)\nonumber \\
	&&
	- 2\CC^{\alpha\ga}\left(  2 \CC_{\ga\para,\para}
	- \CC_{\para\para,\ga}+\AA_{,\ga} 
	-\BB'_{\ga} -2\BB_{[\para,\gamma]}
	- 2\CC_{\para\gamma}'\right)
	\Bigg]\Bigg\rvert_{\bar x_{\rbar}}
	\Bigg\}
	\Biggr)
	\nnn
	&&
	+{\pa\over\pa \pp}\Biggr(
	\frac1{\rbar_z\sin\ttt}\pp_\al\Bigg\{
	-\rbar_z\Bigg[\VV^\al+n^i\bigg(A^\al{}_i+\CC^\al_{i}+\frac12\BB_i\BB^\al-\frac12\VV_i\VV^\al-\frac12\Om_i\Om^\al+\frac12\de^\al_{i}\Om^k\Om_k-\frac32\CC^\be_i\CC^\al_{\be}
	\nnn
	&&
	-\CC^\be_{(i}\varepsilon^{\al)}{}_{k\be}\Om^k-\frac12C_{\be,(i}\CC^{\al)\be}
	+\frac12C^\be{}_{,(i}\varepsilon^{\al)}{}_{k\be}\Om^k
	-\frac12C_{(i,\be}\varepsilon^{\al)}{}_{k}{}^\be\Om^k
	+\frac18C^{\be}{}_{,i}C_{\be}{}^{,\al}
	\nnn
	&&
	+\frac18C_{i,}{}^\be C^\al{}_{,\be}+\frac12\CC^\be_{(i}C^{\al)}{}_{,\be}
	-\frac14C^\be{}_{,(i}C^{\al)}{}_{,\be} \bigg)\Bigg]\Bigg\rvert_{\bobs}+\de x_o^\mu(\pa_\mu\de x^\al)\big\rvert_{\bobs}+\left[\DX_{\rbar}^\mu\,\pa_\mu\left( \DX^\al+n^\al\DT\right)\big\rvert_{\bar x_{\rbar}}\right]^z_{\bobs}
	\nnn
	&&
	+\int_0^{\bar t_{\rm o}}dt\,\frac1a\left(\VV^\al+\AA\,\VV^\al-\VV^\be\de x^\al{}_{,\be}\right)\bigg\rvert_{\bar x_t}
	+\int_0^{\rbar_z}d\rbar\Bigg[ -\dea^\be\left(\DX^\al{}_{,\be}+n^\al\DT_{,\be}\right)
	\nnn
	&&
	+\dnu\left(\DX'^\al+n^\al\DT'\right)
	-\left(\rbar_z-\rbar\right)\bigg(-\dnu\dea'^\al+\dea^\be\dea^\al{}_{,\be}+\AA^{,\alpha}-\BB'^{\alpha}
	-  \BB_{\para}^{\;\;,\alpha} + \BB^\alpha_{\;\;,\para}
	\nnn
	&&
	- 2 \CC'^{\alpha}_\para+
	2 \CC^\alpha_{\para,\para} - \CC_{\para\para}^{\;\;\;\;,\alpha}
	+\dnu\left(2\AA^{,\alpha} - 2\BB'^{\alpha}-\BB_\para^{\;\;,\alpha}
	+ \BB^\alpha_{\;\;,\para}- 2 \CC'^{\alpha}_{\para} \right) \nnn
	&&
	-\dea^\ga\left(\BB_\ga^{\;\;,\alpha}
	- \BB^\alpha_{\;\;,\ga}+ 2 \CC'^{\alpha}_\ga 
	-4 \CC^\alpha_{(\para,\gamma)}+2\CC_{\para\gamma}^{\;\;\;\;,\alpha}
	\right)+ \BB^\alpha \left(\AA'+ \BB_{\para,\para} + \CC'_{\para\para}- 2 \AA_{,\para}\right)\nonumber \\
	&&
	- 2\CC^{\alpha\ga}\left(  2 \CC_{\ga\para,\para}
	- \CC_{\para\para,\ga}+\AA_{,\ga} 
	-\BB'_{\ga} -2\BB_{[\para,\gamma]}
	- 2\CC_{\para\gamma}'\right)
	\Bigg]\Bigg\rvert_{\bar x_{\rbar}}
	\Bigg\}
	\Biggr)\nonumber
	\enar
	\bear
	&&
	+{\pa\over\pa \ttt}\dtt{\pa\over\pa \pp}\dpp-{\pa\over\pa \pp}\dtt{\pa\over\pa \ttt}\dpp
	+\cot\ttt\,\dtt\left({\pa\over\pa \ttt}\dtt+{\pa\over\pa \pp}\dpp\right)
	-\frac12\dtt^2
	\nnn
	&&
	+\AA+\frac32 \AA^2 + {1\over 2}
	\VV^\alpha \VV_\alpha-\VV_{\al}\BB^{\al}
	+\VV_\para+3\,\dz\dg+3~\dz^2
	-\AA\left(2\,{\drr\over\rbar_z}-2\kappa
	+H_z{\pa\over\pa z}\drr\right)
	\nnn
	&&
	+{\drr^2\over\rbar^2_z}+2{\drr\over\rbar_z}
	\left(H_z{\pa\over\pa z}\drr-2\kappa\right)
	-2\kappa H_z{\pa\over\pa z}\drr
	-H_z{\pa\over\pa z}\de\ttt{\pa\over\pa \ttt}\drr
	-H_z{\pa\over\pa z}\dpp{\pa\over\pa \pp}\drr+\VV_\theta\dtt
	\nnn
	&&
	+\VV_\para\left(2{\drr\over\rbar_z}
	-2\kappa-H_z{\pa\over\pa z}\DT\right)
	-{1\over\rbar_z}\left(\VV_\theta{\pa\over\pa \ttt}
	+{\VV_\phi\over\sin\ttt}{\pa\over\pa \pp}\right)
	\left(\drr+\DT\right)+\VV_\phi\sin\ttt\dpp
	\nonumber \\
	&&
	+(\dg+3~\dz)
	\left(2{\drr\over\rbar_z}-2\kappa+H_z{\pa\over\pa z}\drr
	-\AA+\VV_\para\right)
	+\DX^\mu\pa_\mu\left( 3\,\dz+\dg
	-\AA+\VV_\para \right)~.
	\enar
	The explicit expressions of the linear-order terms are given in Eqs.\,($\ref{dxlin}$), ($\ref{lapseconf}$), ($\ref{dk1s}$), ($\ref{dk1o}$), ($\ref{dhnu2}$), ($\ref{Deta1}$), ($\ref{Dx1}$), ($\ref{dz2}$), ($\ref{deltar}$), ($\ref{deltat}$), ($\ref{deltap}$), ($\ref{deg2}$).
	
	Perturbation in the physical area $\de A$ occupied by the source in Eq.\,($\ref{DA2}$)
	\bear
	\de A&=& 2\Bigg[ \HH_{\bobs}\Bigg(\de x^\mu_o(\pa_\mu\de t)\big\rvert_{\bobs}-\frac12\HH_{\bobs}\de t_o^2+\int_0^{\bar t_{\rm o}}dt\left(-\AA+\frac12\AA^2
	+\frac12\VV^\al\VV_\al-\VV^\al\BB_\al-\frac1a\VV^\al\de t_{,\al}\right)\bigg\rvert_{\bar x_t} \Bigg)
	\nnn
	&&
	+\bigg(- \AA-\VV_\para+\BBP+\frac32\AA^2-\CC_{\al\para}\left( \VV^\al+\BB^\al \right)+\AA\left( \VV_\para-2\BBP\right)-\VV^\al\BB_\al
	\nnn
	&&
	+\frac12\VV_\al\VV^\al+\left(\VV^\al-\BB^\al\right)C_{[\al,\para]}+\varepsilon_{\al i j}\left( \VV^\al-\BB^\al \right)n^i\Om^j\bigg)_{\bobs}+\int_0^{\rbar_z}d\rbar \bigg[\dnu\dnu'-\dea^\al\dnu_{,\al}
	\nnn
	&&
	+\AA'-2\AAP+{\BB}_{\para,\para}+\CCPP'+2\dnu\left(\AA' - \AAP\right)
	- 2\AA\left(\AA'-2\AAP+\BB_{\para,\para}
	+\CC'_{\para\para}\right)
	\nnn
	&&
	-2\dea^\alpha\left(\AA_{,\alpha}
	-\BB_{(\alpha,\para)} -\CC_{\alpha\para}'\right)
	-\BB^\al \left(\AA_{,\alpha}-\BB'_{\alpha}+2\BB_{[\al,\para]} -2\CC'_{\al\para}
	+ 2 \CC_{\alpha \para,\para}
	-\CC_{\para\para,\al}\right)
	\bigg]\bigg\rvert_{\bar x_{\rbar}}
	\nnn
	&&
	+\AA+\VV_\para-\BB_\para+\dnu\AA-\frac12\AA^2+\AA\BB_\para+\dea_\al\left( \VV^\al-\BB^\al \right)
	+\frac12 \VV^\al\VV_\al
	+2\CC_{\al\para}\VV^\al
	\nnn
	&&
	+\frac12\left(\HH^2_{\bobs}+\HH'_{\bobs}\right)\de\eta_o^2
	+\HH_{\bobs}\de\eta_o\dhnu
	+\DX^\mu_s(\pa_\mu\dhnu)\big\rvert_{\bar x_z} \Bigg]- \AA+\frac32 \AA^2 + {1\over 2}
	\VV^\alpha \VV_\alpha-\VV_{\al}\BB^{\al}
	\nnn
	&&
	+\AA+\CC^\alpha_\alpha-{1\over2}\AA^2+\frac12\BB^\alpha\BB_\alpha
	+\AA~\CC^\alpha_\alpha+{1\over2}\CC^\alpha_\alpha\CC^\be_\be-
	\CC^{\alpha\beta}\CC_{\alpha\beta}+\dtt^{2}+\sin^2\ttt~\dpp^{2}
	\nnn
	&&
	+{2\over\rbar_z}\Bigg\{
	-\rbar_z\Bigg[\VV_\para+n^in_j\de^j_\al\bigg(A^\al{}_i+\CC^\al_{i}+\frac12\BB_i\BB^\al-\frac12\VV_i\VV^\al-\frac12\Om_i\Om^\al+\frac12\de^\al_{i}\Om^k\Om_k-\frac32\CC^\be_i\CC^\al_{\be}
	\nnn
	&&
	-\CC^\be_{(i}\varepsilon^{\al)}{}_{k\be}\Om^k-\frac12C_{\be,(i}\CC^{\al)\be}
	+\frac12C^\be{}_{,(i}\varepsilon^{\al)}{}_{k\be}\Om^k
	-\frac12C_{(i,\be}\varepsilon^{\al)}{}_{k}{}^\be\Om^k
	+\frac18C^{\be}{}_{,i}C_{\be}{}^{,\al}
	\nnn
	&&
	+\frac18C_{i,}{}^\be C^\al{}_{,\be}+\frac12\CC^\be_{(i}C^{\al)}{}_{,\be}
	-\frac14C^\be{}_{,(i}C^{\al)}{}_{,\be} \bigg) - \AA-\VV_\para+\BBP+\frac32\AA^2
	\nnn
	&&
	-\CC_{\al\para}\left( \VV^\al+\BB^\al \right)+\AA\left( \VV_\para-2\BBP\right)-\VV^\al\BB_\al+\frac12\VV_\al\VV^\al+\left(\VV^\al-\BB^\al\right)C_{[\al,\para]}
	\nnn
	&&
	+\varepsilon_{\al i j}\left( \VV^\al-\BB^\al \right)n^i\Om^j\Bigg]\Bigg\rvert_{\bobs}+n_\al\de x_o^\mu(\pa_\mu\de x^\al)\big\rvert_{\bobs}+n_\al\int_0^{\bar t_{\rm o}}dt\,\frac1a\left(\VV^\al+\AA\,\VV^\al-\VV^\be\de x^\al{}_{,\be}\right)\bigg\rvert_{\bar x_t}
	\nnn
	&&
	-\frac{1}{\HH_z}\Bigg[\bigg(- \AA-\VV_\para+\BBP+\frac32\AA^2-\CC_{\al\para}\left( \VV^\al+\BB^\al \right)+\AA\left( \VV_\para-2\BBP\right)-\VV^\al\BB_\al
	\nonumber
	\enar
	\bear\quad
	&&
	+\frac12\VV_\al\VV^\al+\left(\VV^\al-\BB^\al\right)C_{[\al,\para]}+\varepsilon_{\al i j}\left( \VV^\al-\BB^\al \right)n^i\Om^j\bigg)_{\bobs}+\int_0^{\rbar_z}d\rbar \bigg(\dnu\dnu'-\dea^\al\dnu_{,\al}
	\nnn
	&&
	+\AA'-2\AAP+{\BB}_{\para,\para}+\CCPP'+2\dnu\left(\AA' - \AAP\right)
	- 2\AA\left(\AA'-2\AAP+\BB_{\para,\para}
	+\CC'_{\para\para}\right)
	\nnn
	&&
	-2\dea^\alpha\left(\AA_{,\alpha}
	-\BB_{(\alpha,\para)} -\CC_{\alpha\para}'\right)
	-\BB^\al \left(\AA_{,\alpha}-\BB'_{\alpha}+2\BB_{[\al,\para]} -2\CC'_{\al\para}
	+ 2 \CC_{\alpha \para,\para}
	-\CC_{\para\para,\al}\right)
	\bigg)\bigg\rvert_{\bar x_{\rbar}}
	\nnn
	&&
	+\AA+\VV_\para-\BB_\para+\dnu\AA-\frac12\AA^2+\AA\BB_\para+\dea_\al\left( \VV^\al-\BB^\al \right)
	+\frac12 \VV^\al\VV_\al
	+2\CC_{\al\para}\VV^\al
	\nnn
	&&
	+\frac12\left(\HH^2_{\bobs}+\HH'_{\bobs}\right)\de\eta_o^2
	+\HH_{\bobs}\de\eta_o\dhnu
	+\DX^\mu_s(\pa_\mu\dhnu)\big\rvert_{\bar x_z} \Bigg]
	+\frac12\frac{\HH^2_z+\HH'_z}{\HH_z}\left(\frac{\de z}{\HH_z}\right)^2
	\nnn
	&&
	-\Bigg(\frac{\HH_{\bobs}}{\HH_z}-1\Bigg)\Bigg[\de x^\mu_o(\pa_\mu\de t)\big\rvert_{\bobs}+\int_0^{\bar t_{\rm o}}dt\left(-\AA+\frac12\AA^2
	+\frac12\VV^\al\VV_\al-\VV^\al\BB_\al-\frac1a\VV^\al\de t_{,\al}\right)\bigg\rvert_{\bar x_t} \Bigg]
	\nnn
	&&
	+\frac12\HH_{\bobs}\de t_o^2+n_\al\left[\DX_{\rbar}^\mu\,\pa_\mu\left( \DX^\al+n^\al\DT\right)\big\rvert_{\bar x_{\rbar}}\right]^z_{\bobs}
	+n_\al\int_0^{\rbar_z}d\rbar\Bigg[ -\dea^\be\left(\DX^\al{}_{,\be}+n^\al\DT_{,\be}\right)\nonumber
	\nnn
	&&
	+\dnu\left(\DX'^\al+n^\al\DT'\right)
	-\left(\rbar_z-\rbar\right)\bigg(-\dnu\dea'^\al+\dea^\be\dea^\al{}_{,\be}+\AA^{,\alpha}-\BB'^{\alpha}
	-  \BB_{\para}^{\;\;,\alpha} + \BB^\alpha_{\;\;,\para}
	\nnn
	&&
	- 2 \CC'^{\alpha}_\para+
	2 \CC^\alpha_{\para,\para} - \CC_{\para\para}^{\;\;\;\;,\alpha}
	+\dnu\left(2\AA^{,\alpha} - 2\BB'^{\alpha}-\BB_\para^{\;\;,\alpha}
	+ \BB^\alpha_{\;\;,\para}- 2 \CC'^{\alpha}_{\para} \right) \nnn
	&&
	-\dea^\ga\left(\BB_\ga^{\;\;,\alpha}
	- \BB^\alpha_{\;\;,\ga}+ 2 \CC'^{\alpha}_\ga 
	-4 \CC^\alpha_{(\para,\gamma)}+2\CC_{\para\gamma}^{\;\;\;\;,\alpha}
	\right)+ \BB^\alpha \left(\AA'+ \BB_{\para,\para} + \CC'_{\para\para}- 2 \AA_{,\para}\right)\nonumber \\
	&&
	- 2\CC^{\alpha\ga}\left(  2 \CC_{\ga\para,\para}
	- \CC_{\para\para,\ga}+\AA_{,\ga} 
	-\BB'_{\ga} -2\BB_{[\para,\gamma]}
	- 2\CC_{\para\gamma}'\right)+n^\al\bigg(\dnu\dnu'-\dea^\al\dnu_{,\al}
	\nnn
	&&
	+\AA'-2\AAP+{\BB}_{\para,\para}+\CCPP'+2\dnu\left(\AA' - \AAP\right)
	- 2\AA\left(\AA'-2\AAP+\BB_{\para,\para}
	+\CC'_{\para\para}\right)
	\nnn
	&&
	-2\dea^\alpha\left(\AA_{,\alpha}
	-\BB_{(\alpha,\para)} -\CC_{\alpha\para}'\right)
	-\BB^\al \left(\AA_{,\alpha}-\BB'_{\alpha}+2\BB_{[\al,\para]} -2\CC'_{\al\para}
	+ 2 \CC_{\alpha \para,\para}
	-\CC_{\para\para,\al}\right)\bigg)
	\Bigg]\Bigg\rvert_{\bar x_{\rbar}}
	\Bigg\}
	\nnn
	&&
	+\left(\cot\ttt+{\pa\over\pa \ttt}\right)\Biggr(
	\frac1\rbar_z\ttt_\al\Bigg\{
	-\rbar_z\Bigg[\VV^\al+n^i\bigg(A^\al{}_i+\CC^\al_{i}+\frac12\BB_i\BB^\al-\frac12\VV_i\VV^\al-\frac12\Om_i\Om^\al+\frac12\de^\al_{i}\Om^k\Om_k
	\nnn
	&&
	-\frac32\CC^\be_i\CC^\al_{\be}
	-\CC^\be_{(i}\varepsilon^{\al)}{}_{k\be}\Om^k-\frac12C_{\be,(i}\CC^{\al)\be}
	+\frac12C^\be{}_{,(i}\varepsilon^{\al)}{}_{k\be}\Om^k
	-\frac12C_{(i,\be}\varepsilon^{\al)}{}_{k}{}^\be\Om^k
	+\frac18C^{\be}{}_{,i}C_{\be}{}^{,\al}
	\nnn
	&&
	+\frac18C_{i,}{}^\be C^\al{}_{,\be}+\frac12\CC^\be_{(i}C^{\al)}{}_{,\be}
	-\frac14C^\be{}_{,(i}C^{\al)}{}_{,\be} \bigg)\Bigg]\Bigg\rvert_{\bobs}+\de x_o^\mu(\pa_\mu\de x^\al)\big\rvert_{\bobs}+\left[\DX_{\rbar}^\mu\,\pa_\mu\left( \DX^\al+n^\al\DT\right)\big\rvert_{\bar x_{\rbar}}\right]^z_{\bobs}
	\nnn
	&&
	+\int_0^{\bar t_{\rm o}}dt\,\frac1a\left(\VV^\al+\AA\,\VV^\al-\VV^\be\de x^\al{}_{,\be}\right)\bigg\rvert_{\bar x_t}
	+\int_0^{\rbar_z}d\rbar\Bigg[ -\dea^\be\left(\DX^\al{}_{,\be}+n^\al\DT_{,\be}\right)
	\nnn
	&&
	+\dnu\left(\DX'^\al+n^\al\DT'\right)
	-\left(\rbar_z-\rbar\right)\bigg(-\dnu\dea'^\al+\dea^\be\dea^\al{}_{,\be}+\AA^{,\alpha}-\BB'^{\alpha}
	-  \BB_{\para}^{\;\;,\alpha} + \BB^\alpha_{\;\;,\para}
	\nnn
	&&
	- 2 \CC'^{\alpha}_\para+
	2 \CC^\alpha_{\para,\para} - \CC_{\para\para}^{\;\;\;\;,\alpha}
	+\dnu\left(2\AA^{,\alpha} - 2\BB'^{\alpha}-\BB_\para^{\;\;,\alpha}
	+ \BB^\alpha_{\;\;,\para}- 2 \CC'^{\alpha}_{\para} \right) \nnn
	&&
	-\dea^\ga\left(\BB_\ga^{\;\;,\alpha}
	- \BB^\alpha_{\;\;,\ga}+ 2 \CC'^{\alpha}_\ga 
	-4 \CC^\alpha_{(\para,\gamma)}+2\CC_{\para\gamma}^{\;\;\;\;,\alpha}
	\right)+ \BB^\alpha \left(\AA'+ \BB_{\para,\para} + \CC'_{\para\para}- 2 \AA_{,\para}\right)\nonumber \\
	&&
	- 2\CC^{\alpha\ga}\left(  2 \CC_{\ga\para,\para}
	- \CC_{\para\para,\ga}+\AA_{,\ga} 
	-\BB'_{\ga} -2\BB_{[\para,\gamma]}
	- 2\CC_{\para\gamma}'\right)
	\Bigg]\Bigg\rvert_{\bar x_{\rbar}}
	\Bigg\}
	\Biggr)
	\nnn
	&&
	-n_\al\bigg(\VV^\al+A^\al{}_i+\CC^\al_{i}+\frac12\BB_i\BB^\al-\frac12\VV_i\VV^\al-\frac12\Om_i\Om^\al+\frac12\de^\al_{i}\Om^k\Om_k-\frac32\CC^\be_i\CC^\al_{\be}-\CC^\be_{(i}\varepsilon^{\al)}{}_{k\be}\Om^k
	\nonumber
	\enar
	\bear
	&&
	-\frac12C_{\be,(i}\CC^{\al)\be}
	+\frac12C^\be{}_{,(i}\varepsilon^{\al)}{}_{k\be}\Om^k
	-\frac12C_{(i,\be}\varepsilon^{\al)}{}_{k}{}^\be\Om^k
	+\frac18C^{\be}{}_{,i}C_{\be}{}^{,\al}+\frac18C_{i,}{}^\be C^\al{}_{,\be}+\frac12\CC^\be_{(i}C^{\al)}{}_{,\be}
	\nnn
	&&
	-\frac14C^\be{}_{,(i}C^{\al)}{}_{,\be} \bigg)_{\bobs}n^i-n_\al\int_0^{\rbar_z}d\rbar \bigg[-\dnu\dea'^\al+\dea^\be\dea^\al{}_{,\be}+\AA^{,\alpha}-\BB'^{\alpha}
	-  \BB_{\para}^{\;\;,\alpha} + \BB^\alpha_{\;\;,\para}
	\nnn
	&&
	- 2 \CC'^{\alpha}_\para+
	2 \CC^\alpha_{\para,\para} - \CC_{\para\para}^{\;\;\;\;,\alpha}
	+\dnu\left(2\AA^{,\alpha} - 2\BB'^{\alpha}-\BB_\para^{\;\;,\alpha}
	+ \BB^\alpha_{\;\;,\para}- 2 \CC'^{\alpha}_{\para} \right) \nnn
	&&
	-\dea^\ga\left(\BB_\ga^{\;\;,\alpha}
	- \BB^\alpha_{\;\;,\ga}+ 2 \CC'^{\alpha}_\ga 
	-4 \CC^\alpha_{(\para,\gamma)}+2\CC_{\para\gamma}^{\;\;\;\;,\alpha}
	\right)+ \BB^\alpha \left(\AA'+ \BB_{\para,\para} + \CC'_{\para\para}- 2 \AA_{,\para}\right)\nonumber \\
	&&
	- 2\CC^{\alpha\ga}\left(  2 \CC_{\ga\para,\para}
	- \CC_{\para\para,\ga}+\AA_{,\ga} 
	-\BB'_{\ga} -2\BB_{[\para,\gamma]}
	- 2\CC_{\para\gamma}'\right)\bigg]\bigg\rvert_{\bar x_{\rbar}}
	\nnn
	&&
	+\VV_\para
	-\Bigg[ \bigg(- \AA-\VV_\para+\BBP+\frac32\AA^2-\CC_{\al\para}\left( \VV^\al+\BB^\al \right)+\AA\left( \VV_\para-2\BBP\right)-\VV^\al\BB_\al
	\nnn
	&&
	+\frac12\VV_\al\VV^\al+\left(\VV^\al-\BB^\al\right)C_{[\al,\para]}+\varepsilon_{\al i j}\left( \VV^\al-\BB^\al \right)n^i\Om^j\bigg)_{\bobs}+\int_0^{\rbar_z}d\rbar \bigg(\dnu\dnu'-\dea^\al\dnu_{,\al}
	\nnn
	&&
	+\AA'-2\AAP+{\BB}_{\para,\para}+\CCPP'+2\dnu\left(\AA' - \AAP\right)
	- 2\AA\left(\AA'-2\AAP+\BB_{\para,\para}
	+\CC'_{\para\para}\right)
	\nnn
	&&
	-2\dea^\alpha\left(\AA_{,\alpha}
	-\BB_{(\alpha,\para)} -\CC_{\alpha\para}'\right)
	-\BB^\al \left(\AA_{,\alpha}-\BB'_{\alpha}+2\BB_{[\al,\para]} -2\CC'_{\al\para}
	+ 2 \CC_{\alpha \para,\para}
	-\CC_{\para\para,\al}\right)
	\bigg)\bigg\rvert_{\bar x_{\rbar}}
	\nnn
	&&
	+\AA+\VV_\para-\BB_\para+\dnu\AA-\frac12\AA^2+\AA\BB_\para+\dea_\al\left( \VV^\al-\BB^\al \right)
	+\frac12 \VV^\al\VV_\al
	+2\CC_{\al\para}\VV^\al \Bigg]
	\nnn
	&&
	-\AA\left(-2\kappa+2\frac{\drr}{\rbar_z}-\CCPP  \right)
	+\VV_\para\left(\BBP-\VV_\para\right)
	\nnn
	&&
	-\frac{1}{\rbar_z}\left( \VV_\ttt{\pa\over\pa \ttt}+\frac{\VV_\pp}{\sin\ttt}{\pa\over\pa \pp} \right)\DT-2\kappa\left(2\frac{\drr}{\rbar_z}-\CCPP\right)-2\frac{\drr}{\rbar_z}\CCPP
	+\frac{\drr^2}{\rbar^2_z}-\frac12\dtt^2-\frac12(\sin\ttt\,\dpp)^2
	\nnn
	&&
	+\dhnu\left(\VV_\para+\CCPP\right)+\dtt\left(\VV_\ttt+\dea_\ttt\right)
	-\frac{1}{\rbar_z}\left[\left(\VV_\ttt+\dea_\ttt\right){\pa\over\pa \ttt}+\frac{1}{\sin\ttt}\left(\VV_\pp+\dea_\pp\right){\pa\over\pa \pp}\right]\drr
	\nnn
	&&
	+\frac{1}{\rbar_z}\dtt{\pa\over\pa \ttt}\drr
	+\sin\ttt\,\dpp\left(\VV_\pp+\dea_\pp\right)+\frac{1}{\rbar_z}\dpp{\pa\over\pa \pp}\drr+\de z^2+2\,\de z\dg
	\nnn
	&&
	+\left( 2\,\de z+\dg\right)\left(  -\AA+2\frac{\drr}{\rbar_z}-2\kappa-\CCPP\right)
	+\DX^\mu\pa_\mu\left( 2\,\de z+\dg-\AA+\VV_\para+\dea_\para-\dhnu \right)\,.\qquad\,\,\,
	\enar
	The explicit expressions of the linear-order terms are given in Eqs.\,($\ref{dxlin}$), ($\ref{lapseconf}$), ($\ref{dk1s}$), ($\ref{dk1o}$), ($\ref{dhnu2}$), ($\ref{Deta1}$), ($\ref{Dx1}$), ($\ref{dz2}$), ($\ref{deltar}$), ($\ref{deltat}$), ($\ref{deltap}$), ($\ref{deg2}$).

	\bibliographystyle{unsrt}
	\bibliography{2ndorder.bib}

\end{document}

%% file: disegnoC.pdf_tex
%% Creator: Inkscape 1.1 (c68e22c387, 2021-05-23), www.inkscape.org
%% Accompanies image file 'disegnoC.pdf' (pdf, eps, ps)
\begingroup%
  \makeatletter%
  \providecommand\color[2][]{%
    \errmessage{(Inkscape) Color is used for the text in Inkscape, but the package 'color.sty' is not loaded}%
    \renewcommand\color[2][]{}%
  }%
  \providecommand\transparent[1]{%
    \errmessage{(Inkscape) Transparency is used (non-zero) for the text in Inkscape, but the package 'transparent.sty' is not loaded}%
    \renewcommand\transparent[1]{}%
  }%
  \providecommand\rotatebox[2]{#2}%
  \newcommand*\fsize{\dimexpr\f@size pt\relax}%
  \newcommand*\lineheight[1]{\fontsize{\fsize}{#1\fsize}\selectfont}%
  \ifx\svgwidth\undefined%
    \setlength{\unitlength}{841.88976378bp}%
    \ifx\svgscale\undefined%
      \relax%
    \else%
      \setlength{\unitlength}{\unitlength * \real{\svgscale}}%
    \fi%
  \else%
    \setlength{\unitlength}{\svgwidth}%
  \fi%
  \global\let\svgwidth\undefined%
  \global\let\svgscale\undefined%
  \makeatother%
  \begin{picture}(1,0.70707071)%
    \lineheight{1}%
    \setlength\tabcolsep{0pt}%
    \put(0,0){\includegraphics[width=\unitlength,page=1]{disegnoC.pdf}}%
    \put(0.63546003,0.58497568){\color[rgb]{0,0,0}\makebox(0,0)[lt]{\lineheight{1.25}\smash{\begin{tabular}[t]{l}$(\mathcal{M},g)$\end{tabular}}}}%
    \put(0.16139791,0.44020379){\color[rgb]{0,0,0}\makebox(0,0)[lt]{\lineheight{1.25}\smash{\begin{tabular}[t]{l}$S$\end{tabular}}}}%
    \put(0.76106766,0.26189795){\color[rgb]{0,0,0}\makebox(0,0)[lt]{\lineheight{1.25}\smash{\begin{tabular}[t]{l}$O$\end{tabular}}}}%
    \put(0.64674377,0.13249849){\color[rgb]{0,0,0}\makebox(0,0)[lt]{\lineheight{1.25}\smash{\begin{tabular}[t]{l}$\mathcal O(\tau)$\end{tabular}}}}%
    \put(0.45756695,0.39790041){\color[rgb]{0,0,0}\makebox(0,0)[lt]{\lineheight{1.25}\smash{\begin{tabular}[t]{l}$\gamma(\Lambda)$\end{tabular}}}}%
    \put(0.25728815,0.53451795){\color[rgb]{0,0,0}\makebox(0,0)[lt]{\lineheight{1.25}\smash{\begin{tabular}[t]{l}$\mathcal S(\tau')$\end{tabular}}}}%
    \put(0,0){\includegraphics[width=\unitlength,page=2]{disegnoC.pdf}}%
    \put(0.69363742,0.44103534){\color[rgb]{0,0,0}\makebox(0,0)[lt]{\lineheight{1.25}\smash{\begin{tabular}[t]{l}$U$\end{tabular}}}}%
    \put(0,0){\includegraphics[width=\unitlength,page=3]{disegnoC.pdf}}%
  \end{picture}%
\endgroup%